# Data Coding Means and Event Coding Means Multiplexed Over the 1000BASE-T PMA Sublayer

Alexander Ivanov

***Abstract*—There is a work studying the linguistic multiplexing case related to the 100BASE-X physical layer, intended to resolve an extra asynchronous flag-like passing channel inside an already deployed Fast Ethernet point-to-point link. In this paper, we are starting a branch of that work, focusing on the Gigabit Ethernet twisted-pair physical layer, type 1000BASE-T.**

## INTRODUCTION

GIGABIT Ethernet, type 1000BASE-T[1] [1] implements a more sophisticated coding, see Table I, comparing with one its Fast Ethernet predecessor, 100BASE-X, considered in [2] and then [3]. The 1000BASE-T physical layer comprises two entities stacked vertically, where the higher one services the PHY client a media independent manner while the lower one directly interacts the physical media.

The service of that lower entity, which is the 1000BASE-T PMA sublayer,[2] is accessible via the PMA service interface, supporting its user with a memory-less, continuous transport stream of four-dimensional symbols. Three bits per a symbol select one-of-eight-possible code subset we refer to as a page of the transport dictionary, and the rest six bits select—within that selected code subset—the certain code point we refer to as a word within the page. Therefore, we assume we employ a means transporting a continuous text word-by-word, one-of-many-possible ($1$ of $625$) symbol-like word a time.

Similar to the higher one, which is the 1000BASE-T PCS,[3] we leverage the lower entity as is, preserving its behavior with no change, by keeping both the rate and the structure of the words seen in between of the entities the same. Further in the paper, we refer to the bits pointing on a page, participating in the multiplexing, and doing for nothing thereof, as the prefix, root, and postfix morphemes of a word, respectively.

The rest of this paper describes the solution set iteratively fitted over the linguistic multiplexing model introduced in [2], building upon the streaming rules introduced in [3].

A manuscript of this work was submitted to IEEE Communications Letters November 26, 2022 and rejected as not being in the scope of the journal.

Please sorry for the author has no time to find this work a new home, peer reviewed or not, except of arXiv, and just hopes there it meets its reader, one or maybe various, whom the author beforehand thanks for their regard.

A. Ivanov is with JSC Continuum, Yaroslavl, the Russian Federation.

[1]Physical Layer for 1000 Mbps networking, introduced and also specified in [1], Clause 40, operates over four Category 5 twisted pairs in parallel.

[2]Physical Medium Attachment sublayer within the 1000BASE-T definition, embodies the respective media dependent interface, 1000BASE-T MDI.

[3]Physical Coding Sublayer on the top of the 1000BASE-T PMA sublayer, provides a gigabit media independent interface, e.g., GMII or RGMII.

TABLE I
ORIGINAL CODING PROPERTIES

<table>
<tr><th>Subset</th><th>Parity</th><th>Contents</th><th>Points Total</th><th></th><th>Data</th><th></th><th>Control</th><th></th><th>Free</th></tr>
<tr><td>S0</td><td>Even</td><td>XXXX+YYYY</td><td>16 + 81 = 97</td><td>=</td><td>64</td><td>+</td><td>4+(2|5)+16</td><td>+</td><td>26</td></tr>
<tr><td>S2</td><td>Even</td><td>XXYY+YYXX</td><td>36 + 36 = 72</td><td>=</td><td>64</td><td>+</td><td>4+(0|0)+00</td><td>+</td><td>4</td></tr>
<tr><td>S4</td><td>Even</td><td>XYYX+YXXY</td><td>36 + 36 = 72</td><td>=</td><td>64</td><td>+</td><td>4+(0|0)+00</td><td>+</td><td>4</td></tr>
<tr><td>S6</td><td>Even</td><td>XYXY+YXYX</td><td>36 + 36 = 72</td><td>=</td><td>64</td><td>+</td><td>4+(0|0)+00</td><td>+</td><td>4</td></tr>
<tr><td>S1</td><td>Odd</td><td>XXXY+YYYX</td><td>24 + 54 = 78</td><td>=</td><td>64</td><td>+</td><td>4+(0|0)+00</td><td>+</td><td>10</td></tr>
<tr><td>S3</td><td>Odd</td><td>XXYX+YYXY</td><td>24 + 54 = 78</td><td>=</td><td>64</td><td>+</td><td>4+(0|0)+00</td><td>+</td><td>10</td></tr>
<tr><td>S5</td><td>Odd</td><td>XYYY+YXXX</td><td>54 + 24 = 78</td><td>=</td><td>64</td><td>+</td><td>4+(0|0)+00</td><td>+</td><td>10</td></tr>
<tr><td>S7</td><td>Odd</td><td>XYXX+YXYY</td><td>24 + 54 = 78</td><td>=</td><td>64</td><td>+</td><td>4+(0|0)+00</td><td>+</td><td>10</td></tr>
</table>

NOTE – Count and purpose of the Control points in a subset are as the following:

[ 4+(_|_)+__ ] = { S#.xmt_err, S#.CSExtend_Err, S#.CSExtend, S#.CSReset }
[ _+(2|5)+__ ] = { SSD1|ESD1, SSD2|ESD2_Ext_0, ..._1, ..._2, ESD_Ext_Err }
[ _+(_|_)+16 ] = { Idle/CarrExt$_0$, ..., Idle/CarrExt$_{15}$ } = { S0.Data$_0$, ..., S0.Data$_{15}$ }

NOTE – Within S0, some Control and Data points share the same (2+16) images.

TABLE II
POSSIBLE MULTIPLEXING VARIANTS

| Data Words Used $N_C$ | $N_R$ | $N_E$ | GCD $(N_E,N_C)$ | $E_{max}$ $(N_C,N_R)$ | $n_e$, not less than, for an echo modulus of $E$=2 | $E$=4 | $E$=8 | $E$=16 | $E$=32 | $E$=64 |
|---|---|---|---|---|---|---|---|---|---|---|
| 64 + | 1 = | 65 | 1 | 64 | 45 | 90 | 135 | 179 | 224 | 267 |
| 64 + | 2 = | 66 | 2 | 32 | 23 | 46 | 68 | 91 | 113 | — |
| 64 + | 4 = | 68 | 4 | 16 | 12 | 23 | 35 | 46 | — | — |
| 64 + | 6 = | 70 | 2 | 16 | 8 | 16 | 24 | 31 | — | — |
| 64 + | 8 = | 72 | 8 | 8 | 6 | 12 | 18 | — | — | — |
| 64 + | 10 = | 74 | 2 | 8 | 5 | 10 | 15 | ← not applicable within S2, S4, S6 | | |
| 64 + | 12 = | 76 | 4 | 8 | 5 | 9 | 13 | ← not applicable within S2, S4, S6 | | |
| 64 + | 14 = | 78 | 2 | 8 | 4 | 8 | 11 | ← not applicable within S2, S4, S6 | | |

NOTE – We consider each variant and its corresponding numbers within a subset.
NOTE – $N_C$+$N_R$=$N_E$ are numbers of clear + noted = echo-driven noted Data words.
NOTE – GCD(a,b) is the greatest common divider of the integer numbers a and b.
NOTE – $E_{max}$=$N_C$/$N_{R^*}$, $N_{R^*}$ is the nearest integral power of two, not greater than $N_R$.
NOTE – $n_e$ defines the min. echo duration, in words, for a given echo modulus, $E$.
NOTE – $n_e$ depends on $E$, $N_E$, and $N_C$ as the following: $n_e \geq \ln E / (\ln N_E - \ln N_C)$.

TABLE III
PROPOSED CONSTRAINTS

| Subset | Data | Data* | Ctrl+Ctrl* | Subset | Data | Data* | Ctrl+Ctrl* |
|---|---|---|---|---|---|---|---|
| S0 | 64 | 8 | up to 25 | S1 | 64 | 8 | up to 6 |
| S2 | 64 | 8 | *absent* | S3 | 64 | 8 | up to 6 |
| S4 | 64 | 8 | *absent* | S5 | 64 | 8 | up to 6 |
| S6 | 64 | 8 | *absent* | S7 | 64 | 8 | up to 6 |

NOTE – In images, not context, Data** = Data + Data* within each enlisted subset.

## THE PLACE OF THE MULTIPLEXING

Between the possible multiplexing variants, see Table II, we choose the one we consider the most suitable in our situation because it provides the greatest common divider (GCD) of $8$ and a uniform multiplexing ratio ($64 : 8$) among all the code

TABLE IV
DATA CODING SCHEME

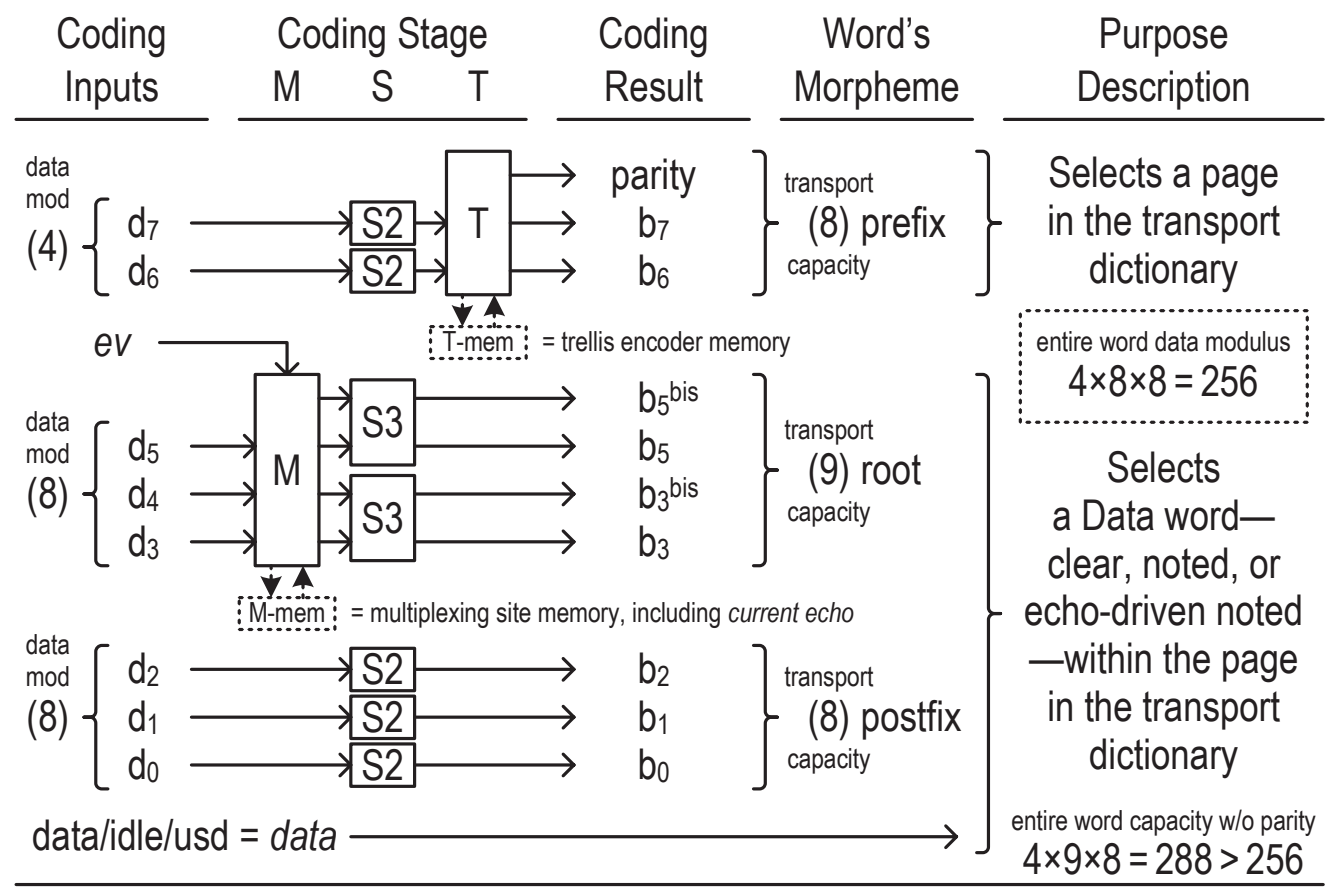

NOTE – Corresponds to the clear Data, Event over Data, Echo over Data cases.
NOTE – Coding stages are Multiplexing, Scrambling, and Trellis-based encoding.
NOTE – Bits $b_3$, $b_3^{bis}$, $b_5$, $b_5^{bis}$ of the result hold a binary-coded nonary numeral.
NOTE – Numbers in parentheses (...) denotes transport capacities of morphemes.

TABLE VI
IDLE CODING SCHEME

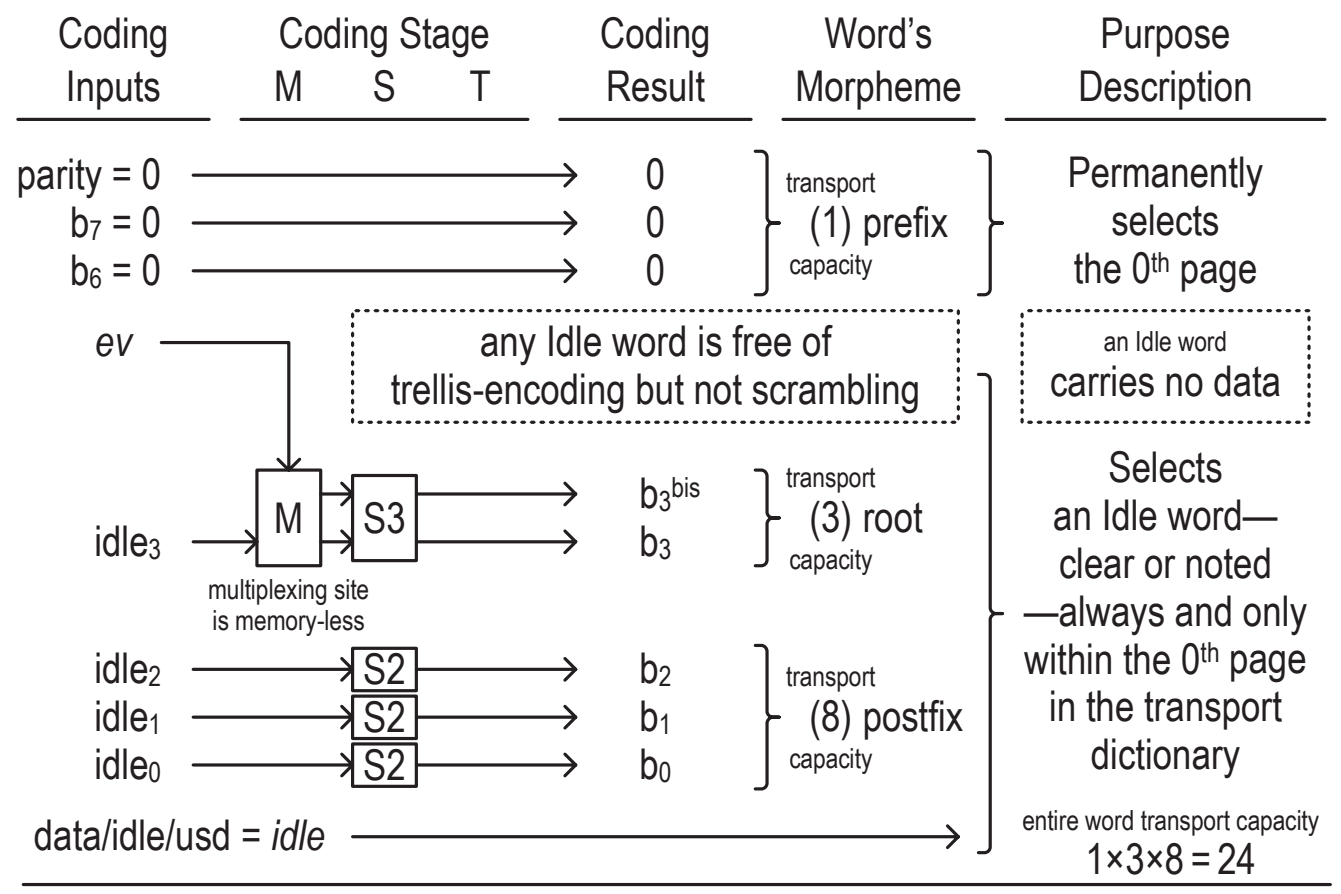

NOTE – Corresponds to the clear Idle, Event over Idle cases (Idle is a Control).
NOTE – Trellis-based encoding is bypassed, the prefix is driven permanently.
NOTE – Bits $b_3$ and $b_3^{bis}$ of the coding result hold a binary-coded ternary numeral.
NOTE – Idle occupies 24 out of 25 possible Control words. [see also USD Coding Scheme]

TABLE V
EVENT-FREE ESC CODING SCHEME

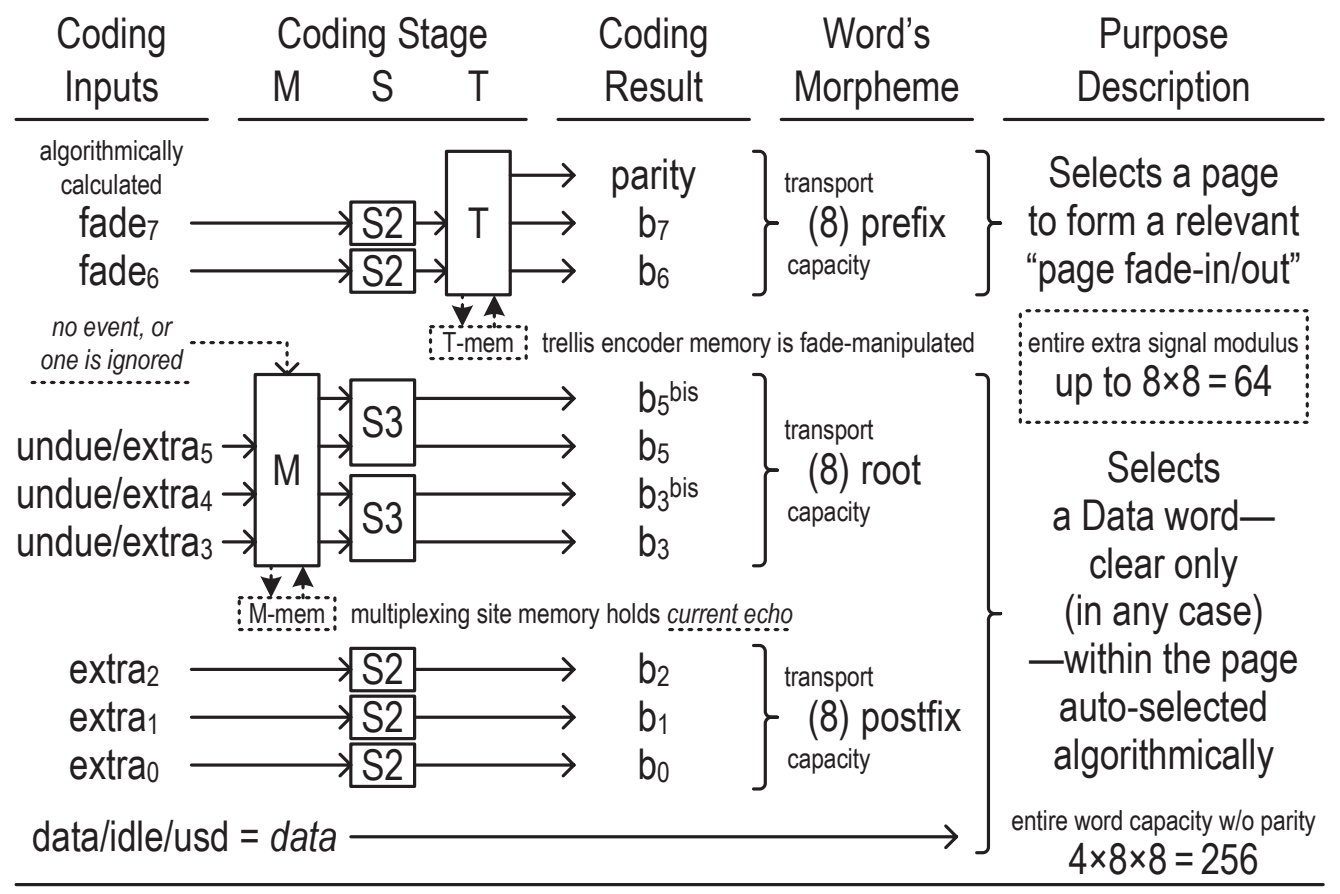

NOTE – Corresponds to the clear ESC, Echo over ESC cases (ESC is a Control).
NOTE – During echo cancellation (echo ≠ 1/1), the coding ignores any new event.
NOTE – Up to six extra signal bits are conveyed, depending on the current echo.
NOTE – Being a Control, ESC however employs the images same with Data.

TABLE VII
EVENT-AFFECTED ESC CODING SCHEME

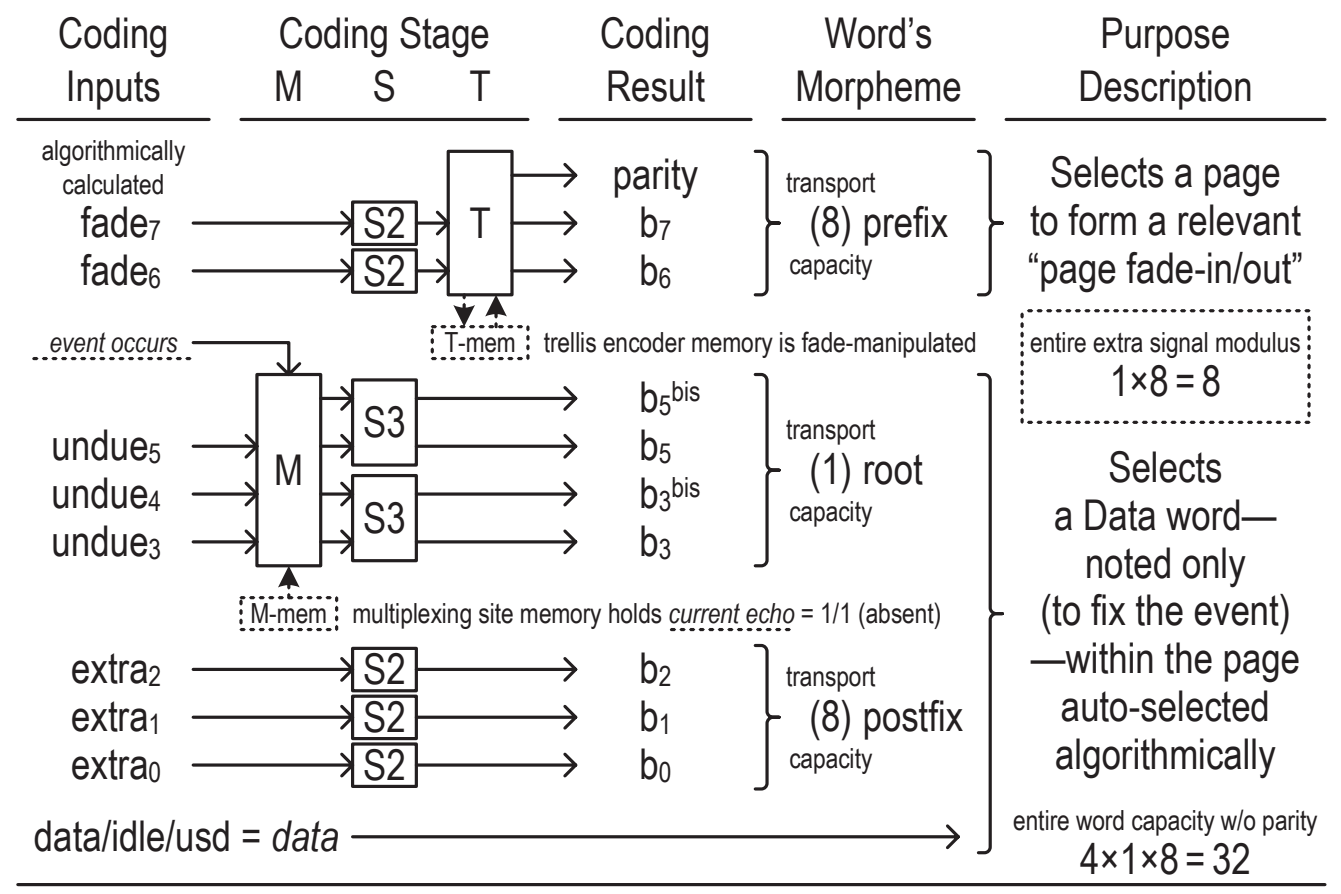

NOTE – Corresponds only to the Event over ESC case (ESC is a Control).
NOTE – Noting on an ESC as well as on any other Control generates no echo.
NOTE – Only three extra bits are conveyed, three undue bits are simply dropped.
NOTE – Muxing for ESC and Data is the same except the undue bits and no echo.

subsets. With the selected variant, we restrict the design to use $64$ clear, $8$ noted, and $64 + 8 = 72$ echo-driven noted Data words available per any transport page, see Table III.

Because $64/\mathrm{GCD} = 8 = 2^3$ and $72/\mathrm{GCD} = 9 = 3^2 < 2^4$, we need only three independent and four dependent in pairs bits to present the value of the root morpheme of a Data word before and after the multiplexing, respectively.

We refer to the number of unique (distinguishable) values of a word's morpheme before and after the coding, including the multiplexing, as its data modulus and transport capacity, respectively. For the root morpheme of a Data word, the data modulus is $2^3 = 8$ while the transport capacity is $3^2 = 9 > 8$. For the prefix and postfix morphemes of a Data word, which are participating in the coding but never in the multiplexing, the corresponding data modulus and transport capacity are considered the same. Based on that, we can employ an echo multiplexing process consisting of three rounds ($k = 3$) each consisting of six words ($n_i = 6$), see Table XII.

Assuming the described above a direct or similar way, we carefully extend the original coding flow of the 1000BASE-T PCS, defining the small, very strict place of the multiplexing stage in all reasonable cases related to the Data, Idle, USD, and ESC words,[4] see Table IV, Table VI, Table VIII, and then together Tables V and VII, respectively.

Besides the multiplexing site designated in those tables by the M box, there is just one another new to the 1000BASE-T PCS, designated by the S3 boxes, that is a base-3 scrambling site present alongside a well-known base-2 one designated in the same tables by the S2 boxes.

[4]USD (universal stream delimiter) and ESC (extra signal code) are the data streaming phases bounding the Payload and Idle ones, ordered USD then ESC before and ESC then USD after the Payload phase, respectively, see [3].

TABLE VIII
USD CODING SCHEME

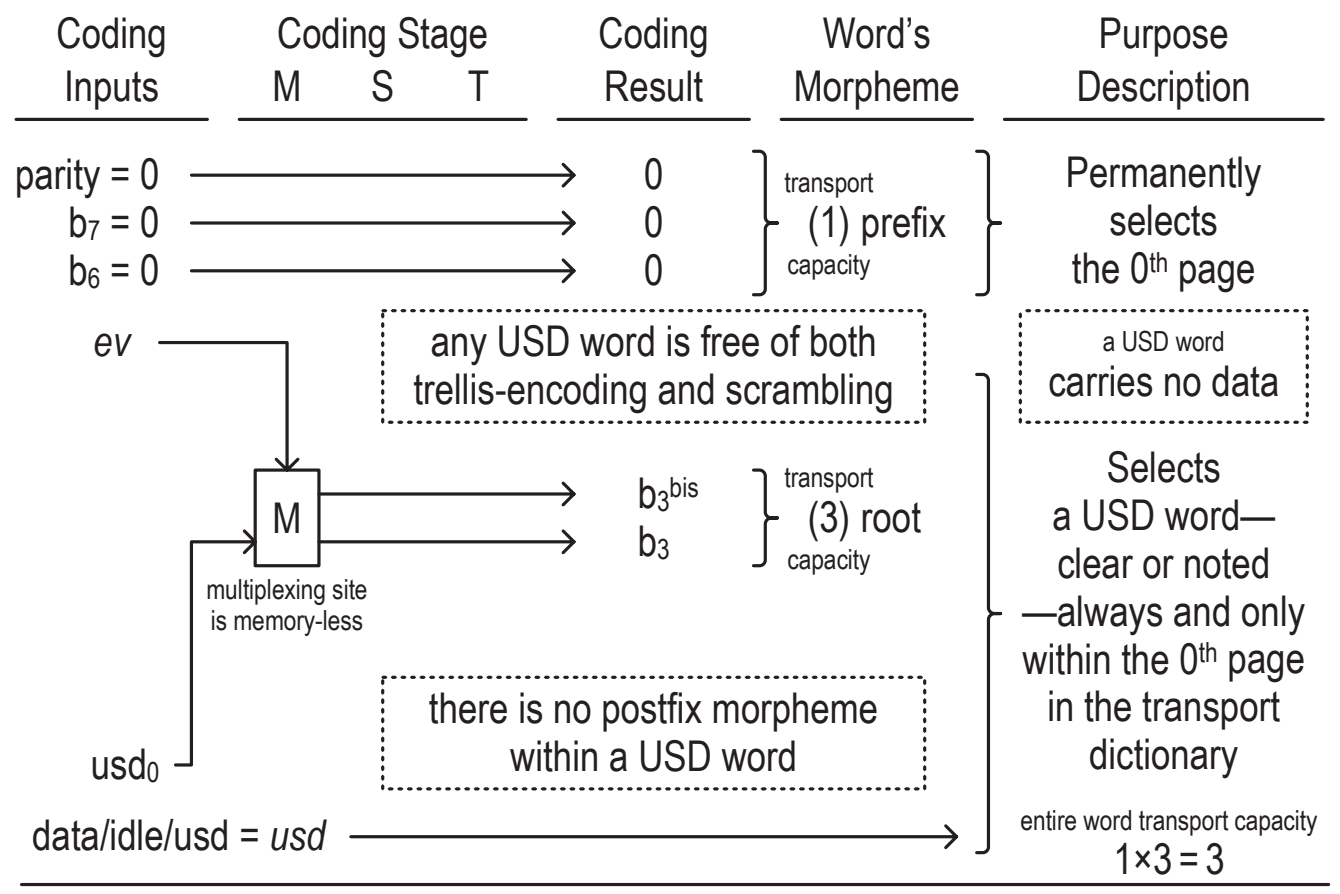

NOTE – Corresponds to the clear USD, Event over USD cases (USD is a Control).
NOTE – Bits $b_3$ and $b_3^{bis}$ of the coding result hold a binary-coded ternary numeral.
NOTE – USD occupies 3 out of 25 possible Control words. [see also Idle Coding Scheme]
NOTE – We need to share at least two images between the Idle and USD words.

TABLE IX
MULTIPAGE TRANSPORT DICTIONARY

| Page | Parity | Total | Data | Data* | Data** | Ctrl | Ctrl* | Free |
|---|---|---|---|---|---|---|---|---|
| P0 | Even | 97 | 64 | 8 | 72 | under consideration | | |
| P2 | Even | 72 | 64 | 8 | 72 | — | — | — |
| P4 | Even | 72 | 64 | 8 | 72 | — | — | — |
| P6 | Even | 72 | 64 | 8 | 72 | — | — | — |
| P1 | Odd | 78 | 64 | 8 | 72 | — | — | 6 |
| P3 | Odd | 78 | 64 | 8 | 72 | — | — | 6 |
| P5 | Odd | 78 | 64 | 8 | 72 | — | — | 6 |
| P7 | Odd | 78 | 64 | 8 | 72 | — | — | 6 |
| Words Summary | | Σ 625 | Σ 512 | Σ 64 | Σ 576 | under consideration | | |

NOTE – Pages correspond to the subsets of the original coding. [see Original Coding Props]
NOTE – In words and these images, a page does not intersect with any each other.
NOTE – Within the dictionary, there are three Data and two Control word sets.
NOTE – Except total/free/Σ, a number counts the words in a word set within a page.
NOTE – Ctrl comprises Idle and USD (16+2), Ctrl* comprises Idle* and USD* (8+1).
NOTE – In images and context, Idle and Idle* do not intersect with each other.
NOTE – In images and context, USD and USD* do not intersect with each other.
NOTE – USD and Idle and/or USD* and Idle* need to share at least two images.
NOTE – In images, Data and Data* are non-intersecting subsets of Data**.
NOTE – By design, ESC = Data, ESC* = Data*, ESC** = Data**. [in images, not context]

TABLE X
KEY FEATURES OF THE PROPOSED WAY

| Parameter | Description |
|---|---|
| Multiplexing Method | Linguistic Multiplexing |
| Multiplexing Delay | 48 ns = 6 words for encoder (transmitting side) |
| Demultiplexing Delay | 56 ns = 7 words for decoder (receiving side) |
| Mux/Demux Delay Behavior | constant, independent of the transferred content |
| No. of Event-affected Words | constant, 1 word/cycle |
| No. of Echo-driven Words | variable, up to 18 words/cycle, or absent |
| No. of Silence Words | at least 1 word/cycle after echo is fully cancelled |
| Event Cycle Typical Period | 20 words/cycle (event + max echo + min silence) |
| Data Transfer Rate | 1000 Mbit/s = 125M words/s × 8 bit/word |
| Data Transfer Behavior | Fully independent of the Event Transfer Service |
| Data Transfer Method | Asynchronous, packet, same as in 1000BASE-T |
| Data Transfer Values | 255...0 (256), like in 1000BASE-T, *except error prop.* |
| Event Transfer Rate | 6.25 Mev/s = 125M words/s × 1/20 ev/word |
| Event Transfer Behavior | Fully independent of the Data Transfer Service |
| Event Transfer Method | Asynchronous, single |
| Event Transfer Values | 1 or 0 = an event or no event |
| Event Fixation Uncertainty ... | (for transport word/symbol time period of 8 ns) |
| on Asynchronous Tx/Rx | ± 4 ns = ± word/symbol transfer half period |
| on Synchronous Tx/Rx | ± 0 ns, if synchronized with word/symbol tx/rx clock |
| Suitable Transport Rate | 500M PAM5 symbols/s in total, via 4 twisted pairs |
| Suitable Transport Behavior | Compatible with the 1000BASE-T PMA sublayer |
| Suitable Transport Method | Synchronous, serial, continuous, memory-less |
| Suitable Transport Values | 4D-PAM5 constellation pts, symbol-wide stream of |
| Data Path Client | IEEE 802.3 Gigabit MAC, *except carrier extension* |
| Data Path Client Examples | MPU, MCU, FPGA, ASIC, IP core |
| Event Path Client | OAM, precise phase/time synchronization |
| Event Path Client Examples | UART, HDLC, PPP, IRIG, PPS, PTP, Ethernet |
| Companion Technologies | SyncE, PTP, GNSS, AVB, Audio over Ethernet |

TABLE XI
EXAMPLE EVENT TRAIN FOR EXTRA RESOLUTION GOALS

| 0*m* | 1*m* | 2*m* | 3*m* | Number-related | Position-related = Offset within the Word |
|---|---|---|---|---|---|
| ev | — | — | — | event type 0 | plus 0/8 of word period = 0 ns ± 0.5 ns |
| ev | — | — | ev | event type 1 | plus 1/8 of word period = 1 ns ± 0.5 ns |
| ev | — | ev | — | event type 2 | plus 2/8 of word period = 2 ns ± 0.5 ns |
| ev | — | ev | ev | event type 3 | plus 3/8 of word period = 3 ns ± 0.5 ns |
| ev | ev | — | — | event type 4 | plus 4/8 of word period = 4 ns ± 0.5 ns |
| ev | ev | — | ev | event type 5 | plus 5/8 of word period = 5 ns ± 0.5 ns |
| ev | ev | ev | — | event type 6 | plus 6/8 of word period = 6 ns ± 0.5 ns |
| ev | ev | ev | ev | event type 7 | plus 7/8 of word period = 7 ns ± 0.5 ns |

NOTE – Word period is 8 ns; single event cycle duration, *m*, is at least 20 words.
NOTE – The 0th [word of the 0th] cycle fixes while the next cycles clarify the event.
NOTE – Some eight-times ($8 = 2^3$) case is shown; other times are also possible.

## PAGE FADING

Because (at least) within three even pages of the transport dictionary—P2, P4, and P6, see Table IX—there are no free images left to carry the control code points responsible for the code subset reset in the original coding scheme,[5] we introduce a means, we will call the page fading, to compensate that lost function. Computationally tricking on the responsible coding inputs, knowing and/or "predicting" (precalculating) the side-stream scrambler state bits, the means manages a "smoothed" but controlled page transition—using two intermediate steps, one per each word of an ESC—generating a so-called "page fade-in" during the ESC before the payload, and a so-called "page fade-out" during the ESC after the payload,[6] altering their prefixes, see Table XIII then Tables V and VII.

The purpose of a "fade-in" is to toggle the memory of the trellis encoder a way native to the encoder, changing it from the related to the zeroth page (P0) into the related to the page corresponding to the prefix of the payload's first word (any page), to prevent an interrupt in the forward error correction function, making that function restlessly continuous.

The purpose of a "fade-out" is oppositely similar and is to do the same but changing the memory from the related to the page corresponding to the prefix of the payload's final word (any page), into the related to the zeroth page (P0).[7]

[5]In 1000BASE-T, there are just four control codes, including the mentioned, accessible within each code subset, see Table I. Because we assume full duplex operation only and no error labeling enabled per word, we substitute only the mentioned function while simply drop the rest.

[6]For the ESC and USD words, we follow the agreement set in the original framing, on that the critical control codes are paired.

[7]It is the only page in our design, within which the coding can access (all) the control words responsible for data packetization, see Table XIII.

TABLE XII
ARITHMETIC REDUNDANCY ANALYSIS

| $n$ | Transport Capacity = $9^n$ | Initial Echo = 1/8 → $8\times8^n$ → 1/4;1/2;1 | 1/4 → $4\times8^n$ → 1/2;1 | 1/2 → $2\times8^n$ → 1 |
|---|---|---|---|---|
| 4 | 6,561 = $9^4$ | 32,768 = $8\times8^4$ > $9^4$ | 16,384 = $4\times8^4$ > $9^4$ | 8,192 = $2\times8^4$ > $9^4$ |
| | | 65,536 = /4 > $9^5$ | | |
| | | 131,072 = /2 > $9^5$ | 65,536 = /2 > $9^5$ | |
| 5 | 59,049 = $9^5$ | 262,144 = $8\times8^5$ > $9^5$ | 131,072 = $4\times8^5$ > $9^5$ | 65,536 = $2\times8^5$ > $9^5$ |
| | | **524,288 = /4 < $9^6$** ① | | |
| | | 1,048,576 = /2 > $9^6$ | **524,288 = /2 < $9^6$** ① | |
| ① **6** | **531,441 = $9^6$** ① | 2,097,152 = **$8\times8^6$** > $9^6$ | 1,048,576 = **$4\times8^6$** > $9^6$ | **524,288 = $2\times8^6$ < $9^6$** ① |
| 10 | 3,486,784,401 = $9^{10}$ | 8,589,934,592 = $8\times8^{10}$ > $9^{10}$ | 4,294,967,296 = $4\times8^{10}$ > $9^{10}$ | |
| | | 17,179,869,184 = /4 < $9^{11}$ | | |
| | | 34,359,738,368 = /2 > $9^{11}$ | 17,179,869,184 = /2 < $9^{11}$ | |
| 11 | 31,381,059,609 = $9^{11}$ | 68,719,476,736 = $8\times8^{11}$ > $9^{11}$ | 34,359,738,368 = $4\times8^{11}$ > $9^{11}$ | |
| | | 137,438,953,472 = /4 < $9^{12}$ | | |
| | | **274,877,906,944 = /2 < $9^{12}$** ② | 137,438,953,472 = /2 < $9^{12}$ | |
| ② **12** | **282,429,536,481 = $9^{12}$** ② | 549,755,813,888 = **$8\times8^{12}$** > $9^{12}$ | **274,877,906,944 = $4\times8^{12}$ < $9^{12}$** ② | |
| 16 | 1,853,020,188,851,841 = $9^{16}$ | 2,251,799,813,685,248 = $8\times8^{16}$ > $9^{16}$ | | |
| | | 4,503,599,627,370,496 = /4 < $9^{17}$ | | |
| | | 9,007,199,254,740,992 = /2 < $9^{17}$ | | |
| 17 | 16,677,181,699,666,569 = $9^{17}$ | 18,014,398,509,481,984 = $8\times8^{17}$ > $9^{17}$ | | |
| | | 36,028,797,018,963,968 = /4 < $9^{18}$ | | |
| | | 72,057,594,037,927,936 = /2 < $9^{18}$ | | |
| ③ **18** | **150,094,635,296,999,121 = $9^{18}$** ③ | **144,115,188,075,855,872 = $8\times8^{18}$ < $9^{18}$** ③ | | |

| PATH | Metrics |
|---|---|
| ① ① ① | $k$=3 **max{$n_i$}=6** |
| ① ② | $k$=2 max{$n_i$}=12 |
| ② ① | $k$=2 max{$n_i$}=12 |
| ③ | $k$=1 max{$n_i$}=18 |

| PATH | k-Round Multiplexing Process | Duration |
|---|---|---|
| ① ① ① | 1/8 → $8\times8^6/4$ → $4\times8^6/2$ → $2\times8^6/1$ → 1/1 | $n_1+n_2+n_3=18$ |
| ① ② | 1/8 → $8\times8^6/4$ → $4\times8^{12}/1$ → 1/1 | $n_1+n_2=18$ |
| ② ① | 1/8 → $8\times8^{12}/2$ → $2\times8^6/1$ → 1/1 | $n_1+n_2=18$ |
| ③ | 1/8 → $8\times8^{18}/1$ → 1/1 | $n_1=n_D=18$ |

TABLE XIII
DATA FRAMING RULE

| Time Period → | . . . | H−5 | H−4 | H−3 | H−2 | H−1 | Head | H+1 | H+2 | . . . | T−2 | T−1 | Tail | T+1 | T+2 | T+3 | T+4 | T+5 | . . . |
|---|---|---|---|---|---|---|---|---|---|---|---|---|---|---|---|---|---|---|---|
| Word Set Used | ... @ | @ | $\$^2$ | $\$^1$ | $\uparrow^{1\blacktriangle}$ | $\uparrow^{2\blacktriangle}$ | # | # | # | . . . | # | # | # | $\uparrow^{2\blacktriangledown}$ | $\uparrow^{1\blacktriangledown}$ | $\$^1$ | $\$^2$ | @ | @ ... |
| Word Set Type | Control [Idle] | Control [Idle] | Control [USD] | Control [USD] | Control [ESC] | Control [ESC] | Data | Data | Data | . . . | Data | Data | Data | Control [ESC] | Control [ESC] | Control [USD] | Control [USD] | Control [Idle] | Control [Idle] |
| Word Set Page | P0 | P0 | P0 | P0 | auto-calculated | | [ any one of P0, P1, P2, P3, P4, P5, P6, P7 after trellis ] | | | | | | | auto-calculated | | P0 | P0 | P0 | P0 |
| Basic Purpose | idle filler | | opening delimiter | | "page fade-in" | | carries data information | | | | | | | "page fade-out" | | closing delimiter | | idle filler | |
| Common Definition | inter-payload glue | | | | | | payload | | | | | | | inter-payload glue | | | | | |
| NOTE – All clear case is shown; during multiplexing, an event-affected Data/ESC/USD/Idle word can carry an event, also an echo-driven Data/ESC word can carry an echo. | | | | | | | | | | | | | | | | | | | |

Comparing with the 1000BASE-T PCS which hardly forces the memory of the trellis encoder into its reset state, the page fading means operates very softly and linearly while provides for an equivalent functionality that makes its integration into the overall coding process a seamless step.

## REST OF THE DESIGN

We synthesize the rest of the design, following the abstract synthesis flow described in [2] but assuming the described in [3] a similar way where it is applicable.

Among the expected properties, see Table X, we would like to pay attention to the event fixation uncertainty which, in its turn, depends on many factors, including the basic resolution, i.e., the number of unique (unambiguously distinguishable) events per a word, that is just one in our case. We consciously choice that resolution as small as possible, because a higher resolution causes a longer multiplexing delay. So, the delay is our only principal factor that limits the resolution.

To rise up over the resolution, we still can use an external means like an event train, see Table XI.

## CONCLUSION

We outlined a way to multiplex a data coding means and an event coding means together over a single transport, grounding on the redundancy of such a transport.

The transport we considered in this paper is a gigabit one, employed in Gigabit Ethernet, type 1000BASE-T, operating over four twisted pairs as the media.

The proposed way preserves the behavior of the underlying transport while keeps the both means functionally orthogonal that would be useful in Ethernet applications.

# Spreading the Constellation Coverage of the Draft Data Coding Means and Event Coding Means Multiplexed Over the 1000BASE-T PMA Sublayer

Alexander Ivanov

***Abstract*—In this paper, we are making a try to develop the linguistic multiplexing branch dedicated to the Gigabit Ethernet twisted-pair physical layer, type 1000BASE-T, now focusing on the coverage of the 4D-PAM5 constellation used as the physical line code space by the 1000BASE-T PMA sublayer.**

***Index Terms*—Ethernet, linguistic multiplexing, multiplexing, 4D-PAM5, constellation coverage, 1000BASE-T.**

## Background

GIGABIT Ethernet, type 1000BASE-T [1] as well as its functionally refined, linguistically multiplexed variant—introduced in [2] and further thru this paper referred to as the draft design—both leverage the 4D-PAM5 physical line code,[1] in which the communication channel uses four simple PAM5 sub-channels in parallel, by one sub-channel per a dimension, each independently assigned with a basic line code—PAM3Y or PAM2X—a given time, see Table I.

Speaking in the terms of the certain power parameters we refer to as the coding slices, see Table II, characterizing the communication channel as a total of its basic line codes, we conclude that neither 1000BASE-T nor the draft design covers the 4D-PAM5 constellation neither evenly nor just completely, whereas the incompleteness—alongside the trellis encoder as the most distorting (disproportioning) factor,[2] see Table III—causes that objectionable unevenness.

We understand that we cannot reach the evenness "native" (natural) for 4D-PAM5 because we cannot avert the distortion (disproportion) until the trellis coded modulation (TCM) site and its associated paging ("code set partitioning") model are employed in the design,[3] but with the completeness we have some chance when we can increase the statistical diversity of the resulting signal in the media, via operating over as much as possible code points of the constellation.

So, the rest of this paper describes the proposed solution as a set of its constellation coverage, coding balance, and power balance characteristics.

A manuscript of this work was submitted to IEEE Communications Letters November 26, 2022 and rejected as not being in the scope of the journal.

Please sorry for the author has no time to find this work a new home, peer reviewed or not, except of arXiv, and just hopes there it meets its reader, one or maybe various, whom the author beforehand thanks for their regard.

A. Ivanov is with JSC Continuum, Yaroslavl, the Russian Federation.

[1] Complete 4D-PAM5 constellation consists of $5^4 = 625$ code points.

[2] In 1000BASE-T, TCM causes the distortion by design.

[3] Such a goal is out of the scope of this paper.

TABLE I
BASIC LINE CODES

| Line Code (Sub)Set | Code Points Within | Normalized Power | Avg. Pwr. |
|---|---|---|---|
| PAM5 | {−2 ; −1 ; 0 ; +1 ; +2} | { 4/4 ; 1/4 ; 0/4 ; 1/4 ; 4/4 } | 2/4 = ½ |
| **PAM3Y** | **{−2 ; 0 ; +2}** | **{ 4/4 ; 0/4 ; 4/4 }** | **$P_Y$ = ⅔** |
| PAM2Y | {−2 ; +2} | { 4/4 ; 4/4 } | 4/4 = 1/1 |
| PAM2Y−, PAM2Y+ | {−2 ; 0 } , { 0 ; +2} | { 4/4 ; 0/4 } , { 0/4 ; 4/4 } | 2/4 = ½ |
| PAM1Y−, PAM1Y+ | {−2} , {+2} | { 4/4 } , { 4/4 } | 4/4 = 1/1 |
| PAM3X | {−1 ; 0 ; +1 } | { 1/4 ; 0/4 ; 1/4 } | 2/12 = ⅙ |
| **PAM2X** | **{−1 ; +1 }** | **{ 1/4 ; 1/4 }** | **$P_X$ = ¼** |
| PAM2X−, PAM2X+ | {−1 ; 0 },{ 0 ; +1 } | { 1/4 ; 0/4 },{ 0/4 ; 1/4 } | ⅛ = ⅛ |
| PAM1X−, PAM1X+ | {−1} , {+1} | { 1/4 } , { 1/4 } | ¼ = ¼ |
| zero (quiet) | { 0 } | { 0/4 } | 0/4 = 0 |

NOTE – $U_{REF} = \pm 2$, $P_{REF} = [U_{REF}]^2 = 4$, therefore, $(\pm 2)^2/P_{REF} = 4/4$ and $(\pm 1)^2/P_{REF} = 1/4$.
NOTE – Average power is an arithmetic mean of the respective normalized values.

TABLE II
CODING SLICES

| Page | Contents | Slices | Page | Contents | Slices |
|---|---|---|---|---|---|
| P0 | XXXX, YYYY | 4X+0Y, 0X+4Y | P1 | XXXY, YYYX | 3X+1Y, 1X+3Y |
| P2 | XXYY, YYXX | 2X+2Y, 2X+2Y | P3 | XXYX, YYXY | 3X+1Y, 1X+3Y |
| P4 | XYYX, YXXY | 2X+2Y, 2X+2Y | P5 | XYYY, YXXX | 3X+1Y, 1X+3Y |
| P6 | XYXY, YXYX | 2X+2Y, 2X+2Y | P7 | XYXX, YXYY | 3X+1Y, 1X+3Y |

NOTE – A slice counts the numbers of X's and Y's in the respective page content.
NOTE – There are just five slices distinct in the numbers of X's ($N_X$) and Y's ($N_Y$).
NOTE – Slice average power $P_{SLICE}(N_X, N_Y) = 1/4 \cdot N_X \cdot P_X + 1/4 \cdot N_Y \cdot P_Y < 1$, $N_X + N_Y = 4$.

TABLE III
PROBABILITY DISTRIBUTION COMPARISON

| ↓Page, *probability* → | "Native" [for 4D-PAM5] | | Given by a Design | Disproportion |
|---|---|---|---|---|
| P0 slices within↓ | 97/625 = 0.1552 | > | ⅛ = 0.125 | −19% given/"native" |
| ├slice 4X+0Y, one/page | 16/625 = 0.0256 | <or> | ⅛ × {16/64; 16/72; 16/96; 1/7} | {+22; +9; −19; −39}% |
| └slice 0X+4Y, one/page | 81/625 = 0.1296 | > | ⅛ × {48/64; 56/72; 80/96; 7/8} | {−28; −25; −20; −16}% |
| P2, P4, or P6 | 72/625 = 0.1152 | < | ⅛ = 0.125 | +8.5% given/"native" |
| └slice 2X+2Y, two/page | 36/625 = 0.0576 | < | ⅛ × {32/64; 36/72; 36/72; 1/2} | { +9; +9; +9; +9}% |
| P1, P3, P5, or P7 | 78/625 = 0.1248 | < | ⅛ = 0.125 | +0.2% given/"native" |
| ├slice 3X+1Y, one/page | 24/625 = 0.0384 | <or> | ⅛ × {24/64; 24/72; 24/78; 1/4} | {+22; +9; +0; −19}% |
| └slice 1X+3Y, one/page | 54/625 = 0.0864 | < | ⅛ × {40/64; 48/72; 54/78; 3/4} | {+22; +9; +0; +19}% |

NOTE – "Native" is a uniform distribution in the 4D-PAM5 space, 1/625 per a point.

## Complete Constellation Coverage

Analyzing the content of the pages of the transport dictionary, see Table II, we conclude that, although there are eight distinct pages, P0 to P7, together comprising 16 distinct code subsets, XXXX to YYYY, there are just five distinct coding

TABLE IV
EXPANDING OVER EVEN PAGES EXCEPT P0

Input | Coding Stages, Blocks, and Flow | PAM | Code Point-to-Pair Mapping

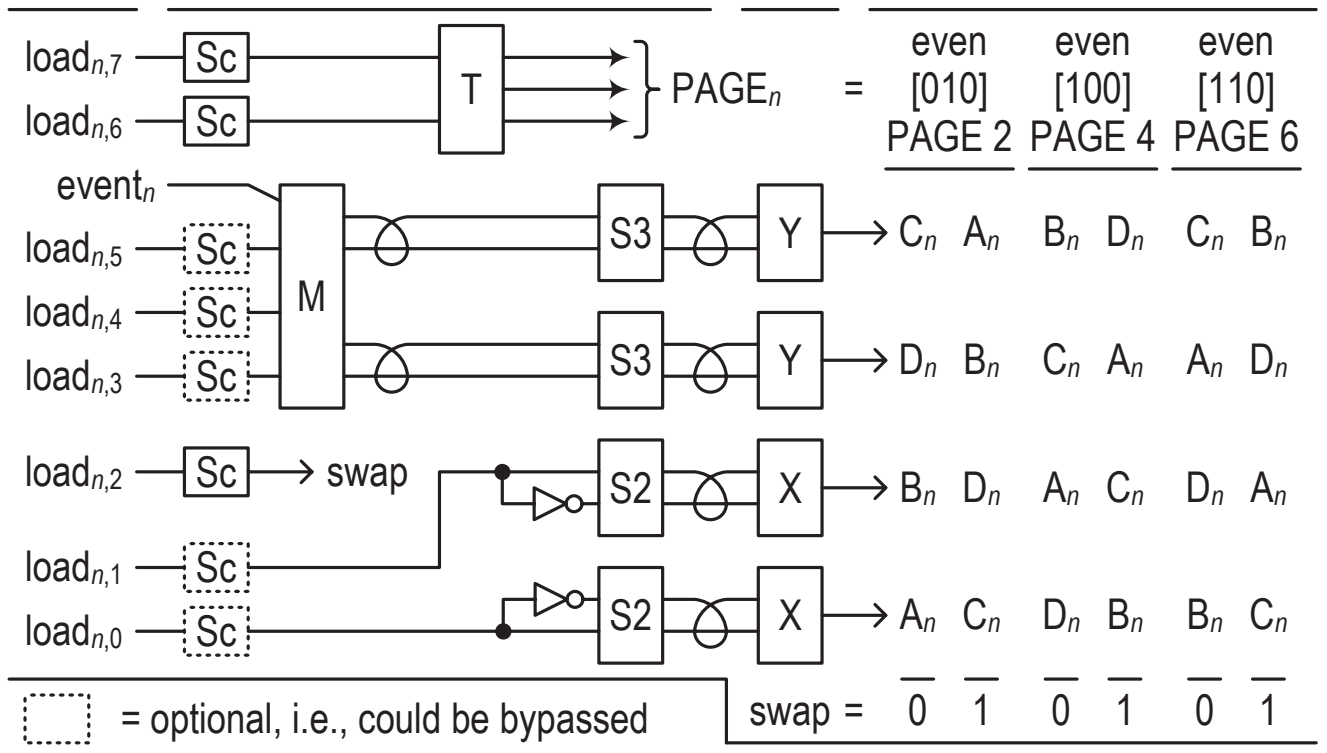

| ↓ref | ↓description | ↓input values | ↓output values |
|---|---|---|---|
| Sc | bit scrambler per IEEE 802.3ab | {0;1} | {0;1} |
| S2 | pre-PAM base-2 sub-scrambler | {1;2} | {1;2} |
| S3 | pre-PAM base-3 sub-scrambler | {1;2;3} | {1;2;3} |
| M | linguistic multiplexing site | {$\overline{ev}$;ev} × $\{0;1\}^3$ | {1;2;3} × {1;2;3} |
| T | trellis coded modulation site | $\{0;1\}^2$ | $\{0;1\}^3$ |
| X | PAM2X line code convertor | {1;2} | {−1 ; +1} |
| Y | PAM3Y line code convertor | {1;2;3} | {−2 ; =0 ; +2} |
| ▷o | bit inverter using Boolean logic | {0;1} | {0;1} |

NOTE – Covers 216 out of 625 possible code points, 36 items per a slice (2X+2Y).

TABLE V
CODING BALANCE WITHIN EVEN PAGES

| Page | Slice | Content | Pair A | Pair B | Pair C | Pair D |
|---|---|---|---|---|---|---|
| P0 | 4X+0Y | XXXX | __+1____ | +1______ | __+1____ | +1______ |
| P2 | 2X+2Y | XXYY | +1______ | __+1____ | ____+1__ | ______+1 |
| P4 | 2X+2Y | XYYX | __+1____ | ____+1__ | ______+1 | +1______ |
| P6 | 2X+2Y | XYXY | ______−1 | −1______ | ____−1__ | __−1____ |
| P0 | 0X+4Y | YYYY | __−1____ | −1______ | __−1____ | −1______ |
| P2 | 2X+2Y | YYXX | −1______ | __−1____ | ____−1__ | ______−1 |
| P4 | 2X+2Y | YXXY | __−1____ | ____−1__ | ______−1 | −1______ |
| P6 | 2X+2Y | YXYX | ______+1 | +1______ | ____+1__ | __+1____ |
| Summation (per pair) → | | | =0=0__=0 | =0=0=0__ | __=0=0=0 | =0=0__=0 |

↓ Pair masking corresponds to the following structural patterns we assume opposite:

| Coding Structures When X's Lead | | | v. | Coding Structures When Y's Lead | | |
|---|---|---|---|---|---|---|
| **Slice** | **Pattern** | **Mask** | **v.** | **Mask** | **Pattern** | **Slice** |
| 2X+2Y swap=0 | S2 X | +1______ | ↔ | −1______ | M S3 Y | 2X+2Y swap=1 |
| 2X+2Y swap=0 | S2 X | __+1____ | ↔ | __−1____ | M S3 Y | 2X+2Y swap=1 |
| 2X+2Y swap=0 | M S3 Y | ____+1__ | ↔ | ____−1__ | S2 X | 2X+2Y swap=1 |
| 2X+2Y swap=0 | M S3 Y | ______+1 | ↔ | ______−1 | S2 X | 2X+2Y swap=1 |
| 4X+0Y | M S2 X | +1______ | ↔ | −1______ | M S3 Y | 0X+4Y |
| 4X+0Y | M S2 X | __+1____ | ↔ | __−1____ | M S3 Y | 0X+4Y |
| 4X+0Y | M S2 X | ____+1__ | ↔ | ____−1__ | S3 Y | 0X+4Y |
| 4X+0Y | M S2 X | ______+1 | ↔ | ______−1 | S3 Y | 0X+4Y |

By design, the 2X+2Y slice occurs ever per a word sent within a non-P0 even page.
By design, the 4X+0Y slice occurs ⅛ times per a word sent within the P0 even page.
By design, the 0X+4Y slice occurs ⅞ times per a word sent within the P0 even page.

Therefore, the [normalized] average power per a word sent within an even page is:

$$P_{2,4,6} = \tfrac{1}{1} \times (\tfrac{2}{4} \cdot P_X + \tfrac{2}{4} \cdot P_Y) = \tfrac{11}{24} = \tfrac{44}{96}\text{, or} \quad (\approx 0.46)$$

$$P_0 = \tfrac{1}{8} \times (\tfrac{4}{4} \cdot P_X + \tfrac{0}{4} \cdot P_Y) + \tfrac{7}{8} \times (\tfrac{0}{4} \cdot P_X + \tfrac{4}{4} \cdot P_Y) = \tfrac{236}{384} = \tfrac{59}{96}. \quad (\approx 0.61)$$

NOTE – During transmission, an even page occurs with the probability $p_{0,2,4,6} = 4/8$.
NOTE – Within the noted above, $p_{0,2,4,6} = p_{2,4,6} + p_0$, where $p_{2,4,6} = 3/8$ and $p_0 = 1/8$.

TABLE VI
EXPANDING OVER EVEN PAGE P0 WHEN X'S PREVAIL

Input | Coding Stages, Blocks, and Flow | PAM | Mapping

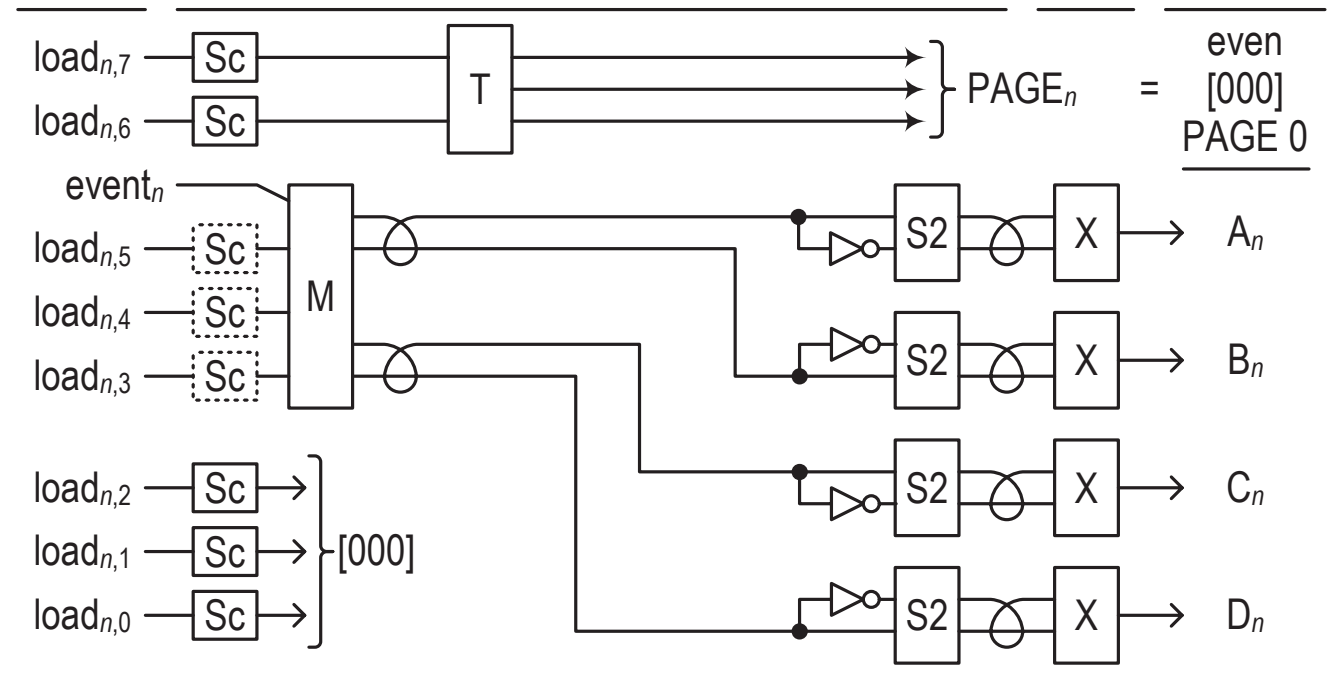

NOTE – Covers 16 out of 625 possible code points, all per the sole slice (4X+0Y).

TABLE VII
EXPANDING OVER EVEN PAGE P0 WHEN Y'S PREVAIL

Input | Coding Stages, Blocks, and Flow | PAM | Mapping

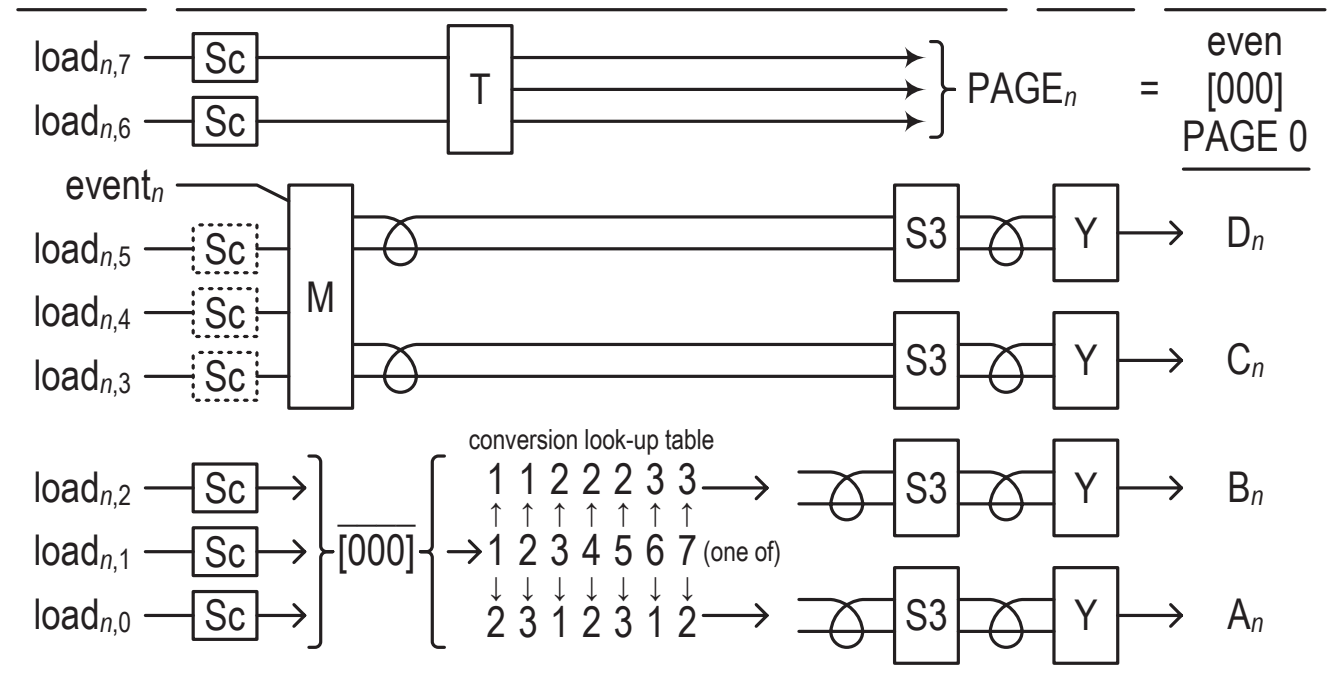

NOTE – Covers 81 out of 625 possible code points, all per the sole slice (0X+4Y).

slices individually varying in the (normalized) average power but collectively covering the whole line code space, i.e., they cover the 4D-PAM5 constellation completely.

Based on this, we construct the coding means compositely, as a collection of five coding schemes each matching to one coding slice corresponding, in its turn, with one or more pages whose selection depends only on the current state of the TCM site,[4] so the means operates across the full transport dictionary, i.e., covers the whole line code space, the following manner, leaving no code points uninvolved.

Within the transport dictionary practiced, there are the three "symmetric" even pages, P2, P4, and P6, each corresponding with the coding slice 2X+2Y twice, see Table II, which one the coding means processes transforming the information bits of a word into a four-dimensional code point comprising two PAM2X and two PAM3Y single-dimensional levels, at a rate of one entire word, i.e., eight bits that is exactly one octet, per a word time period,[5] see Table IV.

Within the same dictionary, there is the only "asymmetric" even page, P0, corresponding with the coding slices 4X+0Y and 0X+4Y, see Table II, which two the coding means processes in different ways, transforming the information bits of

[4] Only two out of eight information bits of a word affect the TCM state, else three participate in the multiplexing and three are simply bypassed.

[5] Denoted by the index $n$ in tables, that period is 8 ns long (125 MHz).

TABLE VIII
EXPANDING OVER ODD PAGES WHEN X'S PREVAIL

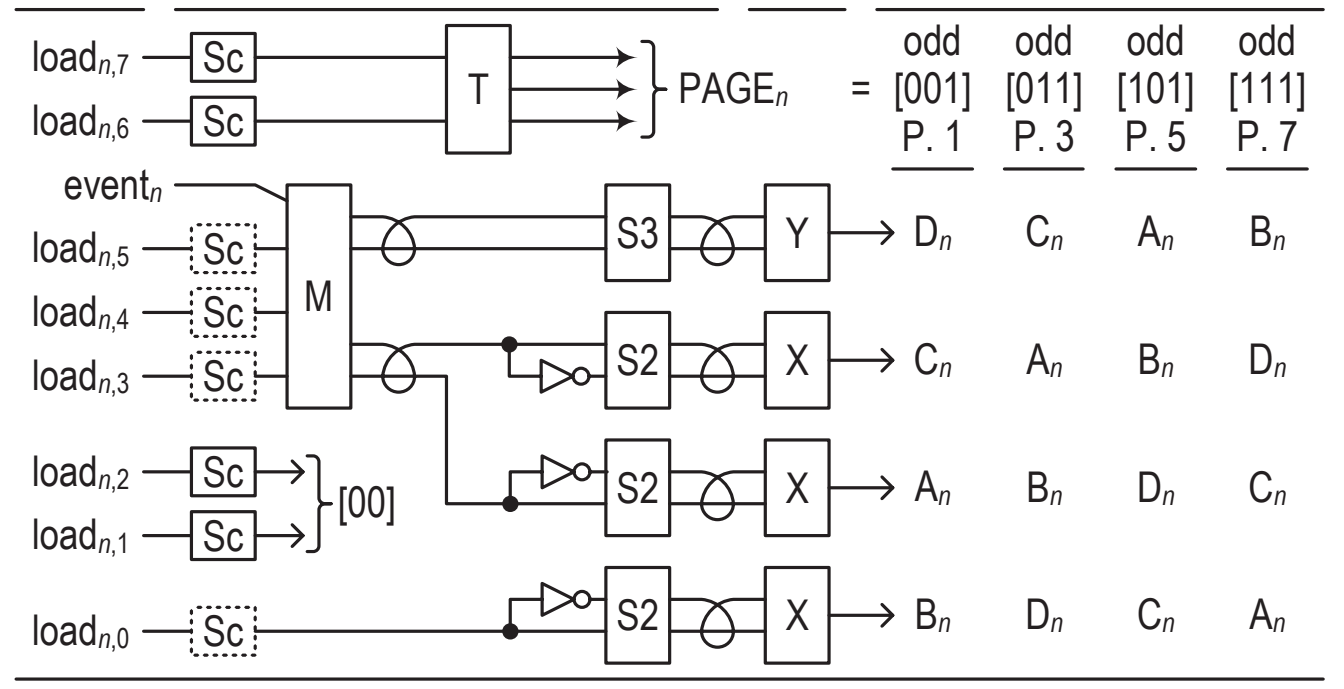

NOTE – Covers 96 out of 625 possible code points, 24 items per a slice (3X+1Y).

TABLE IX
EXPANDING OVER ODD PAGES WHEN Y'S PREVAIL

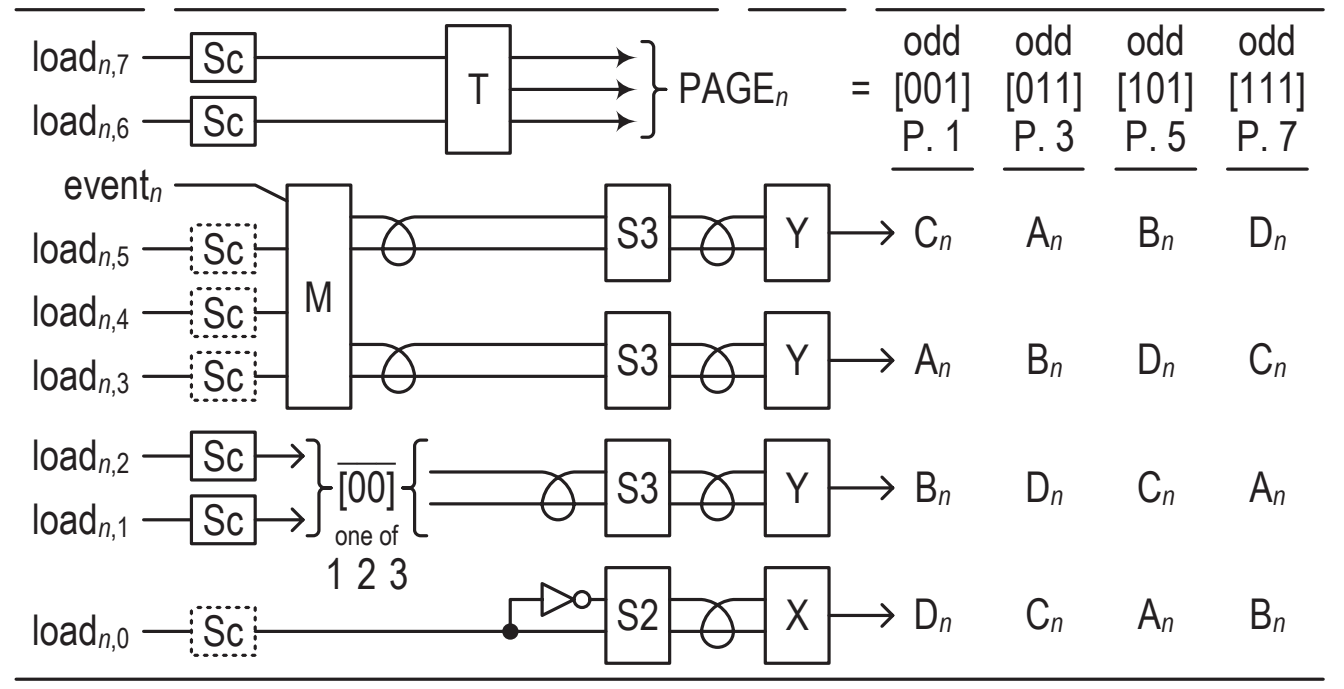

NOTE – Covers 216 out of 625 possible code points, 54 items per a slice (1X+3Y).

TABLE X
CODING BALANCE WITHIN ODD PAGES

| Page | Slice | Content | Pair A | Pair B | Pair C | Pair D |
|---|---|---|---|---|---|---|
| P1 | 3X+1Y | XXXY | ____+1__ | ______+1 | __+1____ | +1______ |
| P3 | 3X+1Y | XXYX | __+1____ | ____+1__ | +1______ | ______+1 |
| P5 | 3X+1Y | YXXX | +1______ | __+1____ | ______+1 | ____+1__ |
| P7 | 3X+1Y | XYXX | ______+1 | +1______ | ____+1__ | __+1____ |
| P1 | 1X+3Y | YXXX | ____−1__ | ______−1 | __−1____ | −1______ |
| P3 | 1X+3Y | YYYX | __−1____ | ____−1__ | −1______ | ______−1 |
| P5 | 1X+3Y | XYYY | −1______ | __−1____ | ______−1 | ____−1__ |
| P7 | 1X+3Y | YXYY | ______−1 | −1______ | ____−1__ | __−1____ |
| Summation (per pair) → | | | =0=0=0=0 | =0=0=0=0 | =0=0=0=0 | =0=0=0=0 |

↓ Pair masking corresponds to the following structural patterns we assume opposite:

| Coding Structures When X's Prevail | | | v. | Coding Structures When Y's Prevail | | |
|---|---|---|---|---|---|---|
| 3X+1Y | M S3 Y | +1______ | ↔ | −1______ | S2 X | 1X+3Y |
| 3X+1Y | M S2 X | __+1____ | ↔ | __−1____ | M S3 Y | 1X+3Y |
| 3X+1Y | M S2 X | ____+1__ | ↔ | ____−1__ | M S3 Y | 1X+3Y |
| 3X+1Y | S2 X | ______+1 | ↔ | ______−1 | S3 Y | 1X+3Y |
| Slice | Pattern | Mask | v. | Mask | Pattern | Slice |

By design, the 3X+1Y slice occurs ¼ times per a word sent within an odd page.
By design, the 1X+3Y slice occurs ¾ times per a word sent within an odd page.
Therefore, the [normalized] average power per a word sent within an odd page is:

$P_{1,3,5,7} = \frac{1}{4} \times (\frac{3}{4} \cdot P_X + \frac{1}{4} \cdot P_Y) + \frac{3}{4} \times (\frac{1}{4} \cdot P_X + \frac{3}{4} \cdot P_Y) = \frac{49}{96}$ . (≈ 0.51)

NOTE – During transmission, an odd page occurs with the probability $p_{1,3,5,7} = \frac{4}{8}$.

TABLE XI
DESIGN COMPARISON

| Parameter | 1000BASE-T | Draft Design | "Extensive" | Proposed |
|---|---|---|---|---|
| Coding means | single | multiplexed | multiplexed | multiplexed |
| └data transfer | └1000 Mbit/s | ├1000 Mbit/s | ├1000 Mbit/s | ├1000 Mbit/s |
| └event transfer | | └6.25 Mev/s | └6.25 Mev/s | └6.25 Mev/s |
| Trellis encoder | eight-state | eight-state | eight-state | eight-state |
| └pages used | └P0÷P7,⅛ | └P0÷P7,⅛ | └P0÷P7,⅛ | └P0÷P7,⅛ |
| Pts per P0 | 64 | 72 | 96 | 97 |
| ├of slice 4X+0Y | ├16, 16/64 | ├16, 16/72 | ├16, 16/96 | ├16, ⅛ |
| └of slice 0X+4Y | └48, 48/64 | └56, 56/72 | └80, 80/96 | └81, ⅞ |
| Pts per P2,4,6 | 64 (192Σ) | 72 (216Σ) | 72 (216Σ) | 72 (216Σ) |
| └of slice 2X+2Y | └64, 64/64 | └72, 72/72 | └72, 72/72 | └72, 1/1 |
| Pts per P1,3,5,7 | 64 (256Σ) | 72 (288Σ) | 78 (312Σ) | 78 (312Σ) |
| ├of slice 3X+1Y | ├24, 24/64 | ├24, 24/72 | ├24, 24/78 | ├24, ¼ |
| └of slice 1X+3Y | └40, 40/64 | └48, 48/72 | └54, 54/78 | └54, ¾ |
| (total) | 512Σ out of 625 | 576Σ out of 625 | 624Σ out of 625 | 625Σ out of 625 |
| Scr. unification | per design | per design | per page group | per slice |
| Scr. base(s) | single: 2 | two: 2, 3 | three: 2, 3, 13 | two: 2, 3 |
| NAP at P0 | $\approx 0.56 \approx \mu_0^{+13\%}$ | $\approx 0.57 \approx \mu_0^{+17\%}$ | $\approx 0.60 \approx \mu_0^{+20\%}$ | $\approx 0.61 \approx \mu_0^{+22\%}$ |
| NAP at P2,4,6 | $\approx 0.46 \approx \mu_0^{-3\%}$ | $\approx 0.46 \approx \mu_0^{-6\%}$ | $\approx 0.46 \approx \mu_0^{-8\%}$ | $\approx 0.46 \approx \mu_0^{-9\%}$ |
| NAP at P1,3,5,7 | $\approx 0.48 = \mu_0$ | $\approx 0.49 \approx \mu_0$ | $\approx 0.50 \approx \mu_0$ | $\approx 0.51 \approx \mu_0^{+1\%}$ |
| NAP mean, $\mu_0$ | ≈0.48 | ≈0.49 | ≈0.50 | ≈0.50 |
| NAP std dev, σ | ≈0.03 | ≈0.04 | ≈0.04 | ≈0.05 |
| NAP variance, $\sigma^2$ | ≈0.001 | ≈0.001 | ≈0.002 | ≈0.002 |
| NAP $\sigma/\mu_0$ | ≈6.56% | ≈7.25% | ≈8.60% | ≈9.58% |
| NAP $\mu_0 \pm 1\sigma$ | 0.48 ± 0.03 | 0.49 ± 0.04 | 0.50 ± 0.04 | 0.50 ± 0.05 |
| NAP $\mu_0 \pm 2\sigma$ | 0.48 ± 0.06 | 0.49 ± 0.07 | 0.50 ± 0.09 | 0.50 ± 0.10 |
| NAP $\mu_0 \pm 3\sigma$ | 0.48 ± 0.10 | 0.49 ± 0.11 | 0.50 ± 0.13 | 0.50 ± 0.14 |

NOTE – Pts = points employed (Σ = group summary); NAP = normalized average power.
NOTE – Probabilities (expressed ita/ic) are within the respective higher essences.
NOTE – Scr = scrambling; 2, 3, 13 are the factors of $96=2^5 \cdot 3$, $72=2^3 \cdot 3^2$, $78=2 \cdot 3 \cdot 13$.
NOTE – Since 97 is a prime, a 625-pts "extensive" needs a base-97 sub-scrambler.

a word into a four-dimensional code point comprising exclusively either PAM2X or PAM3Y single-dimensional levels, respectively, at the same rate, see Tables VI and VII.

Within the same dictionary again, there are finally the four odd pages, P1, P3, P5, and P7, all are "asymmetric" like P0, each corresponding with the coding slices 3X+1Y and 1X+3Y, see Table II, which two the coding means also processes in different ways, transforming the information bits of a word into a four-dimensional code point comprising either three PAM2X but just one PAM3Y or, vice versa, just one PAM2X but three PAM3Y single-dimensional levels, respectively, all at the same rate, too, see Tables VIII and IX.

Defining those coding schemes, we preserve the page fading means introduced in the draft design,[6] but update the corresponding streaming rule to simplify the coding means, adding the stretchable ESCs,[7] see Tables XII and XIII, similarly to the introduced in [3]. We also assume that base-3 scrambling, included in the coding means since the draft design, is implementable similarly to the described in [4] and [5].

## CODING BALANCE

We seek to balance the proposed coding means within the full transport dictionary in first, considering that it definitively consists of the two separate parts—one comprising only even pages and other comprising only odd pages, respectively—we assume opposite, and then, separately, within the even pages together comprising just a little more than a half of the code points,[8] see Table V in conjunction with Tables IV, VI, and VII, and within the odd pages together comprising just a little less than a half of the same space,[9] see Table X in conjunction with Tables VIII and IX.

[6] In Tables XII and XIII, the periods affected by that means are labeled by black triangles pointing on the "fade-in" and "fade-out" series, see [2].

[7] Extra signal codes bounding the payload on its head and tail, see [3].

TABLE XII
TAIL ESC ECHO-DRIVEN STRETCHING PATTERNS INSCRIBED WITHIN MINIMAL INTER-PAYLOAD GLUE

| Remaining Echo ................... | | | | | Payload Tail | $IPG_1$ | $IPG_2$ | $IPG_3$ | $IPG_4$ | $IPG_5$ | $IPG_6$ | $IPG_7$ | $IPG_8$ | $IPG_9$ | $IPG_{10}$ | $IPG_{11}$ | $IPG_{12}$ | Head | <T> |
|---|---|---|---|---|---|---|---|---|---|---|---|---|---|---|---|---|---|---|---|
| ......................... 1/1 (echo is absent) → | | | | | | $\uparrow^{2\blacktriangledown}$ | $\uparrow^{1\blacktriangledown}$ | $\$^{1}$ | $\$^{2}$ | @ | @ | @ | @ | $\$^{2}$ | $\$^{1}$ | $\uparrow^{1\blacktriangle}$ | $\uparrow^{2\blacktriangle}$ | .... | T = 0 |
| 1/8;1/4;1/2 → | $\#^{6}$ ** | $\#^{5}$ ** | $\#^{4}$ ** | $\#^{3}$ ** | $\#^{2}$ ** | $\uparrow^{2\blacktriangledown}$ ** | $\uparrow^{1\blacktriangledown}$ | $\$^{1}$ | $\$^{2}$ | @ | @ | @ | @ | $\$^{2}$ | $\$^{1}$ | $\uparrow^{1\blacktriangle}$ | $\uparrow^{2\blacktriangle}$ | .... | T = 0 |
| ..... 1/8;1/4;1/2 → | | $\#^{6}$ ** | $\#^{5}$ ** | $\#^{4}$ ** | $\#^{3}$ ** | $\uparrow^{2\blacktriangledown}$ ** | $\uparrow^{1\blacktriangledown}$ ** | $\$^{1}$ | $\$^{2}$ | @ | @ | @ | @ | $\$^{2}$ | $\$^{1}$ | $\uparrow^{1\blacktriangle}$ | $\uparrow^{2\blacktriangle}$ | .... | T = 0 |
| ........... 1/8;1/4;1/2 → | | | $\#^{6}$ ** | $\#^{5}$ ** | $\#^{4}$ ** | $\uparrow^{4}$ ** | $\uparrow^{3}$ ** | $\uparrow^{2\blacktriangledown}$ ** | $\uparrow^{1\blacktriangledown}$ | $\$^{1}$ | $\$^{2}$ | @ | @ | $\$^{2}$ | $\$^{1}$ | $\uparrow^{1\blacktriangle}$ | $\uparrow^{2\blacktriangle}$ | .... | T = 1 |
| .................. 1/8;1/4;1/2 → | | | | $\#^{6}$ ** | $\#^{5}$ ** | $\uparrow^{4}$ ** | $\uparrow^{3}$ ** | $\uparrow^{2\blacktriangledown}$ ** | $\uparrow^{1\blacktriangledown}$ ** | $\$^{1}$ | $\$^{2}$ | @ | @ | $\$^{2}$ | $\$^{1}$ | $\uparrow^{1\blacktriangle}$ | $\uparrow^{2\blacktriangle}$ | .... | T = 1 |
| ........................ 1/8;1/4;1/2 → | | | | | $\#^{6}$ ** | $\uparrow^{6}$ ** | $\uparrow^{5}$ ** | $\uparrow^{4}$ ** | $\uparrow^{3}$ ** | $\uparrow^{2\blacktriangledown}$ ** | $\uparrow^{1\blacktriangledown}$ | $\$^{1}$ | $\$^{2}$ | $\$^{2}$ | $\$^{1}$ | $\uparrow^{1\blacktriangle}$ | $\uparrow^{2\blacktriangle}$ | .... | T = 2 |
| .............................. 1/8;1/4;1/2 → | | | | | | $\uparrow^{6}$ ** | $\uparrow^{5}$ ** | $\uparrow^{4}$ ** | $\uparrow^{3}$ ** | $\uparrow^{2\blacktriangledown}$ ** | $\uparrow^{1\blacktriangledown}$ ** | $\$^{1}$ | $\$^{2}$ | $\$^{2}$ | $\$^{1}$ | $\uparrow^{1\blacktriangle}$ | $\uparrow^{2\blacktriangle}$ | .... | T = 2 |
| NOTE – ESC[$] on T=3 causes a processing delay of 9 words, that we consider too long.<br>NOTE – ESC[$] on T=3 is depicted for comparison only, not recommended to implement. | | | | | | $\uparrow^{8}$ | $\uparrow^{7}$ | $\uparrow^{6}$ | $\uparrow^{5}$ | $\uparrow^{4}$ | $\uparrow^{3}$ | $\uparrow^{2\blacktriangledown}$ | $\uparrow^{1\blacktriangledown}$ | $\$^{1}$ | $\$^{1}$ | $\uparrow^{1\blacktriangle}$ | $\uparrow^{2\blacktriangle}$ | .... | T = 3 |

TABLE XIII
HEAD ESC USER-DEMANDED STRETCHING PATTERNS INSCRIBED WITHIN MINIMAL INTER-PAYLOAD GLUE

| <H> | Tail | $IPG_1$ | $IPG_2$ | $IPG_3$ | $IPG_4$ | $IPG_5$ | $IPG_6$ | $IPG_7$ | $IPG_8$ | $IPG_9$ | $IPG_{10}$ | $IPG_{11}$ | $IPG_{12}$ | Payload Head ................... User Extra Bits |
|---|---|---|---|---|---|---|---|---|---|---|---|---|---|---|
| H = 0 | .... | $\uparrow^{2\blacktriangledown}$ | $\uparrow^{1\blacktriangledown}$ | $\$^{1}$ | $\$^{2}$ | @ | @ | @ | @ | $\$^{2}$ | $\$^{1}$ | $\uparrow^{1\blacktriangle}$ | $\uparrow^{2\blacktriangle}$ | .............................. up to 4 + 6 = 10 |
| H = 1 | .... | $\uparrow^{2\blacktriangledown}$ | $\uparrow^{1\blacktriangledown}$ | $\$^{1}$ | $\$^{2}$ | @ | @ | $\$^{2}$ | $\$^{1}$ | $\uparrow^{1\blacktriangle}$ | $\uparrow^{2\blacktriangle}$ | $\uparrow^{3}$ | $\uparrow^{4}$ | ......................... up to 4 + 6 + 8 + 8 = 26 |
| H = 2 | .... | $\uparrow^{2\blacktriangledown}$ | $\uparrow^{1\blacktriangledown}$ | $\$^{1}$ | $\$^{2}$ | $\$^{2}$ | $\$^{1}$ | $\uparrow^{1\blacktriangle}$ | $\uparrow^{2\blacktriangle}$ | $\uparrow^{3}$ | $\uparrow^{4}$ | $\uparrow^{5}$ | $\uparrow^{6}$ | ................... up to 4 + 6 + 8 + 8 + 8 + 8 = 42 |
| H = 3 | .... | $\uparrow^{2\blacktriangledown}$ | $\uparrow^{1\blacktriangledown}$ | $\$^{1}$ | $\$^{1}$ | $\uparrow^{1\blacktriangle}$ | $\uparrow^{2\blacktriangle}$ | $\uparrow^{3}$ | $\uparrow^{4}$ | $\uparrow^{5}$ | $\uparrow^{6}$ | $\uparrow^{7}$ | $\uparrow^{8}$ | NOTE – [$]ESC on H=3 causes a processing delay of 9 words, that we consider too long.<br>NOTE – [$]ESC on H=3 is depicted for comparison only, not recommended to implement. |

Balancing within the dictionary, we implement the so called quantitative criterion established as the goal to reach a probabilistic equality in the use of those two parts, see Table III, while balancing within each part, we implement the so called structural criterion established now as the goal to prevent an arithmetic inequality in the use of oppositely weighted coding structures employed in the coding process, see Tables V and X, per each twisted pair in the link. Remember, however, that the structural criterion is widely variable and, therefore, has a visible subjective dimension.

## POWER BALANCE

Examining the power characteristics of the resulting coding means, see Table XI, we conclude that the proposed way looks not good enough. However, comparing the same power characteristics among the listed designs, including the 1000BASE-T original coding scheme, then the draft design and its possible "extensively" extended variant, and then the proposed way, we conclude that all the addressed approaches look the same bad due to the trait highlighted earlier in this paper, the cause of that is the paging-driven probability disproportions caused, in their turn, by the ("code set partitioning") model chosen to be the ground of the forward error correction means, i.e., TCM, since the original scheme, see also Table III again.

Because and until we preserve the TCM site and its associated paging model, we cannot reach a solution balanced so its (normalized) average power either remains the same, i.e., does not differ at all, or at least changes negligibly, i.e., varies not as much as now, between the provided pages.[10] Therefore, we read that the power balance is not a manageable property during the synthesis flow and accept the corresponding design characteristics just such as they are.

## FURTHER WORK

It is clear for us to go further, we particularly need a fresh design ground that will help us to balance the interesting characteristics, including those of the average power, by providing us with the synthesis basics necessary to enable a "uniform" (distortion-less[11]) mapping of a "binary-metric" (base-2) code space onto a "non-binary-metric" (another base) code space, where the badge "uniform" assumes that such a mapping does not change (not distort) the probability distribution functions those code spaces are initially given with.

As far as we know, that is a gap in coding theory we need to fill up in first. Thus, we plan to do this and then continue to develop the proposed way.

[8] Together exact 313 out of 625 code points are in pages P0, P2, P4, P6.

[9] Together exact 312 out of 625 code points are in pages P1, P3, P5, P7.

[10] We suppose $0\%$ is the ideal case and $1/625 < 1\%$ is an enough level.

[11] Distortion-less is the ideal case, a lesser one is what we seek for at least.

# Rethinking the Constellation Model of the Spread Data Coding Means and Event Coding Means Multiplexed Over the 1000BASE-T PMA Sublayer

Alexander Ivanov

***Abstract*—In this paper, we are developing well the linguistic multiplexing branch dedicated to the Gigabit Ethernet physical layer, type 1000BASE-T, by very accurate matching of its coding, line code, and scrambling spaces, all to neutralize the line code level probability distribution distortion caused by the design of the TCM-based FEC means embodied in the layer.**

## Introduction

DEVELOPING the 1000BASE-T physical layer, the draft design [1] done in the useful try to extend its predecessor with an event transfer function, focusing first on the linguistic multiplexing aspect, demonstrated a good applicability of such a multiplexing even in a design based on TCM.[*→1]

However, the spread design [2] done in the spurious try to develop its predecessor further, focusing next on the constellation coverage aspect, exhibited an unacceptable distortion in the probabilistic and dependent (main of all, power) properties of the underlying line code, also present in the original design, that every time was caused by the contradiction between how the TCM model defines and how the FEC means employs the paged coding space, lasting unresolved until now.

Addressing this in the paper, we describe a way eliminating the mentioned contradiction and its outcomes.

## Renewed Modeling Principles

Adopting the design approach developed in [3], we consider the original 1000BASE-T subset-based page bundle—covering the respective 4D-PAM5 constellation completely—the maternal framework and, therefore, all its elements—which all are the corresponding points of that constellation—the subject of an appropriately applicable reject or repeat operation.[2]

Given the constellation, we evaluate its properties quantitatively, in the form interesting us, see Table I, expressing this as a matrix of hit rates,[3] per-level (row) by per-pair (column), which are the measures behind the (per-pair) level probability distribution picture,[4] the correct (ideal), or at least as close to that as possible, state of which we need to achieve.

TABLE I
4D-PAM5 Constellation Metrics

| Matrix of Level Hit Rates | | | | | values if TCM not applied | | | | | Probability of Occurrence | | | |
|---|---|---|---|---|---|---|---|---|---|---|---|---|---|
| $H_{A,-2}$ | $H_{B,-2}$ | $H_{C,-2}$ | $H_{D,-2}$ | | 125 | 125 | 125 | 125 | | $.2_{exact}$ | $.2_{exact}$ | $.2_{exact}$ | $.2_{exact}$ |
| $H_{A,-1}$ | $H_{B,-1}$ | $H_{C,-1}$ | $H_{D,-1}$ | | 125 | 125 | 125 | 125 | | $.2_{exact}$ | $.2_{exact}$ | $.2_{exact}$ | $.2_{exact}$ |
| $H_{A,=0}$ | $H_{B,=0}$ | $H_{C,=0}$ | $H_{D,=0}$ | = | 125 | 125 | 125 | 125 | → | $.2_{exact}$ | $.2_{exact}$ | $.2_{exact}$ | $.2_{exact}$ |
| $H_{A,+1}$ | $H_{B,+1}$ | $H_{C,+1}$ | $H_{D,+1}$ | | 125 | 125 | 125 | 125 | | $.2_{exact}$ | $.2_{exact}$ | $.2_{exact}$ | $.2_{exact}$ |
| $H_{A,+2}$ | $H_{B,+2}$ | $H_{C,+2}$ | $H_{D,+2}$ | | 125 | 125 | 125 | 125 | | $.2_{exact}$ | $.2_{exact}$ | $.2_{exact}$ | $.2_{exact}$ |

TABLE II
Symmetry Definition

| Level Hit Rate Deltas | | | | *per-pair values meaning* | Other Useful Properties |
|---|---|---|---|---|---|
| $\Delta H_Y$ | $\Delta H_Y$ | $\Delta H_Y$ | $\Delta H_Y$ | hits into Y(L)-level bin | sum of these gives |
| $\Delta H_X$ | $\Delta H_X$ | $\Delta H_X$ | $\Delta H_X$ | hits into X(−)-level bin | hits into each pair |
| $\Delta H_Y$ | $\Delta H_Y$ | $\Delta H_Y$ | $\Delta H_Y$ | hits into Y(z)-level bin | $\Delta H_\Sigma = \Delta H_\Sigma(S)$ |
| $\Delta H_X$ | $\Delta H_X$ | $\Delta H_X$ | $\Delta H_X$ | hits into X(+)-level bin | that is also equal to |
| $\Delta H_Y$ | $\Delta H_Y$ | $\Delta H_Y$ | $\Delta H_Y$ | hits into Y(H)-level bin | hits into the pages |

TABLE III
Constellation Modeling Involving Symmetries

| Matrix of Level Hit Rates | | | | ← both sides → are modeled as a sum of effects of the same symmetries | Bundle of Page Hit Rates |
|---|---|---|---|---|---|
| $H_Y$ | $H_Y$ | $H_Y$ | $H_Y$ | | (even) (even) (even) (even) $H_0$ $H_2$ $H_4$ $H_6$ |
| $H_X$ | $H_X$ | $H_X$ | $H_X$ | $H_Y = \sum_s \Delta H_Y(S)$ | |
| $H_Y$ | $H_Y$ | $H_Y$ | $H_Y$ | $H_X = \sum_s \Delta H_X(S)$ | $H_0+\ldots+H_7 = H_\Sigma = 3H_Y+2H_X$ |
| $H_X$ | $H_X$ | $H_X$ | $H_X$ | $H_P = \sum_s \Delta H_P(S)$, $P=0\div7$ | |
| $H_Y$ | $H_Y$ | $H_Y$ | $H_Y$ | | $H_1$ $H_3$ $H_5$ $H_7$ (odd) (odd) (odd) (odd) |

Along the constellation, subsets, and pages, all going further the same manner as before, we introduce one new composite entity, a symmetry, see Table II,[5] again comprising a number of (distinct) points defining its size,[6] that is intended to be the only subject of all the following manipulations, regenerating all the former entities this new way, see Table III.

Having understood all the ties of all the composite entities, both already given and newly introduced, instead of the given as caused by TCM, we restrict on all the pages to be uniform in their *nominal* size,[7] see Tables IV and V, that, in the turn, restricts on symmetries, guiding us in how we should collect them, see Table VI, and grants us the solid ground enough for performing all our further steps, confidently.

Recalling the fate of submission of many prior works to the peer reviewed journal, such a try with this one also promises no chance, probably.

Please sorry for the author has no time to find this work a new home, peer reviewed or not, except of arXiv, and just hopes there it meets its reader, one or maybe various, whom the author beforehand thanks for their regard.

A. Ivanov is with JSC Continuum, Yaroslavl, the Russian Federation.

[*] All notes, except this, are listed after the References, see page 12.

TABLE IV
MATCHING PRINCIPLE

| $H_P = f\,(\text{TCM}) = \text{varied}$ | Original → Ground Behind → Matched | $H_P = f\,(\text{symmetries}) = const$ |
|---|---|---|
| $H_{P,\text{TCM-caused}} = \sum_s \Delta H_P\,(S)$ | per-symmetry REJECT / REPEAT ($R = 0$) / ($R > 0$) operation ⟶ | $H_{P,\text{balanced}} = \sum_s \Delta H_P\,(S) \cdot R_s$ |

TABLE V
MATCHING PRINCIPLE DEPICTION

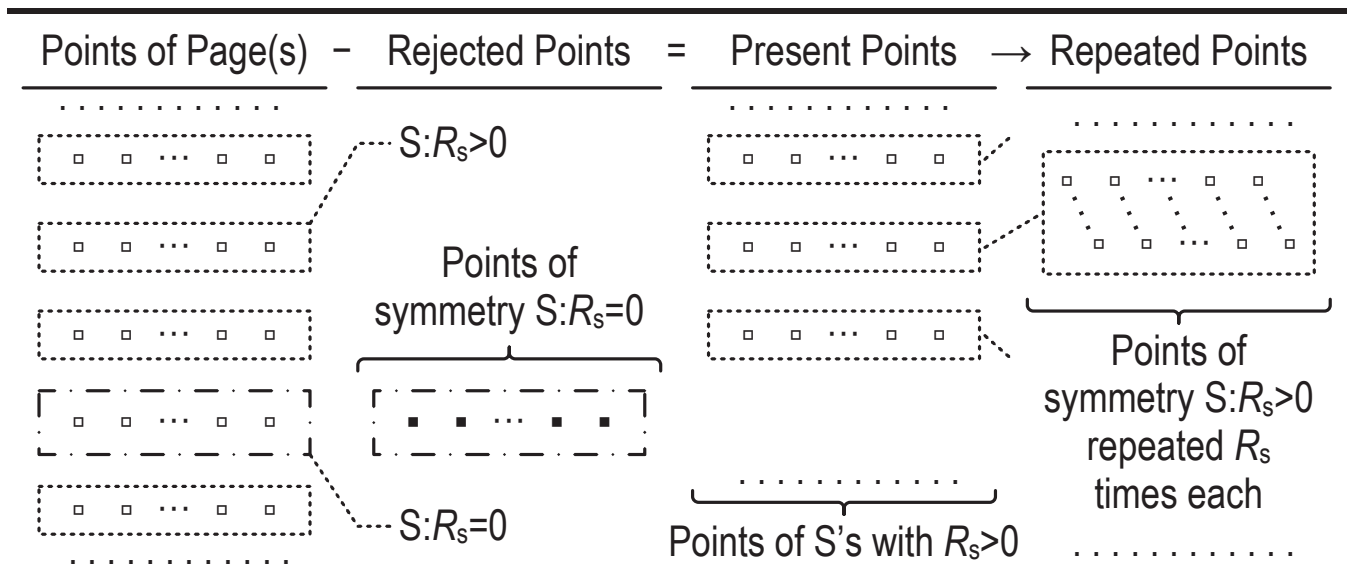

TABLE VI
SYMMETRIES COLLECTING PRINCIPLES

| Effect on Even Pages | Effect on Odd Pages |
|---|---|
| $\Delta H_0(S)$ any, $\Delta H_2(S) = \Delta H_4(S) = \Delta H_6(S)$ | $\Delta H_1(S) = \Delta H_3(S) = \Delta H_5(S) = \Delta H_7(S)$ |

TABLE VII
SYMMETRIES COLLECTED ACROSS EVEN PAGE P0

| Symmetry | $Y_3$ $Y_2$ $Y_1$ $Y_0$ == == == == | $Y_3$ $Y_2$ $Y_1$ $Y_0$ ++ ++ ++ ++ | $Y_3$ $Y_2$ $Y_1$ $Y_0$ −− −− −− −− | Size | Syn. | Reps. |
|---|---|---|---|---|---|---|
| 1-odd-1 | L L L L | z z z z | H H H H | 3 | $^3S_1$ | $^3R_1$ |
| 1-odd-2 | L L L z | z z z H | H H H L | 3 | $^3S_2$ | $^3R_2$ |
| 1-odd-3 | L L L H | z z z L | H H H z | 3 | $^3S_3$ | $^3R_3$ |
| 1-odd-4 | L L z L | z z H z | H H L H | 3 | $^3S_4$ | $^3R_4$ |
| 1-odd-5 | L L z z | z z H H | H H L L | 3 | $^3S_5$ | $^3R_5$ |
| 1-odd-6 | L L z H | z z H L | H H L z | 3 | $^3S_6$ | $^3R_6$ |
| 1-odd-7 | L L H L | z z L z | H H z H | 3 | $^3S_7$ | $^3R_7$ |
| 1-odd-8 | L L H z | z z L H | H H z L | 3 | $^3S_8$ | $^3R_8$ |
| 1-odd-9 | L L H H | z z L L | H H z z | 3 | $^3S_9$ | $^3R_9$ |
| 1-odd-10 | L z L L | z H z z | H L H H | 3 | $^3S_{10}$ | $^3R_{10}$ |
| 1-odd-11 | L z L z | z H z H | H L H L | 3 | $^3S_{11}$ | $^3R_{11}$ |
| 1-odd-12 | L z L H | z H z L | H L H z | 3 | $^3S_{12}$ | $^3R_{12}$ |
| 1-odd-13 | L z z L | z H H z | H L L H | 3 | $^3S_{13}$ | $^3R_{13}$ |
| 1-odd-14 | L z z z | z H H H | H L L L | 3 | $^3S_{14}$ | $^3R_{14}$ |
| 1-odd-15 | L z z H | z H H L | H L L z | 3 | $^3S_{15}$ | $^3R_{15}$ |
| 1-odd-16 | L z H L | z H L z | H L z H | 3 | $^3S_{16}$ | $^3R_{16}$ |
| 1-odd-17 | L z H z | z H L H | H L z L | 3 | $^3S_{17}$ | $^3R_{17}$ |
| 1-odd-18 | L z H H | z H L L | H L z z | 3 | $^3S_{18}$ | $^3R_{18}$ |
| 1-odd-19 | L H L L | z L z z | H z H H | 3 | $^3S_{19}$ | $^3R_{19}$ |
| 1-odd-20 | L H L z | z L z H | H z H L | 3 | $^3S_{20}$ | $^3R_{20}$ |
| 1-odd-21 | L H L H | z L z L | H z H z | 3 | $^3S_{21}$ | $^3R_{21}$ |
| 1-odd-22 | L H z L | z L H z | H z L H | 3 | $^3S_{22}$ | $^3R_{22}$ |
| 1-odd-23 | L H z z | z L H H | H z L L | 3 | $^3S_{23}$ | $^3R_{23}$ |
| 1-odd-24 | L H z H | z L H L | H z L z | 3 | $^3S_{24}$ | $^3R_{24}$ |
| 1-odd-25 | L H H L | z L L z | H z z H | 3 | $^3S_{25}$ | $^3R_{25}$ |
| 1-odd-26 | L H H z | z L L H | H z z L | 3 | $^3S_{26}$ | $^3R_{26}$ |
| 1-odd-27 | L H H H | z L L L | H z z z | 3 | $^3S_{27}$ | $^3R_{27}$ |
| | $X_3$ $X_2$ $X_1$ $X_0$ == == == == | $X_3$ $X_2$ $X_1$ $X_0$ ~~ ~~ ~~ ~~ | Y-LEVEL ROTATION TRUTH TABLE | ↑Σ 81 ↓Σ 16 } Σ 97 pts | | |
| 1-even-1 | − − − − | + + + + | | 2 | $^2S_1$ | $^2R_1$ |
| 1-even-2 | − − − + | + + + − | | 2 | $^2S_2$ | $^2R_2$ |
| 1-even-3 | − − + − | + + − + | | 2 | $^2S_3$ | $^2R_3$ |
| 1-even-4 | − − + + | + + − − | | 2 | $^2S_4$ | $^2R_4$ |
| 1-even-5 | − + − − | + − + + | X-LEVEL ROTATION TRUTH TABLE | 2 | $^2S_5$ | $^2R_5$ |
| 1-even-6 | − + − + | + − + − | | 2 | $^2S_6$ | $^2R_6$ |
| 1-even-7 | − + + − | + − − + | | 2 | $^2S_7$ | $^2R_7$ |
| 1-even-8 | − + + + | + − − − | | 2 | $^2S_8$ | $^2R_8$ |

Y-LEVEL ROTATION TRUTH TABLE

| | | | |
|---|---|---|---|
| Y : | L | z | H |
| X== : | L | z | H |
| X++ : | z | H | L |
| X−− : | H | L | z |

X-LEVEL ROTATION TRUTH TABLE

| | | |
|---|---|---|
| X : | + | − |
| X== : | + | − |
| X~~ : | − | + |

## DECOMPOSITION: FROM SUBSETS INTO SYMMETRIES

There are infinite decomposition options possible, however, any possible factors restricting the underlying search is out of the scope of this paper, so we can perform the decomposition just illustratively, i.e., arbitrarily to some extent, allowing for any (we believe) clear symmetries, including cross-subset and cross-page, to be present in the decomposition result.

Conducting the search for symmetries, S's paired with $R$'s, in a systematic shape via periodic rotations of period-ordered, level-meaning, subset-reflecting variables, X's and Y's, we find $27{+}8+9+6{+}9{+}1{+}4=64$ distinct, non-intersecting symmetries varying in their size and content, all together covering (all the points of) all the pages, see Tables VII, VIII, and IX, and therefore (as well as all the points of) the whole constellation, as one can state, not uniformly, but completely.

Got the decomposition done, we divide the symmetries into seven groups by the details—in the scope of an affected entity, see Table X—of the effect occurring with every one repetition of a symmetry from a given among the groups, see Table XI, that allows us to impersonalize the symmetries and, thanks to this, further operate on a group a generalized manner.

## BALANCING BY BUBBLING

Facing the need to match between the three heterogeneous spaces—which are the coding space consisting of the symmetries, non-regular in its content and non-binary in its size, the line code space, already regular but still non-binary due to its original design, and the scrambling space we leverage, regular and binary due to the original design, too—with no (ideal), or at least as so close to such as acceptable, matching error, we formulate the corresponding search exercise, considering it a balancing problem, see Table XII, also solvable a generalized manner transiting separate $R$'s of S's to $\Sigma R$'s.

During this, we bound all the pages in their least *effective* size to not shrink below the multiplicity of echo-driven noted words, $N_{\text{E}}$, enough to enable a linguistic multiplexing process over the paged coding space [1], see Table XIII, as well as in their most nominal size to not spread outside the performance of the scrambler-backing sources leveraged since the original design with no modification, see Table XIV.

Got the boundaries drawn, we prepare a series of balancing variants, appropriate in action and illustrative in comparison, see Table XV, among those are some requiring a sub-scrambler of a base different than a power of two, but of 5 or 9 instead, which could be anchor-based, like in [4], or interval-mapping, like in [5], we assume implementable when backable from the same sources, see Tables XVI and XVII, respectively.

Except for the variant of $H_{\text{page}} = 72$, see Table XVIII, the balancing routine looks like we bubble (up) the constellation, managing its nominal size and content to match the line code-related properties, all the same time as we bubble (up or down each of) the pages, managing their nominal size and content to match the paged coding-related as well as scrambling-related properties, all thanks to we bubble (up or down, by repeat and reject, respectively, each of) the right symmetries we already collected across all the original composite entities.

TABLE VIII
SYMMETRIES COLLECTED ACROSS EVEN PAGES EXCEPT P0

| Symmetry | Y1 Y0 □ □ (== ==) | Y1 □ Y0 □ (−− ++) | . . . | Y1 □ □ Y0 (++ ++) | . . . | . . . | □ Y1 Y0 □ (−− −−) | . . . | □ Y1 □ Y0 (++ −−) | □ □ Y1 Y0 (== ==) | Size | Syn. | Reps. |
|---|---|---|---|---|---|---|---|---|---|---|---|---|---|
| 1-XX-1 | L L □ □ | H □ z □ | three-row −− sub-rotation | z □ □ z | six-row then three-row ++ sub-rotations | six-row then three-row ++ sub-rotations | □ H H □ | three-row −− sub-rotation | □ z □ H | □ □ L L | 24 | $^{24}S_1$ | $^{24}R_1$ |
| 1-XX-2 | L z □ □ | H □ H □ | | z □ □ L | | | □ H z □ | | □ L □ H | □ □ L z | 24 | $^{24}S_2$ | $^{24}R_2$ |
| 1-XX-3 | L H □ □ | z □ H □ | | H □ □ L | | | □ L z □ | | □ z □ z | □ □ L H | 24 | $^{24}S_3$ | $^{24}R_3$ |
| 1-XX-4 | z L □ □ | H □ L □ | | L □ □ z | | | □ z H □ | | □ H □ H | □ □ z L | 24 | $^{24}S_4$ | $^{24}R_4$ |
| 1-XX-5 | z z □ □ | L □ H □ | | H □ □ H | | | □ L L □ | | □ H □ L | □ □ z z | 24 | $^{24}S_5$ | $^{24}R_5$ |
| 1-XX-6 | z H □ □ | L □ L □ | | H □ □ z | | | □ L H □ | | □ z □ L | □ □ z H | 24 | $^{24}S_6$ | $^{24}R_6$ |
| 1-XX-7 | H L □ □ | z □ z □ | | L □ □ H | | | □ z L □ | | □ H □ z | □ □ H L | 24 | $^{24}S_7$ | $^{24}R_7$ |
| 1-XX-8 | H z □ □ | L □ z □ | | z □ □ H | | | □ H L □ | | □ L □ L | □ □ H z | 24 | $^{24}S_8$ | $^{24}R_8$ |
| 1-XX-9 | H H □ □ | z □ L □ | | L □ □ L | | | □ z z □ | | □ L □ z | □ □ H H | 24 | $^{24}S_9$ | $^{24}R_9$ |
| | 36 pts of page P2 | 36 pts of page P6 | | 36 pts of page P4 | DEFINITIONS: □ = { − ; + } , \|□\| = 2 | | 36 pts of page P4 | | 36 pts of page P6 | 36 pts of page P2 | Σ 216 pts | | |

TABLE IX
SYMMETRIES COLLECTED ACROSS ODD PAGES

| Symmetry | Y2 Y1 Y0 □ (== == ==) | Y1 Y0 □ Y2 (== == ==) | Y0 □ Y2 Y1 (== == ==) | □ Y2 Y1 Y0 (== == ==) |
|---|---|---|---|---|
| 3-X-1.1 | L L L □ | L L □ L | L □ L L | □ L L L |
| 2-X-1.1 | L L z □ | L z □ L | z □ L L | □ L L z |
| 2-X-2.1 | L L H □ | L H □ L | H □ L L | □ L L H |
| 2-X-4.1 | L z L □ | z L □ L | L □ L z | □ L z L |
| 2-X-5.1 | L z z □ | z z □ L | z □ L z | □ L z z |
| 1-X-1 | L z H □ | z H □ L | H □ L z | □ L z H |
| 2-X-8.1 | L H L □ | H L □ L | L □ L H | □ L H L |
| 1-X-2 | L H z □ | H z □ L | z □ L H | □ L H z |
| 2-X-5.2 | L H H □ | H H □ L | H □ L H | □ L H H |
| 2-X-6.1 | z L L □ | L L □ z | L □ z L | □ z L L |
| 2-X-9.1 | z L z □ | L z □ z | z □ z L | □ z L z |
| 1-X-3 | z L H □ | L H □ z | H □ z L | □ z L H |
| 2-X-3.1 | z z L □ | z L □ z | L □ z z | □ z z L |
| 3-X-1.2 | z z z □ | z z □ z | z □ z z | □ z z z |
| 2-X-2.2 | z z H □ | z H □ z | H □ z z | □ z z H |
| 1-X-4 | z H L □ | H L □ z | L □ z H | □ z H L |
| 2-X-8.2 | z H z □ | H z □ z | z □ z H | □ z H z |
| 2-X-6.2 | z H H □ | H H □ z | H □ z H | □ z H H |
| 2-X-7.1 | H L L □ | L L □ H | L □ H L | □ H L L |
| 1-X-5 | H L z □ | L z □ H | z □ H L | □ H L z |
| 2-X-9.2 | H L H □ | L H □ H | H □ H L | □ H L H |
| 1-X-6 | H z L □ | z L □ H | L □ H z | □ H z L |
| 2-X-7.2 | H z z □ | z z □ H | z □ H z | □ H z z |
| 2-X-4.2 | H z H □ | z H □ H | H □ H z | □ H z H |
| 2-X-3.2 | H H L □ | H L □ H | L □ H H | □ H H L |
| 2-X-1.2 | H H z □ | H z □ H | z □ H H | □ H H z |
| 3-X-1.3 | H H H □ | H H □ H | H □ H H | □ H H H |
| | X2 X1 X0 ● (== == ==) | X1 X0 ● X2 (== == ==) | X0 ● X2 X1 (== == ==) | ● X2 X1 X0 (== == ==) |
| 2-Y-1.1 | − − − ● | − − ● − | − ● − − | ● − − − |
| 2-Y-2.1 | − − + ● | − + ● − | + ● − − | ● − − + |
| 2-Y-3.1 | − + − ● | + − ● − | − ● − + | ● − + − |
| 2-Y-4.1 | − + + ● | + + ● − | + ● − + | ● − + + |
| 2-Y-4.2 | + − − ● | − − ● + | − ● + − | ● + − − |
| 2-Y-3.2 | + − + ● | − + ● + | + ● + − | ● + − + |
| 2-Y-2.2 | + + − ● | + − ● + | − ● + + | ● + + − |
| 2-Y-1.2 | + + + ● | + + ● + | + ● + + | ● + + + |
| | 54+24 pts of page P1 | 54+24 pts of page P3 | 54+24 pts of page P7 | 54+24 pts of page P5 |

Same symmetries re-shown with ties between parts, if any → Grouped by

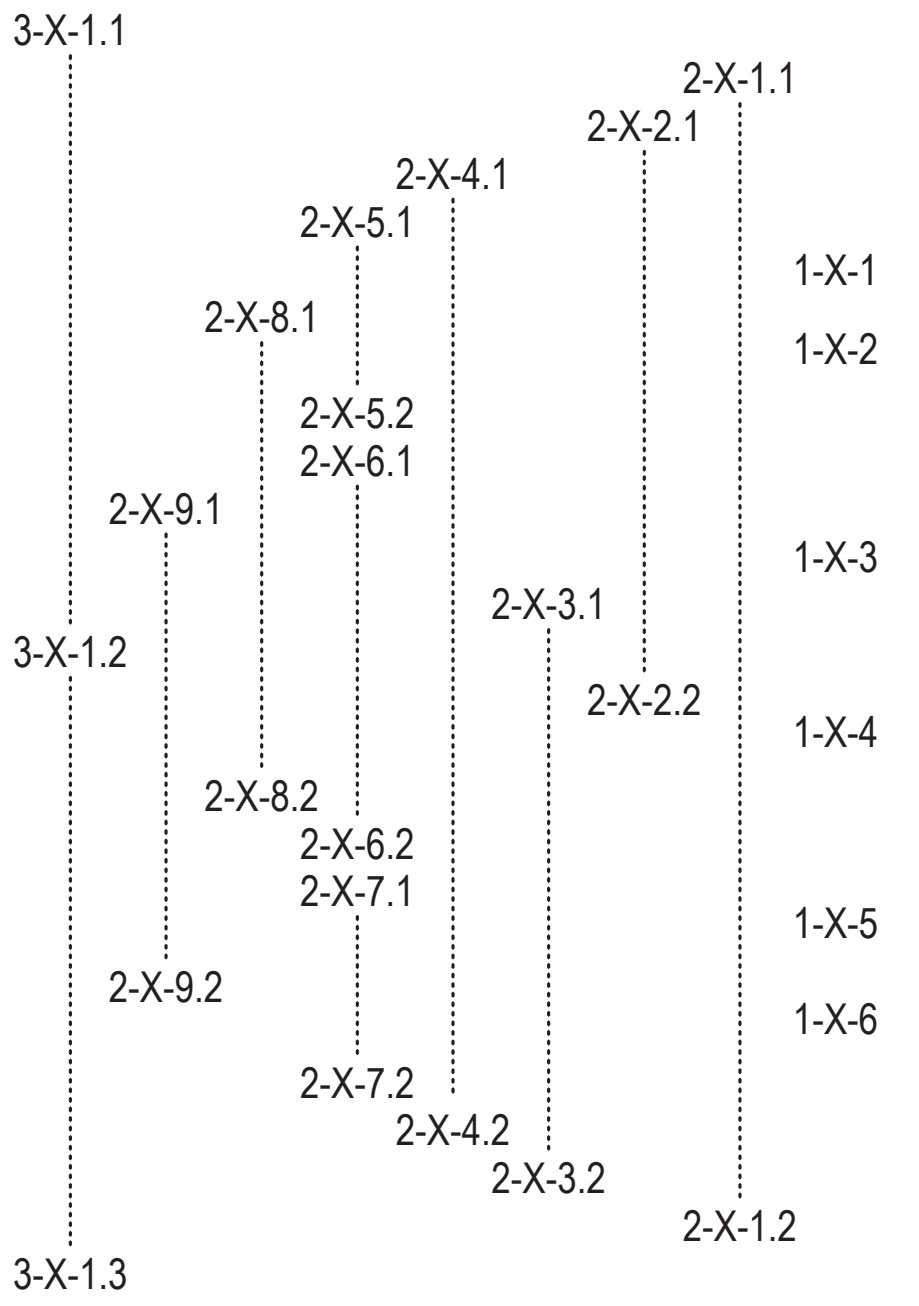

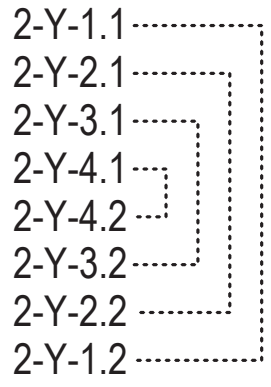

| Grouped | Size | Syn. | Reps. |
|---|---|---|---|
| 1-X-1 | 8 | $^{8}S_1$ | $^{8}R_1$ |
| 1-X-2 | 8 | $^{8}S_2$ | $^{8}R_2$ |
| 1-X-3 | 8 | $^{8}S_3$ | $^{8}R_3$ |
| 1-X-4 | 8 | $^{8}S_4$ | $^{8}R_4$ |
| 1-X-5 | 8 | $^{8}S_5$ | $^{8}R_5$ |
| 1-X-6 | 8 | $^{8}S_6$ | $^{8}R_6$ |
| 2-X-1.1, 2-X-1.2 | 16 | $^{16}S_1$ | $^{16}R_1$ |
| 2-X-2.1, 2-X-2.2 | 16 | $^{16}S_2$ | $^{16}R_2$ |
| 2-X-3.1, 2-X-3.2 | 16 | $^{16}S_3$ | $^{16}R_3$ |
| 2-X-4.1, 2-X-4.2 | 16 | $^{16}S_4$ | $^{16}R_4$ |
| 2-X-5.1, 2-X-5.2 | 16 | $^{16}S_5$ | $^{16}R_5$ |
| 2-X-6.1, 2-X-6.2 | 16 | $^{16}S_6$ | $^{16}R_6$ |
| 2-X-7.1, 2-X-7.2 | 16 | $^{16}S_7$ | $^{16}R_7$ |
| 2-X-8.1, 2-X-8.2 | 16 | $^{16}S_8$ | $^{16}R_8$ |
| 2-X-9.1, 2-X-9.2 | 16 | $^{16}S_9$ | $^{16}R_9$ |
| 3-X-1.1, 3-X-1.2, 3-X-1.3 | 24 | $^{24}S_{10}$ | $^{24}R_{10}$ |
| | ↑Σ 216, ↓Σ 96 → Σ 312 pts | | |
| 2-Y-1.1, 2-Y-1.2 | 24 | $^{24}S_{11}$ | $^{24}R_{11}$ |
| 2-Y-2.1, 2-Y-2.2 | 24 | $^{24}S_{12}$ | $^{24}R_{12}$ |
| 2-Y-3.1, 2-Y-3.2 | 24 | $^{24}S_{13}$ | $^{24}R_{13}$ |
| 2-Y-4.1, 2-Y-4.2 | 24 | $^{24}S_{14}$ | $^{24}R_{14}$ |

TCM PAGES ALLOCATED IN 4D-PAM5 CONSTELLATION SPACE

| P. | Parity | Contents | Points Within |
|---|---|---|---|
| P0 | Even | XXXX+YYYY | 16 + 81 = 97 |
| P2 | Even | XXYY+YYXX | 36 + 36 = 72 |
| P4 | Even | XYYX+YXXY | 36 + 36 = 72 |
| P6 | Even | XYXY+YXYX | 36 + 36 = 72 |
| P1 | Odd | XXXY+YYYX | 24 + 54 = 78 |
| P3 | Odd | XXYX+YYXY | 24 + 54 = 78 |
| P5 | Odd | XYYY+YXXX | 54 + 24 = 78 |
| P7 | Odd | XYXX+YXYY | 24 + 54 = 78 |

DEFINITIONS: ● = { L ; z ; H } , |●| = 3

TABLE X
HIT RATE MAP

| PAM-5 LEVEL | LEVEL Type | Hits per Level for Pair <A> | <B> | <C> | <D> | Hits per Even Page P0 | P2 | P4 | P6 |
|---|---|---|---|---|---|---|---|---|---|
| −2/2 u | Y(L) | $H_Y$ | $H_Y$ | $H_Y$ | $H_Y$ | $H_0$ | $H_2$ | $H_4$ | $H_6$ |
| −1/2 u | X(−) | $H_X$ | $H_X$ | $H_X$ | $H_X$ | Σ of Hits per Pages | | | |
| 0 | Y(z) | $H_Y$ | $H_Y$ | $H_Y$ | $H_Y$ | | | | |
| +1/2 u | X(+) | $H_X$ | $H_X$ | $H_X$ | $H_X$ | $H_1$ | $H_3$ | $H_5$ | $H_7$ |
| +2/2 u | Y(H) | $H_Y$ | $H_Y$ | $H_Y$ | $H_Y$ | P1 | P3 | P5 | P7 |
| Σ of Hits per Levels = | | $H_\Sigma$ = | $H_\Sigma$ = | $H_\Sigma$ = | $H_\Sigma$ = | Hits per Odd Page | | | |

TABLE XI
REPEAT OPERATION EFFECT

| ↓Repeating gives→ | $\Delta H_Y$ | $\Delta H_X$ | $\Delta H_0$ | $\Delta H_1$ | $\Delta H_2$ | $\Delta H_3$ | $\Delta H_4$ | $\Delta H_5$ | $\Delta H_6$ | $\Delta H_7$ | $\Delta H_\Sigma$ |
|---|---|---|---|---|---|---|---|---|---|---|---|
| anyone of $^{2}S_{1\div8}$ | — | +1 | +2 | — | — | — | — | — | — | — | +2 |
| anyone of $^{3}S_{1\div27}$ | +1 | — | +3 | — | — | — | — | — | — | — | +3 |
| anyone of $^{8}S_{1\div6}$ | +2 | +1 | — | +2 | — | +2 | — | +2 | — | +2 | +8 |
| anyone of $^{16}S_{1\div9}$ | +4 | +2 | — | +4 | — | +4 | — | +4 | — | +4 | +16 |
| anyone of $^{24}S_{1\div9}$ | +4 | +6 | — | — | +8 | — | +8 | — | +8 | — | +24 |
| the one of $^{24}S_{10}$ | +6 | +3 | — | +6 | — | +6 | — | +6 | — | +6 | +24 |
| anyone of $^{24}S_{11\div14}$ | +2 | +9 | — | +6 | — | +6 | — | +6 | — | +6 | +24 |

TABLE XII
BALANCING PROBLEM DEFINITION

| Aspect | Contents | | Remark |
|---|---|---|---|
| Given | $H_\Sigma : H_\Sigma \bmod 8 = 0,\ 8{\bullet}N_E \le H_\Sigma \le$ rand. source performance | | hits/pair |
| Relation | $3{\bullet}H_Y + 2{\bullet}H_X = H_\Sigma = 8{\bullet}H_{page}$ | | hits/level to hits/page matching |
| Target | $\Delta H_{level} = \lvert H_Y - H_X \rvert \rightarrow \min$ | | hits/y-level to hits/x-level matching |
| Con-straints | ${}^2R_{1\div8},\ {}^3R_{1\div27},\ {}^8R_{1\div6},\ {}^{16}R_{1\div9},\ {}^{24}R_{1\div14} \ge 0$ | | reject or repeat |
| | $H_0 = H_1 = H_2 = H_3 = H_4 = H_5 = H_6 = H_7 = H_{page}$ | | uniform pages |
| where $P(R) = 1$ if $R > 0$ and 0 otherwise | $\lvert{}^2S_1\rvert \bullet P({}^2R_1) + \ldots + \lvert{}^2S_1\rvert \bullet P({}^2R_1) + \lvert{}^3S_1\rvert \bullet P({}^3R_1) + \ldots + \lvert{}^3S_{27}\rvert \bullet P({}^3R_{27})$ | $\ge 1 \bullet N_E$ | page P0 |
| | $\lvert{}^{24}S_1\rvert \bullet P({}^{24}R_1) + \ldots + \lvert{}^{24}S_9\rvert \bullet P({}^{24}R_9)$ | $\ge 3 \bullet N_E$ | pages P2,4,6 |
| | $\lvert{}^8S_1\rvert \bullet P({}^8R_1) + \ldots + \lvert{}^8S_6\rvert \bullet P({}^8R_6) + \lvert{}^{16}S_1\rvert \bullet P({}^{16}R_1) + \ldots + \lvert{}^{16}S_9\rvert \bullet P({}^{16}R_9) + \lvert{}^{24}S_{10}\rvert \bullet P({}^{24}R_{10}) + \ldots + \lvert{}^{24}S_{14}\rvert \bullet P({}^{24}R_{14})$ | ↕ effective no. of pts in ↕ $\ge 4 \bullet N_E$ | pages P1,3,5,7 |
| Find | $\Sigma^2R_{1\div8},\ \Sigma^3R_{1\div27},\ \Sigma^8R_{1\div6},\ \Sigma^{16}R_{1\div9},\ \Sigma^{24}R_{1\div9},\ {}^{24}R_{10},\ \Sigma^{24}R_{11\div14}$ | | times/group |

TABLE XIII
IMPLIED LINGUISTIC MULTIPLEXING PROFILE

| $N_C$ | $N_R$ | $N_E$ | GCD($N_E$,$N_C$) | $N_E$/GCD | $E$ | $n_e$ | $k$ | Root Bin. Spaces |
|---|---|---|---|---|---|---|---|---|
| 64 | 8 | 72 | $8 = 2^3 = N_R$ | $2^3 < 9 < 2^4$ | $2^1$ | 6 | 3 | $2^{19} + 2^{12} + 2^{11} + \ldots$ |
| modulus → tr. capacity | | | postfix mortheme | root mortheme | echo cancellation params | | | 531,441 echo samples total |

TABLE XIV
AVAILABLE RANDOMITY

| Period | Sources | PRNG-sourced | (Extra) | Total |
|---|---|---|---|---|
| $n \ge 0$ | PRNG($n$), ($n$) | $sx_n$[0:3], $sy_n$[0:3], $sg_n$[0:3] | $n$ mod 2 | 12 (13) |
| Remark→ | updated once/word | four bits + four bits + four bits | single bit | bits/word |

TABLE XV
BALANCE VARIANTS

| $\lvert s\cdots_n \rvert$ | Extra+ | $H_\Sigma$ | $H_Y^{unbalance}$ | | $H_X^{unbalance}$ | $H_{page}$ | ⅓ $p$(Y) | ½ $p$(X) |
|---|---|---|---|---|---|---|---|---|
| $2^6/2$ | $n$ mod 9, or base-9 sub-scr. | 576 | $116^{+1.\%}$ | > | $114^{-1.\%}$ | 72 | .201 | .198 |
| | | | $114^{-1.\%}$ | < | $117^{+2.\%}$ | 72 | .198 | .203 |
| $2^7/2$ | same as above | 1,152 | $230^{-.2\%}$ | < | $231^{+.3\%}$ | 144 | .1997 | .2005 |
| | | | $232^{+1.\%}$ | > | $228^{-1.\%}$ | 144 | .201 | .198 |
| $2^8/2$ | same as above | 2,304 | $460^{-.2\%}$ | < | $462^{+.3\%}$ | 288 | .1997 | .2005 |
| | | | $462^{+.3\%}$ | > | $459^{-.4\%}$ | 288 | .2005 | .1992 |
| $2^9/2$ | same as above | 4,608 | $922^{+.04\%}$ | > | $921^{-.07\%}$ | 576 | .20009 | .19987 |
| | | | $920^{-.2\%}$ | < | $924^{+.3\%}$ | 576 | .1997 | .2005 |
| $2^{10}/2$ | — | 1,024 | $204^{-.4\%}$ | < | $206^{+.6\%}$ | 128 | .1992 | .2012 |
| | | | $206^{+.6\%}$ | > | $203^{-.9\%}$ | 128 | .2012 | .1982 |
| $2^{11}/2$ | — | 2,048 | $410^{+.1\%}$ | > | $409^{-.1\%}$ | 256 | .2002 | .1997 |
| | | | $408^{-.4\%}$ | < | $412^{+.6\%}$ | 256 | .1992 | .2012 |
| $2^{12}/2$ | — | 4,096 | $820^{+.1\%}$ | > | $818^{-.1\%}$ | 512 | .2002 | .1997 |
| | | | $818^{-.1\%}$ | < | $821^{+.2\%}$ | 512 | .1997 | .2004 |
| $2^{13}/2$ | — | 8,192 | $1{,}638^{-.02\%}$ | < | $1{,}639^{+.04\%}$ | 1,024 | .19995 | .20007 |
| | | | $1{,}640^{+.1\%}$ | > | $1{,}636^{-.1\%}$ | 1,024 | .2002 | .1997 |
| $2^{\ge7}/2$ | $n$ mod 5, or base-5 sub-scr. | $2^{\ge7}{\bullet}5$ | $(2^{\ge7})^{equal}$ | = | $(2^{\ge7})^{equal}$ | $2^{\ge4}{\bullet}5$ | $.2_{exact}$ | $.2_{exact}$ |

TABLE XVI
BASE-5 SUB-SCRAMBLER DESIGN

| ↓Feature / Implementation→ | Anchor-based, $A_5(n)$ | Mapping $\lvert s\cdots_n \rvert \rightarrow 5$ | | | |
|---|---|---|---|---|---|
| Demand in the use of $\lvert s\cdots_n \rvert$ | $2^3$ (due to $2^2 < 5 < 2^3$) | $2^3$ | $2^4$ | $2^5$ | $2^6$ |
| Error arising due to base conversion, $\lvert\varepsilon\rvert$ | 0% | <40% | <30% | <10% | <7% |

TABLE XVII
BASE-9 SUB-SCRAMBLER DESIGN

| ↓Feature / Implementation→ | Anchor-based, $A_9(n)$ | Mapping $\lvert s\cdots_n \rvert \rightarrow 9$ | | | |
|---|---|---|---|---|---|
| Demand in the use of $\lvert s\cdots_n \rvert$ | $2^4$ (due to $2^3 < 9 < 2^4$) | $2^4$ | $2^5$ | $2^6$ | $2^7$ |
| Error arising due to base conversion, $\lvert\varepsilon\rvert$ | 0% | <50% | <20% | <15% | <6% |

TABLE XVIII
BALANCING EXAMPLES

| $H_\Sigma$ | $H_Y : H_X$ | $H_{page}$ | $\Sigma^2R_{\ldots}$ | $\Sigma^3R_{\ldots}$ | $\Sigma^8R_{\ldots}$ | $\Sigma^{16}R_{\ldots}$ | $\Sigma^{24}R_{<}$ | ${}^{24}R_{10}$ | $\Sigma^{24}R_{>}$ |
|---|---|---|---|---|---|---|---|---|---|
| Original Constellation → | | | 8 | 27 | 6 | 9 | 9 | 1 | 4 |
| (625) | (125) : (125) | (81;78;72) | (all once) | (all once) | (all once) | (all once) | (all once) | (all once) | (all once) |
| $2^6{\bullet}9$ | 116 : 114 | 72 | 8−8 | 27−3 | 6 | 9 | 9 | 1−1 | 4 |
| $2^7{\bullet}9$ | 230 : 231 | 144 | 8−5 | 27+19 | 2×6 | 2×9 | 2×9 | — | 2×4 |
| $2^8{\bullet}9$ | 460 : 462 | 288 | 2×3 | 2×46 | 4×6 | 4×9 | 4×9 | — | 4×4 |
| $2^9{\bullet}9$ | 922 : 921 | 576 | 8+1 | 27+159 | 8×6 | 8×9 | 8×9 | — | 8×4 |
| $2^{10}$ | 204 : 206 | $2^7$ | 8+2 | 27+9 | 6+6 | 9+5 | 9+7 | 1+1 | 4+2 |
| $2^{11}$ | 410 : 409 | $2^8$ | 8+9 | 27+47 | 2×12 | 2×14 | 2×16 | 2×2 | 2×6 |
| $2^{12}$ | 820 : 818 | $2^9$ | 2×17 | 2×74 | 4×12 | 4×14 | 4×16 | 4×2 | 4×6 |
| $2^{13}$ | 1638 : 1639 | $2^{10}$ | 8+63 | 27+267 | 8×12 | 8×14 | 8×16 | 8×2 | 8×6 |
| $2^7{\bullet}5$ | 128 : 128 | 80 | 8−4 | 27−3 | 6 | 9+2 | 9+1 | 1−1 | 4 |
| $2^8{\bullet}5$ | 256 : 256 | 160 | 2×4 | 2×24 | 2×6 | 2×11 | 2×10 | — | 2×4 |
| $2^9{\bullet}5$ | 512 : 512 | 320 | 4×4 | 4×24 | 4×6 | 4×11 | 4×10 | — | 4×4 |
| $2^{10}{\bullet}5$ | 1024 : 1024 | 640 | 8×4 | 8×24 | 8×6 | 8×11 | 8×10 | — | 8×4 |

## CONCLUSION

We described a way triple matching between a non-regular and non-binary coding space, a regular but yet non-binary line code space, and a regular and binary scrambling space.

Describing the way, we solved the problem of very accurate matching of the 1000BASE-T physical layer's FEC specifics, 4D-PAM5 constellation, and PRNG output, respectively.

Due to the way we chose, every solution we met is capable to neutralize the distortion caused by the use of TCM, as well as to run an appropriate linguistic multiplexing process.

[1]TCM is for Trellis Coded Modulation, other abbreviations encountered in this paper are FEC for Forward Error Correction, PAM for Pulse Amplitude Modulation, and PRNG for Pseudo Random Number Generator.

[2]In the scope of an entity, e.g., a symmetry, repeat (more precisely, repeat once) and reject (only once anyway, obviously) are operations complement to each other, i.e., equal in the magnitude while different in the sign of their effect, applicable many times and just a time, respectively.

[3]A hit rate is given implicitly related to the respective reference, i.e., a time interval lasting exact $H_\Sigma$ word time periods of $T_0 = 8$ ns each.

[4]We use P (always alone, typed uppercase upright), $P$ (always of a scalar argument, typed uppercase italic), and $p$ (always of an event argument, typed lowercase italic) to denote a page index (in-set iterator), a symmetry presence function (scalar), and a probability function (scalar), respectively.

[5]We use H (no index, typed upright) and $H$ (always indexed, typed italic) to denote a PAM-5 level (symbol) and a hit rate (scalar), respectively.

[6]We use S (always alone, typed uppercase upright) to denote a symmetry, ${}^{size}\mathrm{S}_{ord}$ (always accompanied, typed the same as S), to denote a synonym of the symmetry whose size is $size = |\mathrm{S}|$ and ordinal is $ord$, unique among all the symmetries of the same size, in the decomposition done, and s (always with the word time period index, $n$, typed lowercase upright) to denote a set of all possible combinations of a number of pseudo random values.

[7]Nominal size of an entity is the number of all its points, counter-balanced, effective size of an entity is the number of its distinct points only.

# Data Coding Means and Event Coding Means Multiplexed Inside the 1000BASE-T PMA Sublayer

Alexander Ivanov

***Abstract*—In the paper, we're starting a new branch dedicated to the Gigabit Ethernet physical layer, type 1000BASE-T. This branch is intended to multiplex the data coding function, already present in, and an event transfer function, new to the layer, both inside the 1000BASE-T PMA sublayer.**

***Index Terms*—Ethernet, MDI output power balancing, power balancing, PAM-5, PAM-17, 1000BASE-T.**

## Introduction

POWER balancing is a prerequisite of the coding synthesis flow, realized during developing of the draft design [1], as the initial try to multiplex the 1000BASE-T-originated data coding means and a pulse-per-an-interval event coding means, into the spread one [2], as the first but spurious try to balance the characteristics of the former, including those of the MDI[1] output power, and then into the bubbled design [3], as the next and now successful step achieving the power balance as close to the ideal as it can be possible, taking the 4D-PAM5 model, all done over the 1000BASE-T PMA sublayer [4].

In this paper, we move the point the implementation of the proposed coding, that purposely services the user, touches the rest of the 1000BASE-T essentials, that directly interacts the media, locating such the split point one nominal foot deeper, now not over but inside the mentioned sublayer.

We begin with a clear, two-dimensional basis, see Tables I, III, and IV briefly, and consecutively come close to the target, four-dimensional reference, see Table XX briefly.

We assume a simple streaming rule, see Table XI briefly, as the pattern of the coding output, enough to reflect any shape of a data as well as of an event, using preemption.

During our further description, we will consider the details completely so as they are seen in the view of the transmitting side, in the form of a set of different measures undertaken by the side when it encodes the physical signal, and just assume the receiving side involving, when it decodes the signal, in the turn, a set of the respective means capable to benefit on such the measures into the favor of the whole transmission channel those both sides terminate on its opposite ends.[2]

Recalling the fate of submission of many prior works to the peer reviewed journal, such a try with this one also promises no chance, probably.

Please sorry for the author has no time to find this work a new home, peer reviewed or not, except of arXiv, and just hopes there it meets its reader, one or maybe various, whom the author beforehand thanks for their regard.

A. Ivanov is with JSC Continuum, Yaroslavl, the Russian Federation.

[1]It is the 1000BASE-T Media Dependent Interface for the four-pair media under the 1000BASE-T Physical Medium Attachment sublayer, see [4].

[2]In the case of 1000BASE-T, e.g., such a measure is the trellis encoding scheme and such a means is the Viterbi decoding algorithm, respectively.

TABLE I
CONVENTIONAL 2D-PAM5 CONSTELLATION

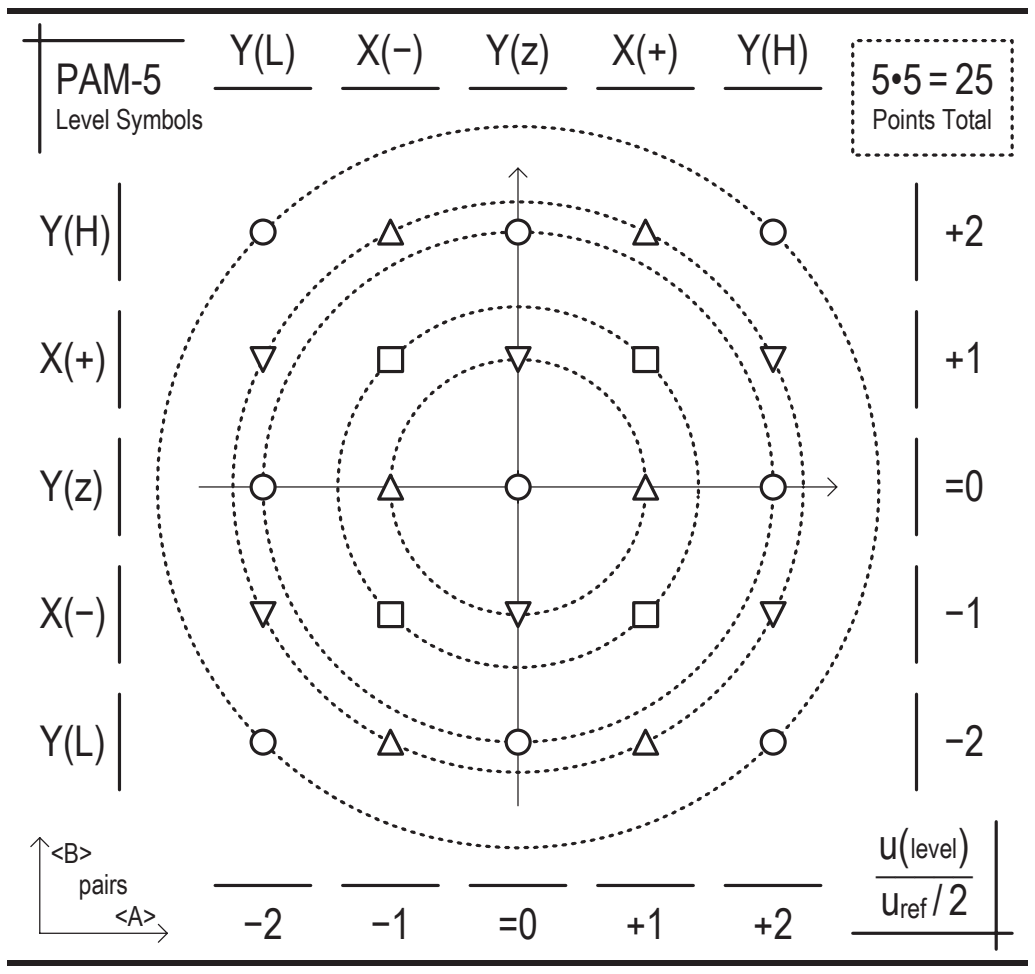

TABLE II
POWER ORBITS

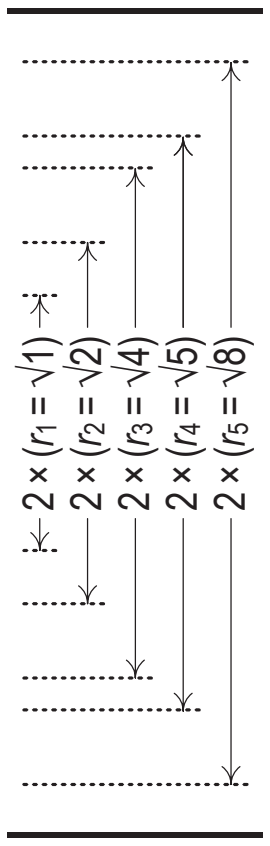

NOTE – $r_0 = 0$.

TABLE III
2D-PAM5 BASIC SUBSETS

| Alias | Sym. | Pts of $r_0$ | $r_1$ | $r_2$ | $r_3$ | $r_4$ | $r_5$ | ΣPts | Rem. |
|---|---|---|---|---|---|---|---|---|---|
| YY | ○ {3×3} | 1 | — | — | 4 | — | 4 | 9 | Σ13 |
| XX | □ {2×2} | — | — | 4 | — | — | — | 4 | |
| YX | ▽ {3×2} | — | 2 | — | — | 4 | — | 6 | Σ12 |
| XY | △ {2×3} | — | 2 | — | — | 4 | — | 6 | |

TABLE IV
2D-PAM5 GEOMETRIC PARTITIONING PROPERTIES

| Selection | Content(s) | Power Orbit(s) | $D_\delta$ | $\Delta_{\delta,\min}$ | $G_\delta$ |
|---|---|---|---|---|---|
| Silence | Ø (single point) | $r_0$ | √16 | — | — |
| Pagelet ... | YY−Ø (8 pts) | $r_3$ — $r_5$ | √16 | √4 | +6.0 dB |
| | XX (4 pts) | $r_2$ | √16 | √4 | +6.0 dB |
| | YX (6 pts) | $r_1$ — $r_4$ | √16 | √4 | +6.0 dB |
| | XY (6 pts) | $r_1$ — $r_4$ | √16 | √4 | +6.0 dB |
| Page ... (12 pts each) | P0: XX + YY−Ø | $r_2$ $r_3$ — $r_5$ | √16 | √2 | +3.8 dB |
| | P1: XY + YX | $r_1$ — $r_4$ | √16 | √2 | +3.8 dB |
| Dictionary | P0 + P1 | $r_1$ $r_2$ $r_3$ $r_4$ $r_5$ | √16 | √1 | +2.5 dB |

## Supposed Metric Measure

An added into the whole coding gain is a metric measure, comparable by because calculable as its contribution:

$$G = 20\log_{10}\frac{D}{D-\Delta_{\min}}\,,$$

where $D$ is the diameter of the measure and $\Delta_{\min}$ is the least same-unit distance it is characterized with, respectively.

TABLE V
2D-PAM5 CONSTELLATION ANALYSIS USING RAYS

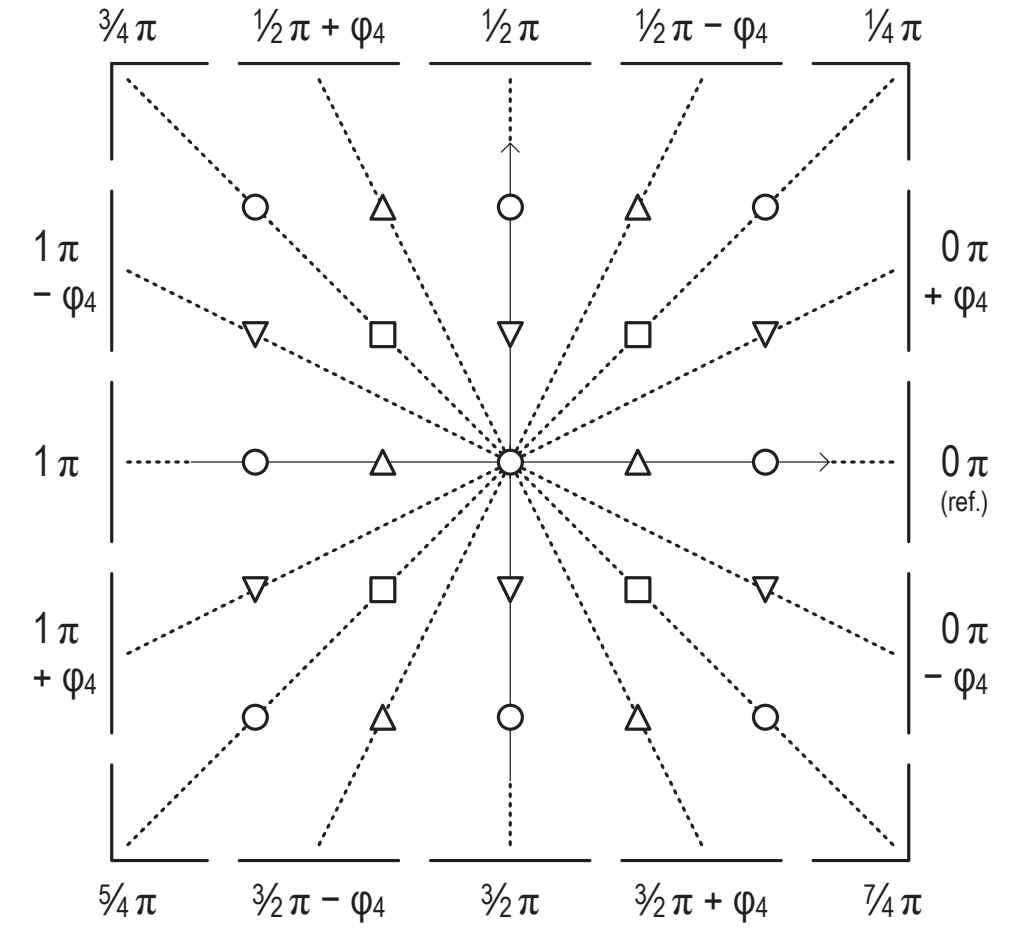

TABLE VI
POWER ANGLES

| Angle | Quadrant |
|---|---|
| 0 π (ref.) | Q1 |
| 0 π + φ4 | |
| ¼ π | |
| ½ π − φ4 | |
| ½ π | Q2 |
| ½ π + φ4 | |
| ¾ π | |
| 1 π − φ4 | |
| 1 π | Q3 |
| 1 π + φ4 | |
| 5⁄4 π | |
| 3⁄2 π − φ4 | |
| 3⁄2 π | Q4 |
| 3⁄2 π + φ4 | |
| 7⁄4 π | |
| 0 π − φ4 | |

TABLE VII
2D-PAM5 CONSTELLATION SUPPLEMENTARY PORTRAYING METRICS

| Angle | | At-ray Pts of $r_0$ | $r_1$ | $r_2$ | $r_3$ | $r_4$ | $r_5$ | ΣPts | Rem. |
|---|---|---|---|---|---|---|---|---|---|
| 0 π | (reference) | ○ (1 of 9) | △ (1 of 6) | — | ○ (1 of 9) | — | — | 1+2 | Q1 |
| 0 π + φ4 | ≈ 0.15 π | ○ (1 of 9) | — | — | — | ▽ (1 of 6) | — | 1+1 | |
| ¼ π | = 0.25 π | ○ (1 of 9) | — | □ (1 of 4) | — | — | ○ (1 of 9) | 1+2 | |
| ½ π − φ4 | ≈ 0.35 π | ○ (1 of 9) | — | — | — | △ (1 of 6) | — | 1+1 | |
| ½ π | = 0.50 π | ○ (1 of 9) | ▽ (1 of 6) | — | ○ (1 of 9) | — | — | 1+2 | Q2 |
| ½ π + φ4 | ≈ 0.65 π | ○ (1 of 9) | — | — | — | △ (1 of 6) | — | 1+1 | |
| ¾ π | = 0.75 π | ○ (1 of 9) | — | □ (1 of 4) | — | — | ○ (1 of 9) | 1+2 | |
| 1 π − φ4 | ≈ 0.85 π | ○ (1 of 9) | — | — | — | ▽ (1 of 6) | — | 1+1 | |
| 1 π | = 1.00 π | ○ (1 of 9) | △ (1 of 6) | — | ○ (1 of 9) | — | — | 1+2 | Q3 |
| 1 π + φ4 | ≈ 1.15 π | ○ (1 of 9) | — | — | — | ▽ (1 of 6) | — | 1+1 | |
| 5⁄4 π | = 1.25 π | ○ (1 of 9) | — | □ (1 of 4) | — | — | ○ (1 of 9) | 1+2 | |
| 3⁄2 π − φ4 | ≈ 1.35 π | ○ (1 of 9) | — | — | — | △ (1 of 6) | — | 1+1 | |
| 3⁄2 π | = 1.50 π | ○ (1 of 9) | ▽ (1 of 6) | — | ○ (1 of 9) | — | — | 1+2 | Q4 |
| 3⁄2 π + φ4 | ≈ 1.65 π | ○ (1 of 9) | — | — | — | △ (1 of 6) | — | 1+1 | |
| 7⁄4 π | = 1.75 π | ○ (1 of 9) | — | □ (1 of 4) | — | — | ○ (1 of 9) | 1+2 | |
| 0 π − φ4 | ≈ 1.85 π | ○ (1 of 9) | — | — | — | ▽ (1 of 6) | — | 1+1 | |

TABLE VIII
2D-PAM5 TRIGONOMETRIC PARTITIONING ANGLE-BOUND PROPERTIES

| Selection | Content(s) | Angle-sharing Orbits | $D_\varphi$ | $\Delta_{\varphi,min}$ | $G_\varphi$ |
|---|---|---|---|---|---|
| Silence | Ø (single point) | $r_0$ ——— | 2 π | — | — |
| Pagelet ... | YY−Ø (8 pts) | ——— | 2 π | ¼ π | +1.2 dB |
| | XX (4 pts) | ——— | 2 π | ½ π | +2.5 dB |
| | YX (6 pts) | ——— | 2 π | φ4+φ4 | +1.4 dB |
| | XY (6 pts) | ——— | 2 π | φ4+φ4 | +1.4 dB |
| Page ... (12 pts each) | P0: XX + YY−Ø | —— $r_2$ —— $r_5$ | 2 π | — | — |
| | P1: XY + YX | ——— | 2 π | φ4 | +0.7 dB |
| Dictionary | P0 + P1 | — $r_1$ $r_2$ $r_3$ — $r_5$ | 2 π | — | — |

TABLE IX
2D-PAM5 TRIGONOMETRIC PARTITIONING RADIUS-BOUND PROPERTIES

| Selection | Content(s) | Most Close Orbits | $D_\rho$ | $\Delta_{\rho,min}$ | $G_\rho$ (assuming $G_\varphi$ exists) |
|---|---|---|---|---|---|
| Pagelet ... | YY−Ø | $r_3$ — $r_5$ | 2 $r_5$ | $r_5 - r_3$ | +1.4 dB |
| NOTE – In XX, there is one orbit. | YX, XY | $r_1$ —— $r_4$ | 2 $r_5$ | $r_4 - r_1$ | +2.1 dB |
| Page ... | P0: XX + YY−Ø | $r_2$ $r_3$ | 2 $r_5$ | $r_3 - r_2$ | +0.9 dB |
| | P1: XY + YX | $r_1$ —— $r_4$ | 2 $r_5$ | $r_4 - r_1$ | +2.1 dB |
| Dictionary | P0 + P1 | $r_3$ $r_4$ | 2 $r_5$ | $r_4 - r_3$ | +0.4 dB |

TABLE X
POSSIBLE 2D-PAM5 CODING VARIANTS

| Scheme | Page P0 | Page P1 | Ø | $G_{TCM}$ | $G_\delta$ | $G_\varphi$ | $G_\rho$ | ΣG's |
|---|---|---|---|---|---|---|---|---|
| 2x8/8 + Ø | YY only | $r_4$ only | involved, too | — | 2.498 | 0.665 | 0.370 | 3.534 |
| 12/12 + Ø | entirely | — | | — | 3.789 | — | n/a | 3.789 |
| | — | entirely | | — | 3.789 | 0.665 | 2.141 | 6.596 |
| 12/24 + Ø | entirely | entirely | | > 0 | 2.498 | — | n/a | >2.498 |
| Pts in P/D | Dict. Resources Involved | | | Expressed floored, as not less than, in dB | | | | |

TABLE XI
IMPLIED PAYLOAD FRAMING PROCESS

| Tr. Phase → | · · · IPG | USD | ESC | Payload | ESC | USD | IPG · · · |
|---|---|---|---|---|---|---|---|
| Sel. Used → | · · · $\overline{Ø}$ | Ø | $\overline{Ø}$ | $\overline{Ø}$ | $\overline{Ø}$ | Ø | $\overline{Ø}$ · · · |

In the design, we suppose one, TCM-sourced measure, that is a metric measure of a given $G_{\text{TCM}} > 0$ as is.[3]

## SUPPOSED GEOMETRIC MEASURE

In the design, we suppose one, plain (or two-dimensional), coordinate geometry-based measure, that is a metric measure whose gain, $G_\delta$, reflects (because it depends on the respective diameter and) the least Euclidean distance between the points of a given selection, see Tables I, III, and IV [4] once.

Thereby, the distance between a given point and the origin reflects (because is proportional, in a direct ratio, to) the MDI output power, as it is fed into the media, whose variations we highlight with the so called, respective, circular power orbits, see Table II once, each of a finite, discrete radius, unique and equal to such the distance and, thus, reflecting such the power in the media, too, see Tables I, II, III, and IV more.

Expectedly, the MDI output power for a point at the origin is absent (zero), therefore we label it as $\emptyset$ ("silence").[5]

## PROPOSED TRIGONOMETRIC MEASURE

In the design, we propose one, orbit-originated, one-sphere-space, see Tables V, VI, and VI, trigonometry-based measure whose angle-related gain part, $G_\varphi$, see Table VIII, and radius-related gain part, $G_\rho$, see Table IX,[6] can extra contribute into (the favor of) the whole coding gain, see Table X, implying a pattern-delimited payload stream, see Table XI.

Strictly speaking, this measure exists when and only when both its gain parts exist (are non-zero), i.e., when there are no overlaying among angles or orbits of the points.

Because the current MDI output power refers to the radius of that power orbit the current point belongs to, the trigonometric view highlights a very visible way to reduce the power variations for a given constellation—this way is to reduce the number of various power orbits present within it.[7]

[3] We keep the Trellis Coded Modulation 8-state convolutional encoder site, see [1] and [4], and the page fading management, see [1], completely.

[4] In this table and further, a gain value expressed "+...dB" is given rounded, otherwise it is floored via truncation of its least digits, anyway in dB.

[5] Equal to no MDI signal, the silence line code symbol acts as the universal stream delimiter (USD), always accompanied by an ESC, see [1] and [2].

[6] Mathematically, equation $\alpha^2 + \beta^2 = \gamma^2$ has up to two real solutions for $\gamma$, which are $\pm|\gamma|$, therefore we reference to the diameter of $2\,r_5$, not $r_5$.

TABLE XII
1000BASE-T SIGNAL SPACE SPREADING

| PMA Service Interface, 4-D, $T_0 = 8$ ns, $f_0 = 1/T_0 = 125$ MHz | 1000BASE-T PMA Sublayer | Gigabit Media Dependent Interface, 1000BASE-T MDI, $T_0^* = 8$ ns |
|---|---|---|
| PAM-5 symbolic code @ pair <A> e.g. $A_m$ = { L ; − ; z ; + ; H } ⟶ | $g(z) = .75\,z^0 + .25\,z^{-1}$ | ⟶ $A^*_m = \{ -^8/_8 ; \cdots ; -^1/_8 ; =0 ; +^1/_8 ; \cdots ; +^8/_8 \}$ i.e. PAM-17 @ pair <A> |
| • • • | • • • | • • • |
| PAM-5 symbolic code @ pair <D> e.g. $D_m$ = { L ; − ; z ; + ; H } ⟶ | $g(z) = .75\,z^0 + .25\,z^{-1}$ | ⟶ $D^*_m = \{ -^8/_8 ; \cdots ; -^1/_8 ; =0 ; +^1/_8 ; \cdots ; +^8/_8 \}$ i.e. PAM-17 @ pair <D> |

TABLE XIII
2D-PAM17-AIMED, BALANCED CONSTELLATION EXAMPLE 1: TWO PAGES, THREE POINTS PER QUADRANT PER PAGE, 24 POINTS TOTAL + SILENCE

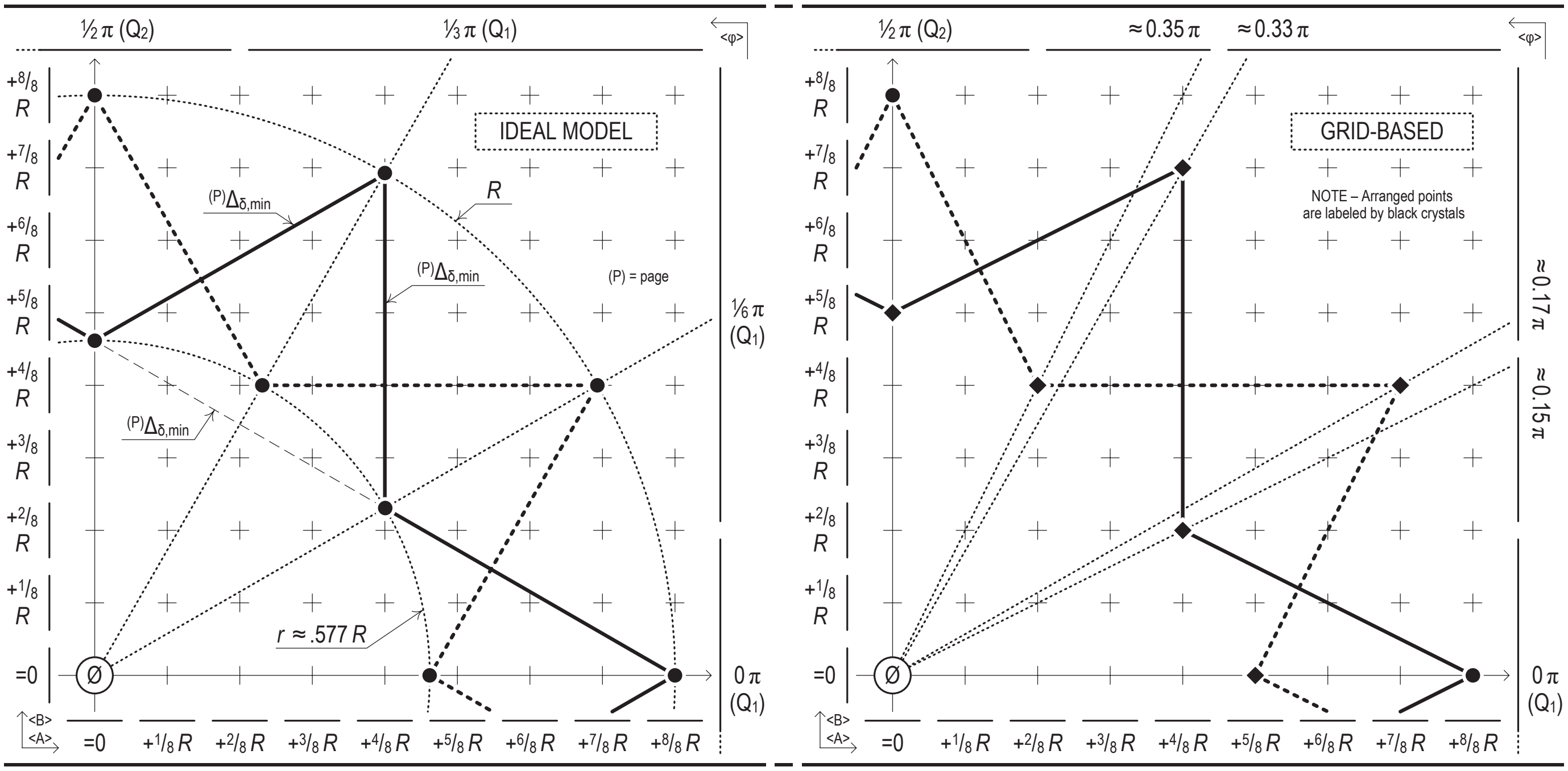

TABLE XIV
2D-PAM17-AIMED, BALANCED CONSTELLATION EXAMPLE 2: TWO PAGES, FOUR POINTS PER QUADRANT PER PAGE, 32 POINTS TOTAL + SILENCE

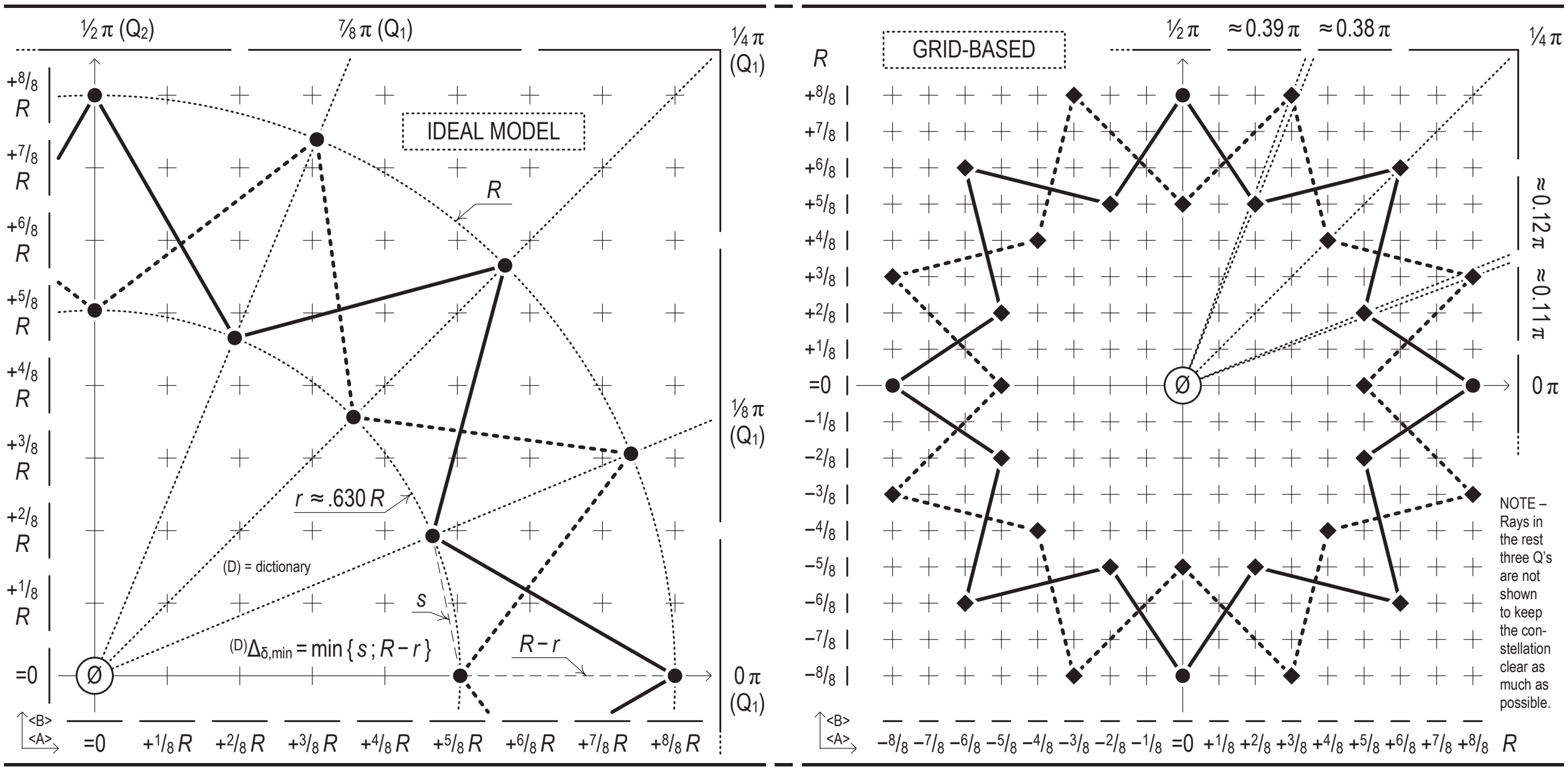

TABLE XV
IDEALISTIC STELLARIAL CONSTELLATION PARTITIONING PROPERTIES

| Pts/D:P:%P | $^{(P)}\Delta_{\varphi,min}$ | $r$ | $^{(P)}\Delta_{\delta,min}$ | $^{(P)}G_\delta$ | $^{(P)}G_\varphi$ | $^{(P)}G_\rho$ | $^{(D)}\Delta_{\delta,min}$ | $^{(D)}G_\delta$ |
|---|---|---|---|---|---|---|---|---|
| 16 : 8 : 2 | ¼ π | .518 R | .732 R | 3.958 | 1.159 | 2.397 | .396 R | 1.917 |
| 24 : 12 : 3 | ⅙ π | .577 R | .577 R | 2.958 | 0.755 | 2.062 | .299 R | 1.405 |
| 32 : 16 : 4 | ⅛ π | .630 R | .482 R | 2.397 | 0.560 | 1.775 | .246 R | 1.139 |
| 40 : 20 : 5 | 1⁄10 π | .673 R | .416 R | 2.024 | 0.445 | 1.551 | .211 R | 0.965 |
| 48 : 24 : 6 | 1⁄12 π | .707 R | .366 R | 1.755 | 0.369 | 1.375 | .185 R | 0.841 |
| 56 : 28 : 7 | 1⁄14 π | .735 R | .327 R | 1.551 | 0.315 | 1.234 | .165 R | 0.746 |
| 64 : 32 : 8 | 1⁄16 π | .758 R | .296 R | 1.390 | 0.275 | 1.118 | .149 R | 0.670 |

TABLE XVI
TWO-BY-TWO PAGES ENABLED OVER TWO COUPLED CONSTELLATIONS

| Coupling Rule Depiction | Pts Q/P | $^*\Delta_{\varphi,min}$ | $^*G_\varphi$ | $^*\Delta_{\rho,min}$ | $^*G_\rho$ | $^*\Delta_{\delta,min}$ | $\Sigma^*G$'s |
|---|---|---|---|---|---|---|---|
| 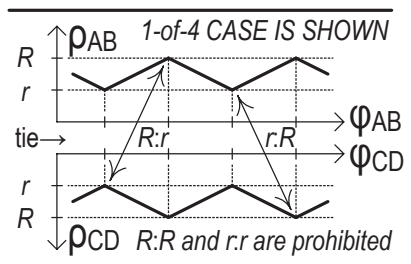  | 2 | .354 π | 1.689 | .682 R | 3.623 | are the same as the respective $^{(P)}\Delta_{\delta,min}$'s | 9.271 |
| | 3 | .236 π | 1.089 | .598 R | 3.083 | | 7.131 |
| | 4 | .177 π | 0.803 | .523 R | 2.632 | | 5.833 |
| | 5 | .141 π | 0.636 | .463 R | 2.285 | | 4.946 |
| | 6 | .118 π | 0.527 | .414 R | 2.015 | | 4.298 |
| $\Delta^2_{\$,AB} + \Delta^2_{\$,CD} \geq 2\Delta^2_{\$,min}$ | 7 | .101 π | 0.450 | .375 R | 1.801 | | 3.803 |
| where \$ is φ or ρ | 8 | .088 π | 0.392 | .342 R | 1.627 | | 3.410 |

TABLE XVII
PAGED DICTIONARY ENABLED OVER TWO COUPLED CONSTELLATIONS

| Possible Coupling Variants | | MDI Power proportional to | Coupling Rule Definition | Pts Q/P | $^{**}\Delta_{\delta,min}$ | $^{**}G_\delta$ | $^{**}G_\varphi$ | $^{**}G_\rho$ |
|---|---|---|---|---|---|---|---|---|
| R | R | $2R^2$ | prohibited | 2 | are the same as the respective $^{(D)}\Delta_{\delta,min}$'s | are the same as the respective $^{(D)}G_\delta$'s | — | n/a |
| R | r | $R^2 + r^2$ | in the use | 3 | | | because $^{**}\Delta_{\varphi,min} = 0$ | |
| r | R | $R^2 + r^2$ | in the use | 4 | | | | |
| r | r | $2r^2$ | prohibited | 5 | | | | |
| Orbit of a Given Point | Orbit of Another Point | set so to keep the power constant $2r^2 < R^2 + r^2 < 2R^2$ | | 6 | | | | |
| | | | | 7 | | | | |
| | | | | 8 | | | | |

TABLE XVIII
PROPERTIES WITH TWO-BY-TWO PAGES REPARTITIONED INTO FOUR

| Pts Q/P | Scheme | $^{*/4}G_\$$ (\$ = δ,φ,ρ) | $^{**/4}G_\delta$ | Internals of Each Page: |
|---|---|---|---|---|
| 2 | 32/128 + Ø | all are the same as | same as | $P_0 = P0\,(8_R + 8_r) : P0\,(8_R + 8_r)$ |
| 3 | 72/288 + Ø | the respective $^*G$'s | $^{**}G_\delta$'s | $P_1 = P0\,(8_R + 8_r) : P1\,(8_R + 8_r)$ |
| 4 | 128/512 + Ø | ........ | ........ | $P_2 = P1\,(8_R + 8_r) : P0\,(8_R + 8_r)$ |
| ... | ........ | ........ | ........ | $P_3 = P1\,(8_R + 8_r) : P1\,(8_R + 8_r)$ |

TABLE XIX
PROPERTIES WITH TWO-BY-TWO PAGES REPARTITIONED INTO EIGHT

| Pts Q/P | Scheme | $^{**/8}G_\delta$ | $^{*/8}G_\varphi$ | $^{*/8}G_\rho$ | $^{*/8}G_\delta$ | $\Sigma^{*/8}G$'s | |
|---|---|---|---|---|---|---|---|
| 2 | 16/128 + Ø | the same as $^{**}G_\delta$'s | 2.498 | the same as $^*G_\rho$'s | the same as $^*G_\delta$'s | 10.08 | $P_0 = P0\,(8_R) : P0\,(8_r)$<br>$P_1 = P0\,(8_r) : P0\,(8_R)$ |
| 3 | 36/288 + Ø | | 1.583 | | | 7.626 | $P_2 = P0\,(8_R) : P1\,(8_r)$ |
| 4 | 64/512 + Ø | | 1.159 | | | 6.189 | $P_3 = P0\,(8_r) : P1\,(8_R)$ |
| 5 | 100/800 + Ø | | 0.915 | | | 5.225 | $P_4 = P1\,(8_R) : P0\,(8_r)$ |
| 6 | 144/1,152 + Ø | | 0.755 | | | 4.527 | $P_5 = P1\,(8_r) : P0\,(8_R)$ |
| 7 | 196/1,568 + Ø | | 0.643 | | | 3.996 | $P_6 = P1\,(8_R) : P1\,(8_r)$ |
| 8 | 256/2,048 + Ø | | 0.560 | | | 3.578 | $P_7 = P1\,(8_r) : P1\,(8_R)$ |

TABLE XX
4D-PAM5 PARTITIONING PROPERTIES

| Selection | Content(s) | | | $D_\delta$ | $\Delta_{\delta,min}$ | $G_\delta$ |
|---|---|---|---|---|---|---|
| Pagelet ... | YYYY (81 pts) | to | XXXX (16 pts) | √16 | √4 | +6.0 dB |
| Page ... | $P_0$ (even) | : YYYY + XXXX | (97 pts) | √16 | √4 | +6.0 dB |
| NOTE – 64 pts/P are used for Data. | $P_7$ (odd) | : YXYY + XYXX | (78 pts) | √16 | √4 | +6.0 dB |
| Dictionary | $P_0 + \ldots + P_7$ | | (625 pts) | √16 | √1 | +2.5 dB |

Diving from over down into inside the 1000BASE-T PMA sublayer, we obtain an access to each per-pair PAM17 coding space [4], see Table XII, enough to map a constellation of just two orbits, $R$ and $r$, see Tables XIII, XIV, and XV.[8]

Coupling (operating on) two instances of such a constellation so, a given time, one of them uses a point at one power orbit and another of them uses a point at the different power orbit, see Tables XVI and XVII,[9] we even can achieve for a coding whose MDI output power is constant.

We consider such coupling conventionally, assuming each couple of effects of the same diameter, i.e., $D_\delta$, $D_\varphi$, $D_\rho$, runs a new, pair$^2$×pair$^2$, but also plain, Euclidean-axiomatic space resulting in the respective metric measure of exactly the same diameter, i.e., again $D_\delta$, $D_\varphi$, $D_\rho$, respectively.

## CONCLUSION

Described in this paper, the trigonometric view and coding design approach propose a way to set up a coding regular in the sense of its output power, pages and dictionary.

Having not less than 512 points in the dictionary allows us to develop a coding means compatible with the 1000BASE-T original duty [4], but preemptable just seamy.

Having more than 512 points, as much as practicable, leads us to extend over the duty with a fitted linguistic multiplexing process [1], for a means preemptable up to seamless.

Also, the described view and approach make now possible a coding without variations of the MDI output power, i.e., of one fixed level, so succeeding after [2] and [3].

We can find an already interesting coding option of 512 and 128 points in the transport dictionary and in each of its four repartitioned pages, respectively, see Table XVIII.

We can find a very interesting coding option of 512 and 64 points in the transport dictionary and in each of its new eight repartitioned pages, respectively, see Table XIX.[10]

We can find an also interesting coding option of 800 > 512 and 100 > 64 points in the dictionary and in each of its eight repartitioned pages, respectively, see Table XIX again.

Anyway, the properties of an interesting coding option are comparable to those of the reference, see Table XX, as it was implemented in [4] and assumed in [3], [2], and [1].

[7] Having two per-pair signals $a$ and $b$ interrelated so $a^2 + b^2 = \text{const}$, we ensure the power fed into the media keeps the same during transmission.

[8] Grid-based constellations are under study today, but in the two illustrated cases, see Tables XIII and XIV, we consider the differences negligible.

[9] It ensures four per-pair signals $a$ with $b$ and $c$ with $d$ are interrelated as $a^2 + b^2 + c^2 + d^2 = \text{const}$, i.e., for a fixed MDI output power level.

[10] Despite there is just one orbit in a so coupled page, even in the case, the receiving side still needs to detect what radius the orbit is, of $R$ or of $r$.

# Design Details, Usable and Useful, Related to the Data Coding Means and Event Coding Means Multiplexed Inside the 1000BASE-T PMA Sublayer

Alexander Ivanov

***Abstract*—In the paper, we're continuing the branch dedicated to the development of the Gigabit Ethernet physical layer, type 1000BASE-T, done inside its PMA sublayer, now focusing on the important design details we left unconsidered initially.**



## Introduction

POWER balance, as a prerequisite of a coding means, and power balancing, as a prerequisite of the design synthesis flow such a coding results from, are ones among the necessary features and tasks of the modern way, respectively.

The design approach [1], shown on the example of coding options operating inside the 1000BASE-T PMA sublayer [2], proposed the trigonometric view and its associated method to balance the MDI output power of the coding means.

That approach leverages the two metric measures used by the transmitting side, called barely metric and geometric, from the original protocol, 1000BASE-T, then introduces the extra one, called trigonometric, all of their gains, $G$'s [1].

In this paper, we consider those measures static—just in the sense of their definition and implementation—and, thus, focus on the so sensed dynamic aspects of such a coding procedure, resulting from the trigonometric view on the design.

## Fuzzy Constellation Model

Accessing each per-pair 1000BASE-T PMA coding space,[1] that is an one-dimensional, PAM17 space [2], see Table I, we assign every distinct transit, possible in that space, with its hit rate [3] as well as with its hop power, see Table II.

Implementing a stellarial constellation [1], that assumes we access a two-dimensional coding space,[2] we need to map an ideal outline of such a constellation onto the two-dimensional, PAM17 by PAM17 space-reflecting grid, see Table III.

Such a mapping results in an expectedly distorted, arranged outline of the analytic, ideal constellation, that consists of the same number of points, each may be of a close but different, fixed or floating position, see Tables IV, V, and VI.[3]

Accessing the coding space of such a constellation, we also assign every transit, possible in that fuzzy space, with a single hit rate, too, but further with a match, one per each dimension, of hop powers, see Table VII in comparison with Table II.

Despite a position-to-position transit can vary in its hit rate from absent, when the transit is excluded, to some maximum, see Table VII again, the room hit rates of a given constellation demonstrate a uniform distribution, see Table VIII.

TABLE I
1000BASE-T Single Output Generation

| $p$ (input) → | ⅕ | ⅕ | ⅕ | ⅕ | ⅕ | ⅕ | ⅕ | ⅕ | ⅕ | ⅕ | $D_\delta$ | Pulse Amplitudes | | | | | |
|---|---|---|---|---|---|---|---|---|---|---|---|---|---|---|---|---|---|
| Input Pulses 1-of-5 by 1-of-5 | −2 | −1 | =0 | +1 | +2 | −2 | −1 | =0 | +1 | +2 | ← 4 | ←0, ±½$u_{ref}$, ±2⁄2$u_{ref}$ | | | | | |
| PMA Filter | $g(z) = .25z^{-1} + .75z^{-0}$ | | | | | | | | | | 16 ↓ | 0, ±⅛$u_{ref}$ ··· ±8⁄8$u_{ref}$ ↓ ↓ ↓ ↓ ↓ | | | | | |
| 1-of-17 (resulting) Output Pulse | −8 | −7 | −6 | −5 | −4 | −3 | −2 | −1 | =0 | +1 | +2 | +3 | +4 | +5 | +6 | +7 | +8 |
| Output Change Min/Max Hops | minimum and often occurring hop, (relative) power = (1×1)/(16×16) = 1/256 ≈ .004; maximum but never occurring hop, (relative) power = (16×16)/(16×16) = 1 | | | | | | | | | | | | | | | | |
| Output Power | 1. | .77 | .56 | .39 | .25 | .14 | .06 | .02 | — | .02 | .06 | .14 | .25 | .39 | .56 | .77 | 1. |
| $p$ (output) → | 1⁄25 | 1⁄25 | 1⁄25 | 2⁄25 | 2⁄25 | 1⁄25 | 2⁄25 | 2⁄25 | 1⁄25 | 2⁄25 | 2⁄25 | 1⁄25 | 2⁄25 | 2⁄25 | 1⁄25 | 1⁄25 | 1⁄25 |

TABLE II
1000BASE-T Single Output Change Properties

| hit rate / hop pwr | −8 | −7 | −6 | −5 | −4 | −3 | −2 | −1 | =0 | +1 | +2 | +3 | +4 | +5 | +6 | +7 | +8 | from / into ↓ | Σ |
|---|---|---|---|---|---|---|---|---|---|---|---|---|---|---|---|---|---|---|---|
| −8 | 1 | 1 | 1 | 1 | 1 | — | — | — | — | — | — | — | — | — | — | — | — | −8 | 5 |
| .77 | — | — | — | 1 | 1 | 1 | 1 | 1 | — | — | — | — | Hit Rate Map | | | | — | −7 | 5 |
| .56 | — | — | — | — | — | — | 1 | 1 | 1 | 1 | 1 | — | — | — | — | — | — | −6 | 5 |
| .39 | .035 | .016 | .004 | 1 | 1 | — | — | — | — | 1 | 1 | 1 | 1 | 1 | — | — | — | −5 | 10 |
| .25 | — | — | — | .004 | 1 | 1 | 1 | 1 | — | — | — | — | 1 | 1 | 1 | 1 | 1 | −4 | 10 |
| .14 | — | — | — | — | — | — | 1 | 1 | 1 | 1 | 1 | — | — | — | — | — | — | −3 | 5 |
| .06 | .141 | .098 | .063 | .035 | .016 | — | — | — | — | 1 | 1 | 1 | 1 | 1 | — | — | — | −2 | 10 |
| .02 | — | — | — | .063 | .035 | .016 | .004 | 1 | — | — | — | — | 1 | 1 | 1 | 1 | 1 | −1 | 10 |
| — | — | — | — | — | — | — | .016 | .004 | 1 | 1 | 1 | — | — | — | — | — | — | =0 | 5 |
| .02 | .316 | .250 | .191 | .141 | .098 | — | — | — | — | 1 | 1 | 1 | 1 | 1 | — | — | — | +1 | 10 |
| .06 | — | — | — | .191 | .141 | .098 | .063 | .035 | — | — | — | — | 1 | 1 | 1 | 1 | 1 | +2 | 10 |
| .14 | — | — | — | — | — | — | .098 | .063 | .035 | .016 | .004 | — | — | — | — | — | — | +3 | 5 |
| .25 | .563 | .473 | .391 | .316 | .250 | — | — | — | — | .035 | .016 | .004 | 1 | 1 | — | — | — | +4 | 10 |
| .39 | — | — | — | .391 | .316 | .250 | .191 | .141 | — | — | — | — | .004 | 1 | 1 | 1 | 1 | +5 | 10 |
| .56 | — | — | — | — | — | — | .250 | .191 | .141 | .098 | .063 | — | — | — | — | — | — | +6 | 5 |
| .77 | — | Hop Power Map | | | | — | — | — | — | .141 | .098 | .063 | .035 | .016 | — | — | — | +7 | 5 |
| 1. | — | — | — | — | — | — | — | — | — | — | — | — | .063 | .035 | .016 | .004 | 1 | +8 | 5 |
| ↑ into / from | 1. | .77 | .56 | .39 | .25 | .14 | .06 | .02 | — | .02 | .06 | .14 | .25 | .39 | .56 | .77 | +8 | ● = sym. pt. | Σ125 |

Recalling the fate of submission of many prior works to the peer reviewed journal, such a try with this one also promises no chance, probably.

Please sorry for the author has no time to find this work a new home, peer reviewed or not, except of arXiv, and just hopes there it meets its reader, one or maybe various, whom the author beforehand thanks for their regard.

A. Ivanov is with JSC Continuum, Yaroslavl, the Russian Federation.



[1] In 1000BASE-T, each of per-pair MDI signals is independent, see [2].

[2] In a stellarial case, its two per-pair signals are interdependent, see [1].

[3] We use up to two vertices per a point, using more is also possible.

## TABLE III
### 5-POINT/P/Q STELLARIAL CONSTELLATION

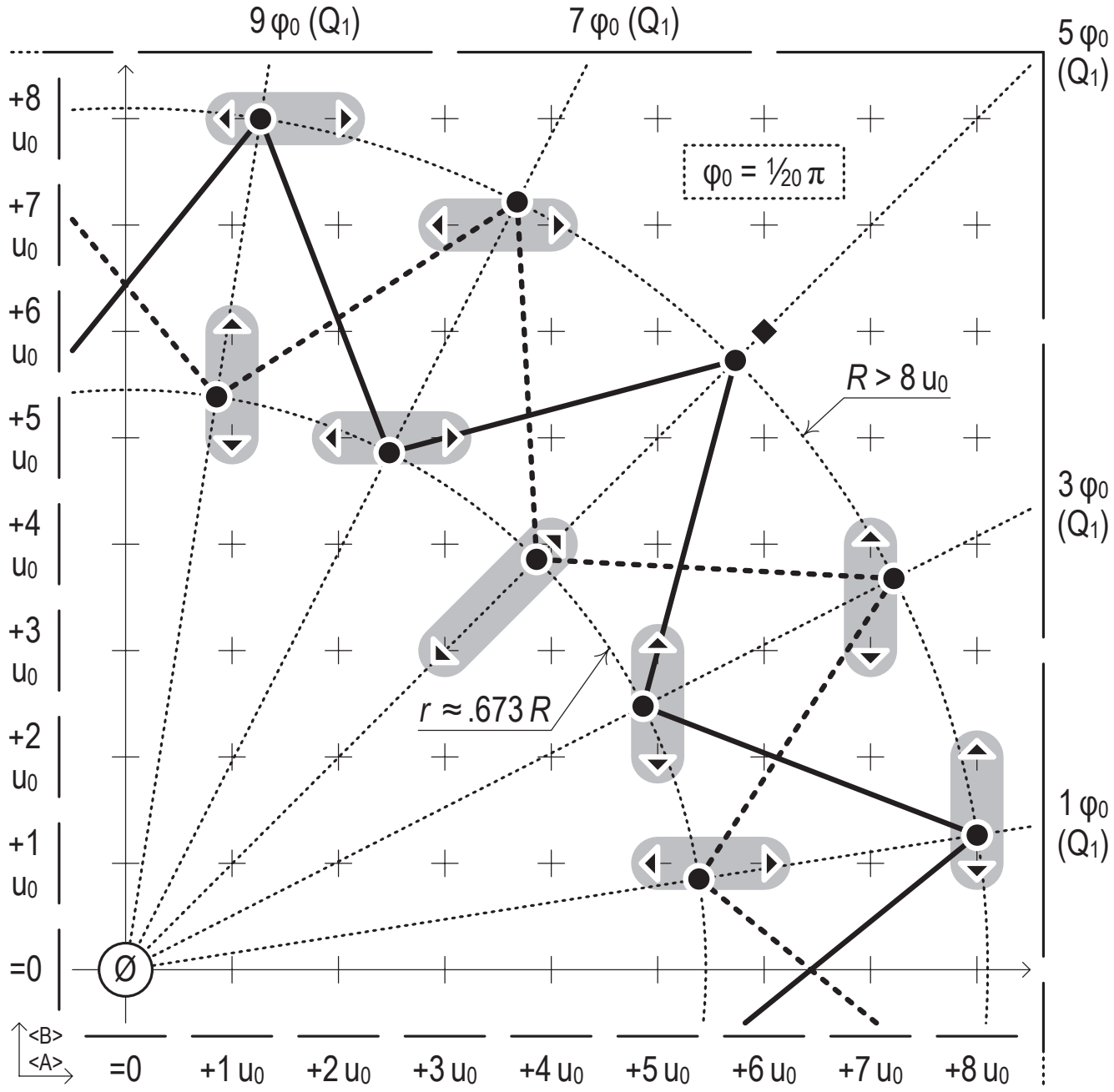


## TABLE IV
### 5-POINT/P/Q ARRANGED CONSTELLATION DEFINITION

| | | | | | | |
|---|---|---|---|---|---|---|
| High Orbit | Room→ | $R\angle 9\varphi_0$ | $R\angle 7\varphi_0$ | $R\angle 5\varphi_0$ | $R\angle 3\varphi_0$ | $R\angle 1\varphi_0$ |
| | <A> | +1 +2 | +3 +4 | +6 | +7 +7 | +8 +8 |
| | <B> | +8 +8 | +7 +7 | +6 | +4 +3 | +2 +1 |
| | (symbol) | (◀)(▶) | (◀)(▶) | (◆) | (▲)(▼) | (▲)(▼) |
| ↕ Points of | | EXCLUDED TRANSITS | | | | |
| Low Orbit | (symbol) | (▲)(▼) | (◀)(▶) | (◣)(◥) | (▲)(▼) | (◀)(▶) |
| | <A> | +1 +1 | +2 +3 | +3 +4 | +5 +5 | +5 +6 |
| | <B> | +6 +5 | +5 +5 | +3 +4 | +3 +2 | +1 +1 |
| | Room→ | $r\angle 9\varphi_0$ | $r\angle 7\varphi_0$ | $r\angle 5\varphi_0$ | $r\angle 3\varphi_0$ | $r\angle 1\varphi_0$ |

DESIGN GOAL IS TO ENSURE $_{(D,grid)}\Delta_{\delta,min} \geq {}_{(D,ideal)}\Delta_{\delta,min} \approx .211\,R$

## TABLE V
### ARRANGED CONSTELLATION CONSTRUCTING PRINCIPLES

| Constraint | Every Room |
|---|---|
| There must be exact FOUR paths, repeated as least as possible each, connecting any to every other room | Describes an unique position at the grid, either **fixed**, defined by a single point only, or **floating**, defined by two boundary points |

## TABLE VI
### IDEALIZED GEOMETRY VERSUS ARRANGED GEOMETRY

| R. | $9\varphi_0 = 81°$ | $7\varphi_0 = 63°$ | $5\varphi_0 = 45°$ | $3\varphi_0 = 27°$ | $1\varphi_0 = 9°$ |
|---|---|---|---|---|---|
| $R$ | 1 ← −.27 ┤├ +.73 → 2<br>1.27<br>◀ .27 ● .73 ▶<br>8.00<br>8 ← ┘└ → 8 | 3 ← −.68 ┤├ +.32 → 4<br>3.68<br>◀ .71 ● .39 ▶<br>7.22<br>7 ← −.22 ┘└ −.22 → 7 | +.27 → 6<br>5.73<br>● .39 ◆<br>5.73<br>+.27 → 6 | 7 ← −.22 ┤├ −.22 → 7<br>7.22<br>▲ .39 ● .71 ▼<br>3.68<br>4 ← +.32 ┘└ −.68 → 3 | 8 ← ┤├ → 8<br>8.00<br>▲ .73 ● .27 ▼<br>1.27<br>2 ← +.73 ┘└ −.27 → 1 |
| $r$ | 1 ← +.15 ┤├ +.15 → 1<br>0.85<br>▲ .63 ● .41 ▼<br>5.38<br>6 ← +.62 ┘└ −.38 → 5 | 2 ← −.47 ┤├ +.53 → 3<br>2.47<br>◀ .50 ● .55 ▶<br>4.86<br>5 ← +.14 ┘└ +.14 → 5 | 3 ← −.85 ┤├ +.15 → 4<br>3.85<br>◣ 1.2 ● .21 ◥<br>3.85<br>3 ← −.85 ┘└ +.15 → 4 | 5 ← +.14 ┤├ +.14 → 5<br>4.86<br>▲ .55 ● .50 ▼<br>2.47<br>3 ← +.53 ┘└ −.47 → 2 | 5 ← −.38 ┤├ +.62 → 6<br>5.38<br>◀ .41 ● .63 ▶<br>0.85<br>1 ← +.15 ┘└ +.15 → 1 |

## TABLE VII
### 5-POINT/P/Q ARRANGED CONSTELLATION OUTPUT CHANGE PROPS.

| ⋯ | Low Orbit, First Quadrant ($Q_1$) | | | | | | | | | | ⋯ | High Orbit ($Q_1$) | | | | | <A><B> from / into <A><B> | |
|---|---|---|---|---|---|---|---|---|---|---|---|---|---|---|---|---|---|---|
| ⋯ | +1 +6 | +1 +5 | +2 +5 | +3 +5 | +3 +3 | +4 +4 | +5 +3 | +5 +2 | +5 +1 | +6 +1 | ⋯ | +6 +6 | +7 +4 | +7 +3 | +8 +2 | +8 +1 | | |
| ⋯ | .191 /1/ .098 | .191 /1/ .063 | .141 /1/ .063 | .098 /1/ .063 | .098 /1/ .016 | .063 /1/ .035 | .035 /1/ .016 | .035 /1/ .004 | .035 /1/ — | .016 /1/ — | ⋯ | .016 /2/ .098 | .004 /1/ .035 | .004 /2/ .016 | — /1/ .004 | — /1/ — | +8 +1 | High Orbit ($Q_1$) |
| ⋯ | .191 /1/ .063 | .191 /1/ .035 | .141 /1/ .035 | .098 /1/ .035 | .098 /1/ .004 | .063 /1/ .016 | .035 /1/ .004 | .035 /1/ — | .035 /1/ .004 | .016 /1/ .004 | ⋯ | .016 /2/ .063 | .004 /1/ .016 | X | — /1/ — | — /1/ .004 | +8 +2 | |
| ⋯ | .141 /1/ .035 | .141 /1/ .016 | .098 /1/ .016 | .063 /1/ .016 | .063 /1/ — | .035 /1/ .004 | .016 /1/ — | .016 /1/ .004 | .016 /1/ .016 | .004 /1/ .016 | ⋯ | .004 /2/ .035 | — /1/ .004 | — /1/ — | X | .004 /1/ .016 | +7 +3 | |
| ⋯ | .141 /1/ .016 | .141 /1/ .004 | .098 /1/ .004 | .063 /1/ .004 | .063 /1/ .004 | .035 /1/ — | .016 /1/ .004 | .016 /1/ .016 | .016 /1/ .035 | .004 /1/ .035 | ⋯ | .004 /2/ .016 | — /1/ — | — /1/ .004 | .004 /2/ .016 | .004 /1/ .035 | +7 +4 | |
| ⋯ | .098 /2/ — | .098 /2/ .004 | .063 /2/ .004 | .035 /2/ .004 | .035 /2/ .035 | .016 /2/ .016 | .004 /2/ .035 | .004 /2/ .063 | .004 /2/ .098 | — /2/ .098 | ⋯ | — /4/ — | .004 /2/ .016 | .004 /2/ .035 | .016 /2/ .063 | .016 /2/ .098 | +6 +6 | |
| ⋯ | ⋯ | ⋯ | ⋯ | ⋯ | ⋯ | ⋯ | ⋯ | ⋯ | ⋯ | ⋯ | ⋯ | ⋯ | ⋯ | ⋯ | ⋯ | ⋯ | ⋯ | |
| ⋯ | .098 /1/ .098 | .098 /1/ .063 | .063 /1/ .063 | .035 /1/ .063 | .035 /1/ .016 | .016 /1/ .035 | .004 /2/ .016 | X | .004 /1/ — | — /1/ — | ⋯ | — /2/ .098 | .004 /1/ .035 | .004 /1/ .016 | .016 /1/ .004 | .016 /1/ — | +6 +1 | Low Orbit ($Q_1$) |
| ⋯ | .063 /1/ .098 | .063 /1/ .063 | .035 /1/ .063 | .016 /1/ .063 | .016 /1/ .016 | .004 /1/ .035 | — /2/ .016 | X | — /1/ — | .004 /1/ — | ⋯ | .004 /2/ .098 | .016 /1/ .035 | .016 /1/ .016 | .035 /1/ .004 | .035 /1/ — | +5 +1 | |
| ⋯ | .063 /1/ .063 | .063 /1/ .035 | .035 /1/ .035 | .016 /1/ .035 | .016 /1/ .004 | .004 /2/ .016 | — /1/ .004 | — /1/ — | X | X | ⋯ | .004 /2/ .063 | .016 /1/ .016 | .016 /1/ .004 | .035 /1/ — | .035 /1/ .004 | +5 +2 | |
| ⋯ | .063 /1/ .035 | .063 /1/ .016 | .035 /1/ .016 | .016 /1/ .016 | .016 /1/ — | X | — /1/ — | — /1/ .004 | — /2/ .016 | .004 /2/ .016 | ⋯ | .004 /2/ .035 | .016 /1/ .004 | .016 /1/ — | .035 /1/ .004 | .035 /1/ .016 | +5 +3 | |
| ⋯ | .035 /1/ .016 | .035 /1/ .004 | .016 /1/ .004 | X | .004 /1/ .004 | — /1/ — | X | .004 /1/ .016 | .004 /1/ .035 | .016 /1/ .035 | ⋯ | .016 /2/ .016 | .035 /1/ — | .035 /1/ .004 | .063 /1/ .016 | .063 /1/ .035 | +4 +4 | |
| ⋯ | .016 /1/ .035 | .016 /1/ .016 | .004 /1/ .016 | — /2/ .016 | — /1/ — | .004 /1/ .004 | .016 /2/ — | .016 /1/ .004 | .016 /1/ .016 | .035 /1/ .016 | ⋯ | .035 /2/ .035 | .063 /1/ .004 | .063 /1/ — | .098 /1/ .004 | .098 /1/ .016 | +3 +3 | |
| ⋯ | .016 /2/ .004 | .016 /2/ — | .004 /1/ — | — /1/ — | — /1/ .016 | X | .016 /1/ .016 | .016 /1/ .035 | .016 /1/ .063 | .035 /1/ .063 | ⋯ | .035 /2/ .004 | .063 /1/ .004 | .063 /1/ .016 | .098 /1/ .035 | .098 /1/ .063 | +3 +5 | |
| ⋯ | X | X | — /1/ — | .004 /1/ — | .004 /1/ .016 | .016 /2/ .004 | .035 /1/ .016 | .035 /1/ .035 | .035 /1/ .063 | .063 /1/ .063 | ⋯ | .063 /2/ .004 | .098 /1/ .004 | .098 /1/ .016 | .141 /1/ .035 | .141 /1/ .063 | +2 +5 | |
| ⋯ | — /1/ .004 | — /1/ — | X | .016 /2/ — | .016 /1/ .016 | .035 /1/ .004 | .063 /1/ .016 | .063 /1/ .035 | .063 /1/ .063 | .098 /1/ .063 | ⋯ | .098 /2/ .004 | .141 /1/ .004 | .141 /1/ .016 | .191 /1/ .035 | .191 /1/ .063 | +1 +5 | |
| ⋯ | — /1/ — | — /1/ .004 | X | .016 /2/ .004 | .016 /1/ .035 | .035 /1/ .016 | .063 /1/ .035 | .063 /1/ .063 | .063 /1/ .098 | .098 /1/ .098 | ⋯ | .098 /2/ — | .141 /1/ .016 | .141 /1/ .035 | .191 /1/ .063 | .191 /1/ .098 | +1 +6 | |
| ⋯ | ⋮ | ⋮ | ⋮ | ⋮ | ⋮ | ⋮ | ⋮ | ⋮ | ⋮ | ⋮ | ⋮ | ⋮ | ⋮ | ⋮ | ⋮ | ⋮ | ⋮ | |
| ⋯ | ⋯ | ⋯ | ⋯ | ⋯ | ⋯ | ⋯ | ⋯ | ⋯ | ⋯ | ⋯ | ⋯ | ⋯ | ⋯ | ⋯ | ⋯ | ⋯ | ⋯ | |
| ⋯ | .191 /1/ .563 | .191 /1/ .473 | .250 /1/ .473 | .316 /1/ .473 | .316 /1/ .316 | .391 /1/ .391 | .473 /1/ .316 | .473 /1/ .250 | .473 /1/ .191 | .563 /1/ .191 | ⋯ | .563 /4/ .563 | .660 /2/ .391 | .660 /2/ .316 | .766 /2/ .250 | .766 /2/ .191 | −6 −6 | High Orbit ($Q_3$) |
| ⋯ | .250 /1/ .391 | .250 /1/ .316 | .316 /1/ .316 | .391 /1/ .316 | .391 /1/ .191 | .473 /1/ .250 | .563 /1/ .191 | .563 /1/ .141 | .563 /1/ .098 | .660 /1/ .098 | ⋯ | .660 /2/ .391 | .766 /1/ .250 | .766 /1/ .191 | .879 /1/ .141 | .879 /1/ .098 | −7 −4 | |
| ⋯ | .250 /1/ .316 | .250 /1/ .250 | .316 /1/ .250 | .391 /1/ .250 | .391 /1/ .141 | .473 /1/ .191 | .563 /1/ .141 | .563 /1/ .098 | .563 /1/ .063 | .660 /1/ .063 | ⋯ | .660 /2/ .316 | .766 /1/ .191 | .766 /1/ .141 | .879 /1/ .098 | .879 /1/ .063 | −7 −3 | |
| ⋯ | .316 /1/ .250 | .316 /1/ .191 | .391 /1/ .191 | .473 /1/ .191 | .473 /1/ .098 | .563 /1/ .141 | .660 /1/ .098 | .660 /1/ .063 | .660 /1/ .035 | .766 /1/ .035 | ⋯ | .766 /2/ .250 | .879 /1/ .141 | .879 /1/ .098 | 1.00 /1/ .063 | 1.00 /1/ .035 | −8 −2 | |
| ⋯ | .316 /1/ .191 | .316 /1/ .141 | .391 /1/ .141 | .473 /1/ .141 | .473 /1/ .063 | .563 /1/ .098 | .660 /1/ .063 | .660 /1/ .035 | .660 /1/ .016 | .766 /1/ .016 | ⋯ | .766 /2/ .191 | .879 /1/ .098 | .879 /1/ .063 | 1.00 /1/ .035 | 1.00 /1/ .016 | −8 −1 | |
| ⋯ | ⋮ | ⋮ | ⋮ | ⋮ | ⋮ | ⋮ | ⋮ | ⋮ | ⋮ | ⋮ | ⋮ | ⋮ | ⋮ | ⋮ | ⋮ | ⋮ | ⋮ | |

<A> and <B> hit rate (middle value of each cell)

shadowed are transits to be excluded for enabling $G_J > 0$

<A> hop power (top value of each cell); <B> hop power (bottom value of each cell)

## TABLE VIII
### ROOM HIT RATES

| ⋯ | | Low Orbit, $r$ ($Q_1$) | | | | | | High Orbit, $R$ ($Q_1$) | | | | | | see with the above↑ table |
|---|---|---|---|---|---|---|---|---|---|---|---|---|---|---|
| Don't Care | ⋯ + | 160 | + | 160 | + | 160 | + ⋯ + | 160 | + | 160 | + | 160 | = | 6,400 |
| | | −2•4 ▪ | | −2•4 ▪ | | −2•4 ▪ | | −2•4 ▪ | | −2•4 ▪ | | −2•4 ▪ | | |
| Enabled | ⋯ + | 152 | + | 152 | + | 152 | + ⋯ + | 152 | + | 152 | + | 152 | = | 6,080 |
| $G_J > 0$ | ⋯ | $5\varphi_0$ | | $3\varphi_0$ | | $1\varphi_0$ | ⋯ | $5\varphi_0$ | | $3\varphi_0$ | | $1\varphi_0$ | | $H_\Sigma$ |

TABLE IX
4.5-POINT/P/Q STELLARIAL CONSTELLATION

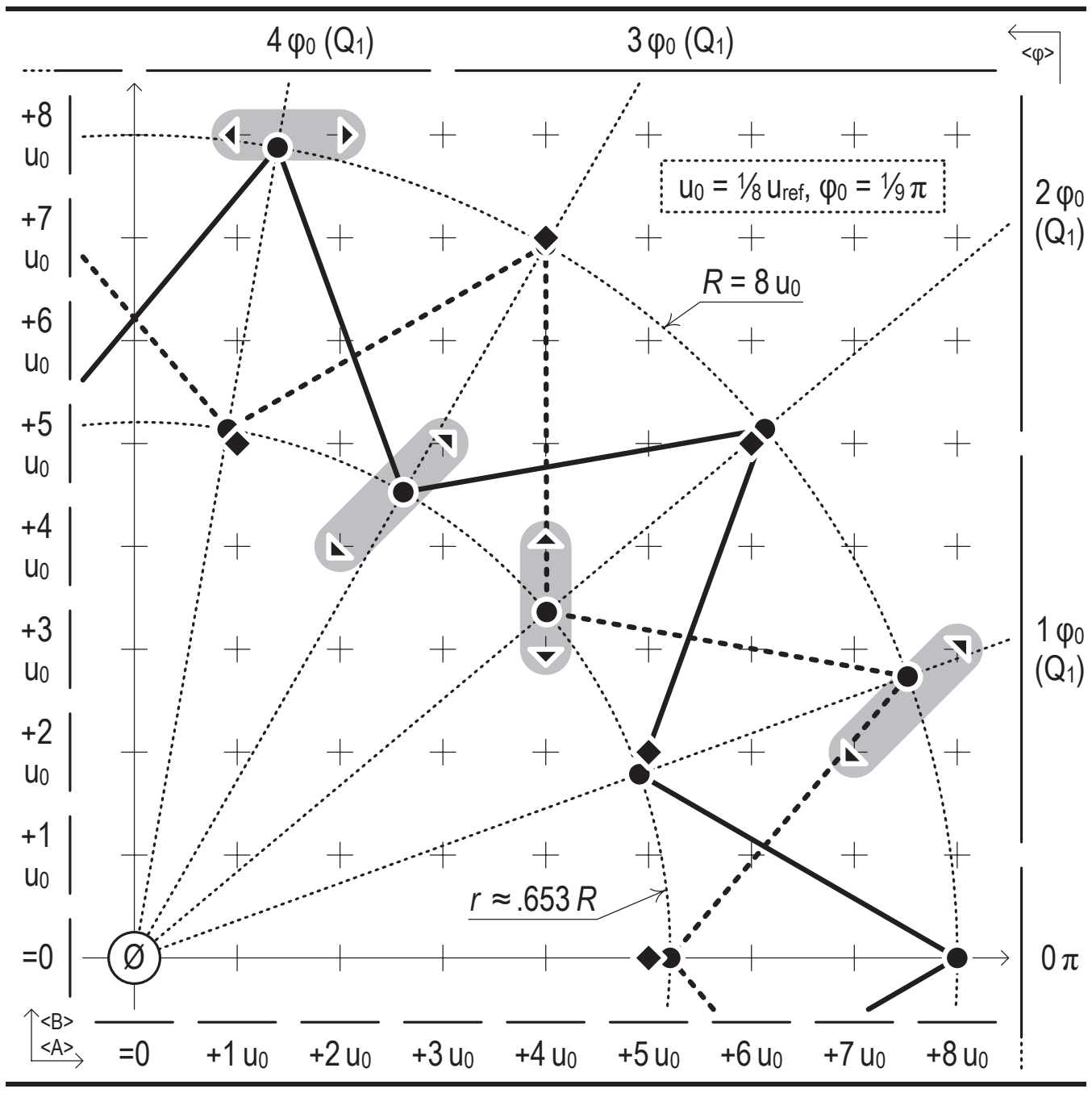


TABLE X
4-POINT/P/Q STELLARIAL CONSTELLATION

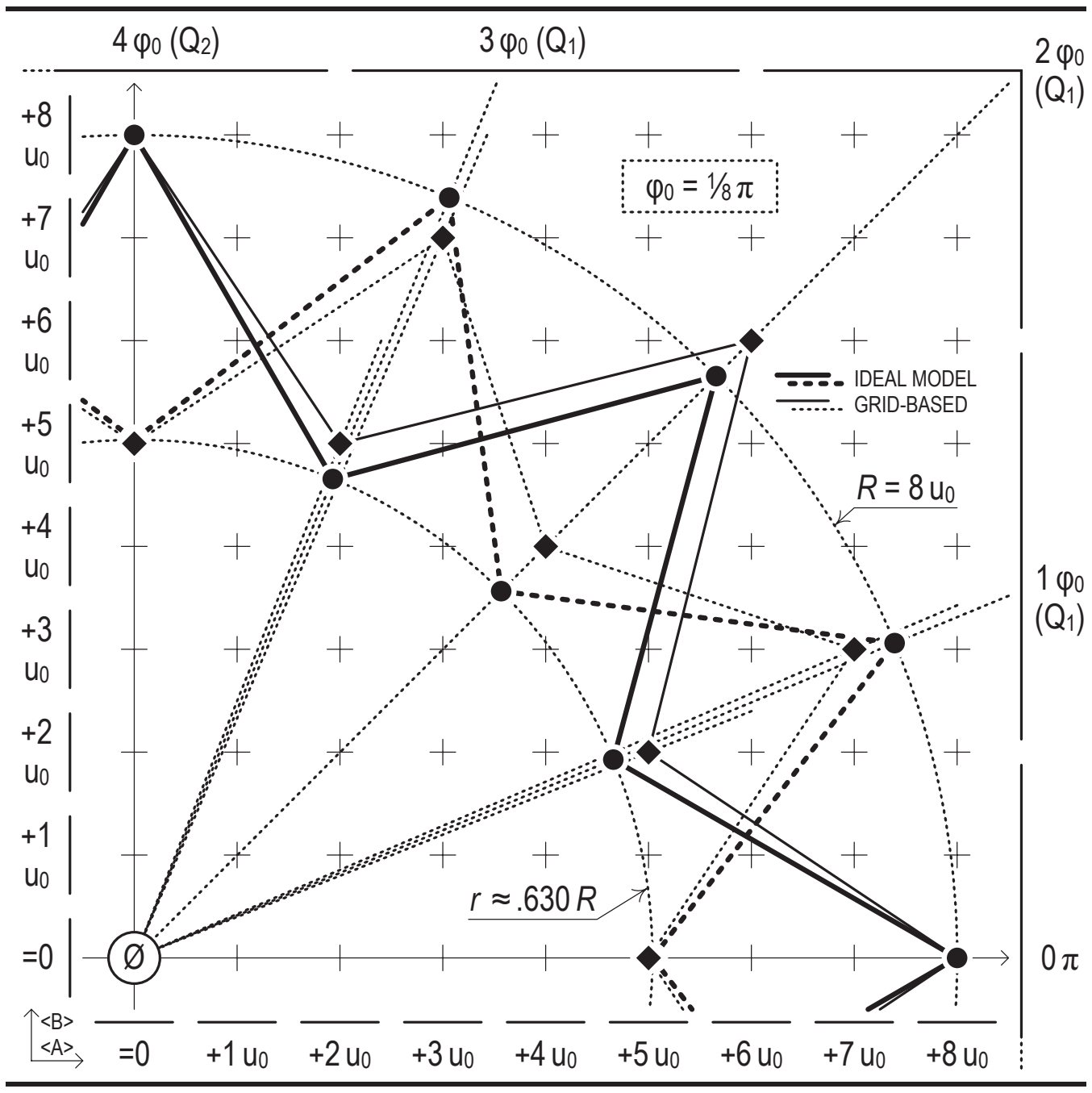


TABLE XI
DESIGN OPTIONS

| ↓Content Description of a Design Option→ | Don't Care | One-way $G_J > 0$ ($G_J$) | Two-way $G_J > 0$ ($G_J$+) |
|---|---|---|---|
| Coding Rules and Repartitioning Restrictions applied { <br> NOTE – Orbit and view reflect the ideal radius and angle of the respective room. | { Orbit chosen, $m+1$ ≠ Orbit another, $m+1$ <br> + Relations, decision period-bounded: chosen = AB or CD, another ≠ chosen, anyone = chosen or another but same on the sides of a line. | { Orbit chosen, $m+1$ ≠ Orbit another, $m+1$ <br> View anyone, $m+1$ ≠ View anyone, $m$ <br> NOTE – $m$ is the current word time period index. | [ Orbit chosen, $m+1$ ≠ Orbit another, $m+1$ <br> View chosen, $m+1$ ≠ View chosen, $m$ <br> View another, $m+1$ ≠ View another, $m$ |
| Repartitioned Page / Multi-RP Dictionary Effective Size: <br> ↓when→ all transits, including diagonal, are possible <br> (Đ) diagonal transits are all excluded (no hops of the most power) | $N \cdot N$ / $8 \cdot N \cdot N$ <br> $(N-1)^2$ / $8 \cdot (N-1)^2$ | $(N-1) \cdot N$ / $8 \cdot (N-1) \cdot N$ <br> $(N-2) \cdot (N-1)$ / $8 \cdot (N-2) \cdot (N-1)$ | $(N-1)^2$ / $8 \cdot (N-1)^2$ <br> $(N-2)^2$ / $8 \cdot (N-2)^2$ |

TABLE XII
COMPLETE MDI OUTPUT PROPERTIES

| Constellation | MDI Output Power | | | | | Output Wobble Power | | | | Output Change Power | | | | |
|---|---|---|---|---|---|---|---|---|---|---|---|---|---|---|
| NOTE – Pk's mean for peak values, min (−), max (+), and span (±), resp. | $\mu_0/w_{max}$ | σ/··· | $Pk^-$/··· | $Pk^+$/··· | $Pk^\pm$/··· | $^w\mu_0/w_{max}$ | $^w\sigma$/··· | $^wPk^-$/··· | $^wPk^+$/··· | $^c\mu_0/\dot{w}_{max}$ | $^c\sigma$/··· | $^cPk^-$/··· | $^cPk^+$/··· | $^cPk^\pm$/··· |
| Original 1000BASE-T | 0.313 | 0.151 | 0 | 1 | 1 | not applicable | | | | 0.109 | 0.064 | 0 | 0.563 | 0.563 |
| 4 Pts/P/Q, Ideal | 0.349 | 0 | | | | completely absent | | | | 0.174 | 0.086 | 0 | 0.349 | 0.349 |
| 4 Pts/P/Q, Grid | 0.358 | 0.024 | 0.324 | 0.406 | 0.082 | 0.001 | 0.001 | 0.00001 | 0.002 | 0.164 | 0.091 | 0 | 0.406 | 0.406 |
| 4.5 Pts/P/Q, Ideal | 0.357 | 0 | | | | completely absent | | | | 0.178 | 0.088 | 0 | 0.357 | 0.357 |
| 4.5 Pts/P/Q, Grid | 0.366 | 0.027 | 0.285 | 0.418 | 0.133 | 0.002 | 0.002 | 0.00011 | 0.006 | 0.183 | 0.092 | 0 | 0.418 | 0.418 |
| 4.5 Pts/P/Q, $G_J$, Ideal | 0.357 | 0 | | | | completely absent | | | | 0.183 | 0.085 | 0.003 | 0.357 | 0.353 |
| 4.5 Pts/P/Q, $G_J$, Grid | 0.366 | 0.027 | 0.285 | 0.418 | 0.133 | 0.002 | 0.002 | 0.00011 | 0.006 | 0.188 | 0.089 | 0.004 | 0.418 | 0.414 |
| 4.5 Pts/P/Q, $G_J$+, Ideal | 0.357 | 0 | | | | completely absent | | | | 0.188 | 0.086 | 0.011 | 0.357 | 0.346 |
| 4.5 Pts/P/Q, $G_J$+, Grid | 0.366 | 0.027 | 0.285 | 0.418 | 0.133 | 0.002 | 0.002 | 0.00011 | 0.006 | 0.192 | 0.089 | 0.008 | 0.418 | 0.410 |
| 4.5 Pts/P/Q, Đ, Ideal | 0.357 | 0 | | | | completely absent | | | | 0.168 | 0.085 | 0 | 0.346 | 0.346 |
| 4.5 Pts/P/Q, Đ, Grid | 0.366 | 0.027 | 0.285 | 0.418 | 0.133 | 0.002 | 0.002 | 0.00011 | 0.006 | 0.172 | 0.089 | 0 | 0.397 | 0.397 |
| 5 Pts/P/Q, Ideal | 0.372 | 0 | | | | completely absent | | | | 0.186 | 0.092 | 0 | 0.372 | 0.372 |
| 5 Pts/P/Q, Grid | 0.367 | 0.033 | 0.266 | 0.426 | 0.160 | 0.005 | 0.008 | 0.00045 | 0.034 | 0.183 | 0.092 | 0 | 0.426 | 0.426 |
| 5 Pts/P/Q, $G_J$+, Ideal | 0.372 | 0 | | | | completely absent | | | | 0.196 | 0.090 | 0.009 | 0.372 | 0.363 |
| 5 Pts/P/Q, $G_J$+, Grid | 0.367 | 0.033 | 0.266 | 0.426 | 0.160 | 0.005 | 0.008 | 0.00045 | 0.034 | 0.192 | 0.090 | 0.008 | 0.426 | 0.418 |
| 5 Pts/P/Q, Đ, Ideal | 0.372 | 0 | | | | completely absent | | | | 0.177 | 0.089 | 0 | 0.363 | 0.363 |
| 5 Pts/P/Q, Đ, Grid | 0.367 | 0.033 | 0.266 | 0.426 | 0.160 | 0.005 | 0.008 | 0.00045 | 0.034 | 0.173 | 0.090 | 0 | 0.403 | 0.403 |
| 5 Pts/P/Q, $G_J$+, Đ, Ideal | 0.372 | 0 | | | | completely absent | | | | 0.186 | 0.087 | 0.009 | 0.363 | 0.354 |
| 5 Pts/P/Q, $G_J$+, Đ, Grid | 0.367 | 0.033 | 0.266 | 0.426 | 0.160 | 0.005 | 0.008 | 0.00045 | 0.034 | 0.184 | 0.087 | 0.008 | 0.403 | 0.396 |

TABLE XIII
DESIGN SIMILE

| ↓Design, Output's→ | | $^{c}\sigma$ | : | $^{c}\mu_0$ | : | $^{c}Pk^{\pm}$ | Ø | Losses due to Wobble (applicable only to the specified case) | | |
|---|---|---|---|---|---|---|---|---|---|---|
| | | | | | | | | $D_w$ | $\Delta_{w,typ}$ | $L_w = -G_w$ |
| 1000BASE-T | ≈ | 1.2 | : | 2 | : | 10 | shared | | | |
| Any Proposed, Ideal | ≈ | 1 | : | 2 | : | 4 | separate | | | |
| Any Proposed, Grid | ≈ | 1 | : | 2 | : | 4.4 | separate | $16\,u_0$ | $<.4\,u_0$ | $-0.24_{typ}$ |

TABLE XIV
CODING SPACE CHOICE

| Pts Q/P | Eff. Pool | Eff. Size | Per-RP Coding Space | $D_J$ | $^{*/8}\Delta_{J,min}$ | $^{**}\Delta_{J,min}$ | $G_J$ |
|---|---|---|---|---|---|---|---|
| 4 | 64/512 → | 64/512 | $\{K, J_1 \cdots J_7\} \times \{K, J_1 \cdots J_7\}$ | — | — | — | — |
| 4.5 | 81/648 → | 81/648 | $\{K, J_1 \cdots J_8\} \times \{K, J_1 \cdots J_8\}$ | — | — | — | — |
| | | 72/576 | $\{J_1 \cdots J_8\} \times \{K, J_1 \cdots J_8\}$ | $2\pi$ | $\sqrt{4 \cdot \pi/9}$ | $\sqrt{1 \cdot \pi/9}$ | >0 |
| | | 64/512 | $\{J_1 \cdots J_8\} \times \{J_1 \cdots J_8\}$ | $2\pi$ | $\sqrt{8 \cdot \pi/9}$ | $\sqrt{2 \cdot \pi/9}$ | >0 |
| | | 64/512 | $\{K, J_1 \cdots J_7\} \times \{K, J_1 \cdots J_7\}$ | — | — | — | — |
| 5 | 100/800 → | 100/800 | $\{K, J_1 \cdots J_9\} \times \{K, J_1 \cdots J_9\}$ | — | — | — | — |
| | | 81/648 | $\{J_1 \cdots J_9\} \times \{J_1 \cdots J_9\}$ | $2\pi$ | $\sqrt{8 \cdot \pi/10}$ | $\sqrt{2 \cdot \pi/10}$ | >0 |
| | NOTE – G's relate to the respective entities, i.e., page or dictionary. | 81/648 | $\{K, J_1 \cdots J_8\} \times \{K, J_1 \cdots J_8\}$ | — | — | — | — |
| | | 64/512 | $\{J_1 \cdots J_8\} \times \{J_1 \cdots J_8\}$ | $2\pi$ | $\sqrt{8 \cdot \pi/10}$ | $\sqrt{2 \cdot \pi/10}$ | >0 |

TABLE XV
SUITABLE LINGUISTIC MULTIPLEXING ROUTINES

| Eff. Size ≥ | Multiplexing Result Space | GCD (Affix) | Pri. Root | Sec. Root | $^{p}N_E$ | $^{p}N_C$ | $^{p}N_R$ | $n_e$ | $k$ | $^{s}N_E$ | $^{s}N_C$ | $^{s}N_R$ |
|---|---|---|---|---|---|---|---|---|---|---|---|---|
| 72/576 = | 72×1/··· = | $2^3$ • | 9 • | 1 | 9 = | 8 + | 1 | 6 | 3 | — | — | — |
| 81/648 = | 9×9/··· = | $2^0$ • | 9 • | 9 | 9 = | 8 + | 1 | 6 | 3 | 9 = | 8 + | 1 |
| = | 81×1/··· = | $2^0$ • | 81 • | 1 | 81 = | 64 + | 17 | 3 | 6 | — | — | — |
| > | 80×1/··· = | $2^4$ • | 5 • | 1 | 5 = | 4 + | 1 | 4 | 2 | — | — | — |
| 100/800 = | 10×10/··· = | $2^2$ • | 5 • | 5 | 5 = | 4 + | 1 | 4 | 2 | 5 = | 4 + | 1 |
| > | 96×1/··· = | $2^5$ • | 3 • | 1 | 3 = | 2 + | 1 | 2 | 1 | — | — | — |
| NOTE ÷2 = | ··· / 20×20 = | 1 • | 20 • | 20 | 20 = | 17 + | 3 | 5 | 5 | 20 = | 17 + | 3 |
| This ↑ > | ··· / 377×1 = | 29 • | 13 • | 1 | 13 = | 9 + | 4 | 2 | 4 | — | — | — |
| accounts > | ··· / 364×1 = | 52 • | 7 • | 1 | 7 = | 5 + | 2 | 3 | 3 | — | — | — |
| for TCM bit > | ··· / 344×1 = | 86 • | 4 • | 1 | 4 = | 3 + | 1 | 3 | 2 | — | — | — |

TABLE XVI
IDEAL CODING GAINS

| Pts Q/P | Eff. Pool | Eff. Size | $^{*/8}G_\delta$ | $^{*/8}G_\varphi$ | $^{*/8}G_\rho$ | $^{*/8}G_J$ | $\Sigma^{*/8}G$ | $^{**}G_\delta$ | $^{**}G_J$ | $\Sigma^{**}G$ |
|---|---|---|---|---|---|---|---|---|---|---|
| 4 | 64/512 → | 64/512 | 2.40 | 1.60 | 2.63 | — | 6.19 | 1.14 | — | 1.14 |
| 4.5 | 81/648 → | 81/648 | 2.19 | 1.02 | 2.45 | — | 5.66 | 1.04 | — | 1.04 |
| | | 72/576 | 2.19 | 1.02 | 2.45 | 1.02 | 6.69 | 1.04 | 0.50 | 1.54 |
| | | 64/512 | 2.19 | 1.02 | 2.45 | 1.48 | 7.15 | 1.04 | 0.71 | 1.76 |
| | | 64/512 | 2.19 | 1.02 | 2.45 | — | 5.66 | 1.04 | — | 1.04 |
| 5 | 100/800 → | 100/800 | 2.02 | 0.92 | 2.29 | — | 5.23 | 0.97 | — | 0.97 |
| | NOTE – *˙˙G's are of pages, rounded, in dB; respectively, **G's are of the whole dictionary. | 81/648 | 2.02 | 0.92 | 2.29 | 1.32 | 6.55 | 0.97 | 0.64 | 1.60 |
| | | 81/648 | 2.02 | 0.92 | 2.29 | — | 5.23 | 0.97 | — | 0.97 |
| | | 64/512 | 2.02 | 0.92 | 2.29 | 1.32 | 6.55 | 0.97 | 0.64 | 1.60 |

## OUTLINE PERFORMANCE OPTIONS

Considering the arranged outlines of the sample constellations, two of the found interesting [1] with an intermediate in between of them, see Tables III again, X, and IX, respectively, we introduce a set of design options to improve the operation time behavior of such a coding means, see Table XI.

Going over, we evaluate the performance of a given among the feasible coding options, as the countable thus comparable properties of its MDI output signal, see Table XII.[4]

Although the grid-based arrangement causes an irreducible wobble in the output of the so born outline, see Tables VI and XII again, respectively, its effect is close to negligible, while the key properties of any proposed option look more balanced than those of the original design, see Table XIII.[5]

## CODES PREVENTING KEEPS

Manipulating on the transport stream as a discrete continuous sequence of words, it is possible to prevent for every two adjacent words to be the same, if the stellarial constellation the design employs features an excessive number of points within itself, see Table XIV in conjunction with Tables VII, VIII, XI, and XII, at the cost of proportional shrinkages in the effective size of the available coding space of such the design.

The size matters because predefines the fashion of payload delimitation, just at least, and preemption, even at most, up to seamless via linguistic multiplexing [4], see Table XV.

The same time, preventing any keeps in the stream gives us the ground to enrich (enforce) the design with a new measure of a comparatively tangible gain, see Table XVI.

This measure is actual not only in the scope of the repartitioned pages, but also in the scope of the transport dictionary, see Tables XIV and XVI again, i.e., where the trigonometric one has some but nothing to add [1], respectively.

## CONCLUSION

Being appropriately applied in the design, the newly introduced, dynamic aspects also arise from the trigonometric view on that design then make its resulting coding space redundant and, thus, more complicated but still managed.[6]

Reaching such redundancy enables for the design to restrict on its behavior during the operation time so as to improve its performance, up to the first prime deep, contributing into the power balance of its output signal driving the MDI.

Reaching such redundancy enables for the design to implement an extra measure, that is based on preventing for keeps to occur in the transport stream during the operation time of the design, contributing into the whole coding gain.

Thereby, we can consider all the underlying design details that together ensure for all the properties mentioned above to be achieved, as it was described in the paper, both practicable and practical, i.e., usable and useful, respectively.

[4] In this table, the respective values looking the same are such only because rounded to the shown number of digits, their real originals are distinct.

[5] In this table and further, gain values are estimated based on the respective metric measures, see [1], given rounded, and always expressed in dB.

[6] E.g., with the arranged outline of the sample constellation of 5 points/P/Q, see Table III, we need to encode the stream (just) two periods ahead.

# Design Limits, Implicit and Explicit, Related to the Data Coding Means and Event Coding Means Multiplexed Inside the 1000BASE-T PMA Sublayer

Alexander Ivanov

***Abstract*—In this paper, we're finalizing the branch dedicated to the development of the Gigabit Ethernet physical layer, type 1000BASE-T, done inside its PMA sublayer, highlighting various implicit and explicit limits of the design approach.**



## INTRODUCTION

POWER-balanced MDI output signal is the key feature of the prior designs, including—as we will further refer to—stellarial [1] and dynamic [2], as well as the main goal of their respective design approaches described in [1] and [2].

Being the initial who formulates the goal and then achieves it, the stellarial design approach [1] introduces the concept of plane (2D) constellations and their couples (4D) whose MDI power-related properties are manageable by design and, thus, can be accurately balanced according to the goal.

Succeeding to the initial and developing on it, the dynamic design approach [2] extends the initial concept, whose behavior is considered static, with a range of extra measures, whose behavior is considered dynamic, that retains the goal and the same time increases the respective coding gains.

In this paper, we plan to highlight the major design limits, including implicit and explicit, all inherent to such the design since and while it aims at such the goal, especially in the view of the underlying MDI signal basis, i.e., 4D-PAM17.

## GEOMETRY-BACKED SUPPOSITIONS

Coupling two instances of a given, two-dimensional (2D) stellarial constellation with two power orbits of radii $R$ and $r$ [1] results in a new, four-dimensional (4D) constellation with a single power orbit of radius $\sqrt{R^2 + r^2}$ [1], that ensures for the complete MDI output power is constant by design.

Such a coupled constellation performs like a set of points located at the hypersphere,[1] that is a three-sphere in our case, representing the surface of a hyperball, that is a four-ball in our case, of the single radius mentioned just above.

Recalling the fate of submission of many prior works to the peer reviewed journal, such a try with this one also promises no chance, probably.

Please sorry for the author has no time to find this work a new home, peer reviewed or not, except of arXiv, and just hopes there it meets its reader, one or maybe various, whom the author beforehand thanks for their regard.

A. Ivanov is with JSC Continuum, Yaroslavl, the Russian Federation.

Digital Object Identifier 10.48550/arXiv.yymm.nnnn (this bundle).

[1]It is an $(n-1)$-dimensional figure in an $n$-dimensional space, $n > 3$.

TABLE I
POWER ORBIT FORMS

| Space | Depiction | Pulse Range | Ideal Signaling Definition |
|---|---|---|---|
| 1D ($n = 1$) | $\text{dim}_1$; $2\lvert u_{\text{ref}}\rvert$ | $u_1 : \lvert u_1\rvert = \lvert u_{\text{ref}}\rvert$ | $\lvert u_1\rvert = \lvert u_{\text{ref}}\rvert$ |
| 2D ($n = 2$) | $\lvert u_{\text{ref}}\rvert$; $\text{dim}_2$, $\text{dim}_1$; one-sphere (circle) | $u_1 : \lvert u_1\rvert \le \lvert u_{\text{ref}}\rvert$<br>$u_2 : \lvert u_2\rvert \le \lvert u_{\text{ref}}\rvert$ | $(u_1)^2 + (u_2)^2 = (u_{\text{ref}})^2$ |
| | | NOTE – Diameter is $D = 2R = 2\lvert u_{\text{ref}}\rvert$ for all $n$'s. | |
| 3D ($n = 3$) | $\lvert u_{\text{ref}}\rvert$; $\text{dim}_2$, $\text{dim}_1$, $\text{dim}_3$; two-sphere (sphere) | $u_1 : \lvert u_1\rvert \le \lvert u_{\text{ref}}\rvert$<br>$u_2 : \lvert u_2\rvert \le \lvert u_{\text{ref}}\rvert$<br>$u_3 : \lvert u_3\rvert \le \lvert u_{\text{ref}}\rvert$ | $(u_1)^2 + (u_2)^2 + (u_3)^2 = (u_{\text{ref}})^2$ |
| >3D ($n > 3$) | not possible<br>($n$–1)-sphere = hypersphere in the { $\text{dim}_1$ ; ... ; $\text{dim}_n$ } basis | $u_1 : \lvert u_1\rvert \le \lvert u_{\text{ref}}\rvert$<br>. . . . . . . . . . . .<br>$u_n : \lvert u_n\rvert \le \lvert u_{\text{ref}}\rvert$ | $(u_1)^2 + ... + (u_n)^2 = (u_{\text{ref}})^2$ |

TABLE II
POWER BALANCE FORMS

| Hyperspherical EQUIVALENT | Single Period | Multiple Periods | |
|---|---|---|---|
| $w_1 = (u_1/u_{\text{ref}})^2$<br>. . . . . . . . . . .<br>$w_n = (u_n/u_{\text{ref}})^2$ | time; pair; $u_1$ ⋮ $u_n$ | time; pair; $u_1$ ⋯ ⋱ $u_n$ | time; pair; $u_1 \cdots u_n$ |
| $w_1 + ... + w_n = 1$ | Multiple Signals | | Single Signal |

Based on this analogy, we can consider a power-balanced constellation, including the coupled designs met in [1] or [2], as a "generalized" hyperspherical constellation whose power orbit as well as power balance are representable in some but one analytic form, see Tables I and II,[2] respectively.

Modeling so an interesting coupled stellarial constellation, see Table III, we locate the vertices of its outline [2], ideal or arranged, see Table IV, all at the surface, that is the accordant ideal hypersphere or its grid-rearranged substitute, of a solid, that is the accordant ideal hyperball or its approximation set up on the same grid, and analyze such figures generalized, as an $(n-1)$-sphere and an $n$-ball, respectively.

[2]In our case, the design drives $n = 4$ multiple signals per single period.

TABLE III
COUPLED STELLARIAL CONSTELLATION MODELING

| Signal | Time Period | Power | Pair | Plane | Complete MDI Output Power | Low Power Orbit + High Power Orbit |
|---|---|---|---|---|---|---|
| $u_1$, $u_2$ | @m | $w_1$, $w_2$ | A, B | <AB> → | $w_{4D} = \begin{bmatrix} w_1 + w_2 \\ + \\ w_3 + w_4 \end{bmatrix} = \begin{bmatrix} w_{2D} \\ + \\ W_{2D} \end{bmatrix}$ or $\begin{bmatrix} W_{2D} \\ + \\ w_{2D} \end{bmatrix}$ | $w_{2D} = \frac{r^2}{R^2 + r^2}$  $W_{2D} = \frac{R^2}{R^2 + r^2}$ |
| $u_3$, $u_4$ | | $w_3$, $w_4$ | C, D | <CD> → | | $w_{4D} = w_{2D} + W_{2D} = w_1 + w_2 + w_3 + w_4 = 1$ |

TABLE IV
OUTLINE MODELING

| Property | Ideal Outline → | Grid-based |
|---|---|---|
| Space Used | analytic, *n*-D, continuous | *n*D-PAM*M*, discrete, finite |
| Vertex coords. | up to exact, strict positions | approximate, may be fuzzy |

TABLE V
EQUIVALENT SURFACE AREA ESTIMATION

| Procedure Progress Description ... | ... and Definition |
|---|---|
| Given parameters of an *n*D-PAM*M* design | $n \geq 1$, $M \geq 2$ |
| Parameters of the given design as an *n*-dimensional grid/space of points | zero-symmetric, finite, regular |
| [known analytically] Surface area of a unit *n*-ball | $s = s(n)$ |
| [known analytically] Body volume of a unit (*n*–1)-ball | $v = v(n-1)$ |
| Volume characterizing the (*n*–1)D-PAM*M* design | $V = V(n-1, M)$ |
| Area of the surface characterizing the design | $S = (s / v) \times V$ |
| NOTE – Surface of a unit *n*-ball solid is an (*n*–1)-sphere, its area is in $(\text{unit})^{n-1}$. | |
| NOTE – Body volume of a design is a model of that of a unit *n*-ball, in $(\text{unit})^n$. | |

TABLE VI
EQUIVALENT BODY VOLUME ESTIMATION

| Procedure Progress Description ... | ... and Definition |
|---|---|
| Given parameters of an *n*D-PAM*M* design | $n \geq 1$, $M \geq 2$ |
| Parameters of the given design, per a dimension | $u_{\text{ref}\pm} = \pm\lvert u_{\text{ref}}\rvert$, $\Delta = \lvert u_{\text{ref}}\rvert / (M-1)$ |
| Highest absolute pulse level and its output power | $u_{\max} = \lvert u_{\text{ref}}\rvert$, $P_{\max} = u_{\max} \times u_{\max}$ |
| Lowest absolute pulse level and its output power | $u_{\min} = \Delta \times (M \bmod 2)$, $P_{\min} = \ldots$ |
| Radius of the (*n*–1)-sphere “bounding” the *n*-ball | $R_{\text{test}} \times R_{\text{test}} = P_{\max} + (n-1) \times P_{\min}$ |
| Number of points on the (*n*–1)-sphere of $R_{\text{test}}$ | $N_{\circ} = \Sigma\,[\,R_{\text{point}} = R_{\text{test}}\,]$ |
| Number of points inside the (*n*–1)-sphere of $R_{\text{test}}$ | $N_{\bullet} = \Sigma\,[\,R_{\text{point}} < R_{\text{test}}\,]$ |
| Volume of the body characterizing the design | $V = (N_{\bullet} + \tfrac{1}{2} N_{\circ}) \times (\Delta / R_{\text{test}})^n$ |

TABLE VII
SURFACE AREA AS A TRANSMIT LIMIT

| Design | n = 1 | n = 2 | n = 3 | n = 4 | n = 5 | n = 6 | n = 7 | n = 8 | n = 9 | n = 10 |
|---|---|---|---|---|---|---|---|---|---|---|
| *n*-Ball: | $(\text{unit})^0$ | $(\text{unit})^1$ | $(\text{unit})^2$ | $(\text{unit})^3$ | $(\text{unit})^4$ | $(\text{unit})^5$ | $(\text{unit})^6$ | $(\text{unit})^7$ | $(\text{unit})^8$ | $(\text{unit})^9$ |
| • analytic | 2 | $2\pi$ | $4\pi$ | $2\pi^2$ | $\frac{8}{3}\pi^2$ | $\pi^3$ | $\frac{16}{15}\pi^3$ | $\frac{1}{3}\pi^4$ | $\frac{32}{105}\pi^4$ | $\frac{1}{12}\pi^5$ |
| • numeric | 2 | 6.28 | 12.57 | 19.74 | 26.32 | 31.01 | 33.07 | 32.47 | 26.69 | 25.50 |
| • percent | ~6% | ~19% | ~38% | ~60% | ~80% | ~94% | **100%** | ~98% | ~90% | ~77% |
| *n*D-PAM2 | **2** | — | — | — | — | — | — | — | — | — |
| *n*D-PAM3 | ~8% | ~26% | ~58% | ~86% | **100%** | ~98% | ~85% | ~66% | ~47% | ~31% |
| *n*D-PAM4 | ~5% | ~16% | ~33% | ~53% | ~73% | ~89% | ~99% | **100%** | ~94% | ~84% |
| *n*D-PAM5 | ~6% | ~19% | ~41% | ~67% | ~88% | ~98% | **100%** | ~96% | ~87% | ~75% |
| *n*D-PAM6 | ~6% | ~18% | ~36% | ~57% | ~77% | ~92% | **100%** | ~100% | ~93% | ~81% |
| *n*D-PAM7 | ~6% | ~19% | ~41% | ~63% | ~81% | ~94% | **100%** | ~100% | ~92% | ~80% |
| *n*D-PAM8 | ~6% | ~19% | ~37% | ~57% | ~77% | ~93% | **100%** | ~99% | ~91% | ~79% |
| ... | | | | | | | | | | |
| *n*D-PAM16 | ~6% | ~19% | ~38% | ~60% | ~80% | ~94% | **100%** | ~98% | ~90% | ~77% |
| *n*D-PAM17 | 2 | 6.28 | 12.62 | 19.80 | 26.47 | 31.08 | 33.00 | 32.33 | 29.57 | 25.45 |
| | ~6% | ~19% | ~38% | ~60% | ~80% | ~94% | **100%** | ~98% | ~90% | ~77% |
| *n*D-PAM18 | ~6% | ~19% | ~37% | ~59% | ~79% | ~93% | **100%** | ~98% | ~90% | ~77% |
| ... | | | | | | | | | | |
| *n*D-PAM32 | ~6% | ~19% | ~38% | ~59% | ~79% | ~94% | **100%** | ~98% | ~90% | ~77% |

TABLE VIII
BODY VOLUME AS A RECEIVE LIMIT

| ··· | n = 1 | n = 2 | n = 3 | n = 4 | n = 5 | n = 6 | n = 7 | n = 8 | n = 9 |
|---|---|---|---|---|---|---|---|---|---|
| | $(\text{unit})^1$ | $(\text{unit})^2$ | $(\text{unit})^3$ | $(\text{unit})^4$ | $(\text{unit})^5$ | $(\text{unit})^6$ | $(\text{unit})^7$ | $(\text{unit})^8$ | $(\text{unit})^9$ |
| ··· | 2 | $\pi$ | $\frac{4}{3}\pi$ | $\frac{1}{2}\pi^2$ | $\frac{8}{15}\pi^2$ | $\frac{1}{6}\pi^3$ | $\frac{16}{105}\pi^3$ | $\frac{1}{24}\pi^4$ | $\frac{32}{945}\pi^4$ |
| ··· | 2 | 3.14 | 4.19 | 4.93 | **5.26** | 5.17 | 4.72 | 4.06 | 3.30 |
| ··· | ~38% | ~60% | ~80% | ~94% | 100% | ~98% | ~90% | ~77% | ~63% |
| ··· | **2** | — | — | — | — | — | — | — | — |
| ··· | 2 | 3.50 | 4.43 | **4.56** | 4.06 | 3.23 | 2.34 | 1.57 | 0.99 |
| ··· | 2 | 3.20 | 4.39 | 5.33 | 5.88 | **5.97** | 5.64 | 5.00 | 4.19 |
| ··· | 2 | 3.40 | 4.69 | 5.41 | **5.49** | 5.15 | 4.60 | 3.94 | 3.21 |
| ··· | 2 | 3.08 | 4.11 | 4.90 | **5.31** | 5.3096 | 4.94 | 4.31 | 3.56 |
| ··· | 2 | 3.30 | 4.36 | 4.95 | **5.16** | 5.07 | 4.70 | 4.10 | 3.35 |
| ··· | 2 | 3.04 | 4.04 | 4.83 | **5.23** | 5.20 | 4.80 | 4.15 | 3.39 |
| ... | | | | | | | | | |
| ··· | 2 | 3.12 | 4.20 | 4.95 | **5.27** | 5.16 | 4.72 | 4.06 | 3.30 |
| ··· | 2 | 3.15 | 4.20 | 4.96 | **5.28** | 5.16 | 4.70 | 4.04 | 3.29 |
| | ~38% | ~60% | ~80% | ~94% | 100% | ~98% | ~89% | ~77% | ~62% |
| ··· | 2 | 3.09 | 4.13 | 4.89 | **5.24** | 5.17 | 4.73 | 4.06 | 3.30 |
| ... | | | | | | | | | |
| ··· | 2 | 3.11 | 4.15 | 4.92 | **5.26** | 5.17 | 4.73 | 4.06 | 3.30 |

Having an ideal $(n-1)$-sphere or its grid-based approximation, we can estimate the area of its surface, that also will be the area of the surface of the respective $n$-ball, as a function of $n$, i.e., the number of dimensions, and $M$, i.e., the number of points per every dimension in the grid, assuming the latter one infinite in the ideal case, see Tables V and VII.[3]

Because equivalently, per every word sent, the transmitting side picks up a point only and only at that surface, its area is the right quantitative reflection to reveal such a limit, e.g., in a from of a local optimum, and here we obtain, see Table VII again, that such the optimum is about $n_{\text{TX}} = 7 \pm 1$.

Having an ideal $n$-ball or its grid-based approximation, we can estimate the volume of its body, see Tables VI and VIII, as a function of $n$ and $M$ in the same sense, too.

[3]To cook those results comparable, we consider a unit $n$-ball and assume each of $n$ per-dimension PAM-$M$ output signals drives a unit load.

Because equivalently, per every word educed, the receiving side faces the need to seek for a point within that body in the whole, instead of just the surface, its volume is the right one to reveal such a limit, e.g., in the same as above form too, and here we observe, see Table VIII again, that such the optimum is about $n_{\text{RX}} = 5 \pm 1$, stating thereby $n_{\text{RX}} < n_{\text{TX}}$.

Based on this, we suppose that the MDI signaling basis of the stellarial design and its successors, that predetermines the 4D-PAM17 grid, i.e., of $n = 4$, primarily, is optimal because is very close to the strictest of the highlighted here, geometry-backed limits, which one is the receive limit.[4]

[4]Remember we consider a design as a set of some measures undertaken by the transmitting side and only assume there exists a set of the respective means the receiving side can implement to benefit on the measures, see [1], therefore, until the respective means are clear and their properties are confirmed, we call the reasoning aspects described here and further just our suppositions, ranging them by what each one is backed with, geometry, trigonometry, etc.

TABLE IX
BASIC TERMS

| Rooms at Plane | | Codes of Rooms | | | | Indexing Rule | |
|---|---|---|---|---|---|---|---|
| $r\angle 1\varphi_0$ | $R\angle 1\varphi_0$ | 1 ab 1 | 1 AB 1 | 1 cd 1 | 1 CD 1 | — 0 | 0 — |
| $r\angle 3\varphi_0$ | $R\angle 3\varphi_0$ | 2 ab 1 | 2 AB 1 | 2 cd 1 | 2 CD 1 | 1 — | — 1 |
| • • • • • • • • • • • | | 3 ab 1 | 3 AB 1 | 3 cd 1 | 3 CD 1 | — 2 | 2 — |
| | | 4 ab 1 | 4 AB 1 | 4 cd 1 | 4 CD 1 | 3 — | — 3 |
| P0 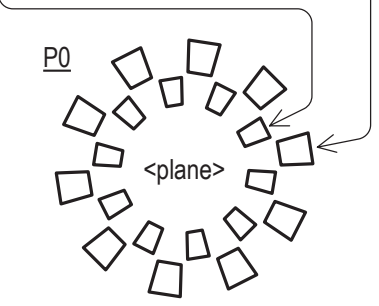 | | 5 ab 1 | 5 AB 1 | 5 cd 1 | 5 CD 1 | — 4 | 4 — |
| | | 1 ab 2 | 1 AB 2 | 1 cd 2 | 1 CD 2 | 5 — | — 5 |
| | | 2 ab 2 | 2 AB 2 | 2 cd 2 | 2 CD 2 | — 6 | 6 — |
| | | 3 ab 2 | 3 AB 2 | 3 cd 2 | 3 CD 2 | 7 — | — 7 |
| | | 4 ab 2 | 4 AB 2 | 4 cd 2 | 4 CD 2 | — 8 | 8 — |
| Pts/P/Q = 5, $\varphi_0 = \pi/20$ | | 5 ab 2 | 5 AB 2 | 5 cd 2 | 5 CD 2 | 9 — | — 9 |
| | | 1 ab 3 | 1 AB 3 | 1 cd 3 | 1 CD 3 | — 10 | 10 — |
| | | 2 ab 3 | 2 AB 3 | 2 cd 3 | 2 CD 3 | 11 — | — 11 |
| | | 3 ab 3 | 3 AB 3 | 3 cd 3 | 3 CD 3 | — 12 | 12 — |
| | | 4 ab 3 | 4 AB 3 | 4 cd 3 | 4 CD 3 | 13 — | — 13 |
| P1 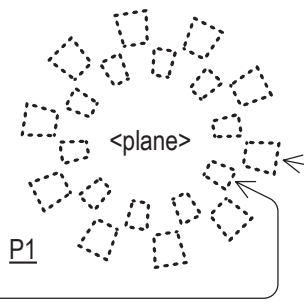 | | 5 ab 3 | 5 AB 3 | 5 cd 3 | 5 CD 3 | — 14 | 14 — |
| | | 1 ab 4 | 1 AB 4 | 1 cd 4 | 1 CD 4 | 15 — | — 15 |
| | | 2 ab 4 | 2 AB 4 | 2 cd 4 | 2 CD 4 | — 16 | 16 — |
| • • • • • • • • • • • | | 3 ab 4 | 3 AB 4 | 3 cd 4 | 3 CD 4 | 17 — | — 17 |
| $r\angle 37\varphi_0$ | $R\angle 37\varphi_0$ | 4 ab 4 | 4 AB 4 | 4 cd 4 | 4 CD 4 | — 18 | 18 — |
| $r\angle 39\varphi_0$ | $R\angle 39\varphi_0$ | 5 ab 4 | 5 AB 4 | 5 cd 4 | 5 CD 4 | 19 — | — 19 |
| Low Orbit | High Orbit | for Plane <AB> | | for Plane <CD> | | Page P0 | Page P1 |

TABLE X
TRANSITS REDUCTION

| Jump to a $2\Delta\varphi$-invariant Room i.e., when $\varphi(\text{Room}_m) \underset{2\Delta\varphi}{\equiv} \varphi(\text{Room}_{m+1})$ | Jump to a non-$2\Delta\varphi$-invariant Room i.e., when $\varphi(\text{Room}_m) \underset{2\Delta\varphi}{\not\equiv} \varphi(\text{Room}_{m+1})$ | |
|---|---|---|
| 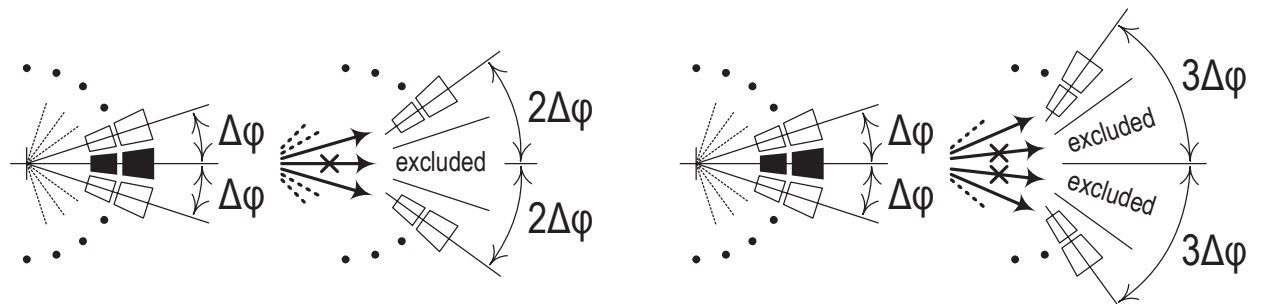 | | |
| Next Period Views: possible: $N = 2\pi / 2\Delta\varphi$; excluded: 1; left: $N-1$ | ... possible: $N$; excluded: 2; left: $N-2$ | WORST CASE $N-2$ |

NOTE: Shown centering $\varphi(\text{Room}_m) \to 0\pi$, $0\pi \le \varphi(\text{Room}) = \Delta\varphi \cdot \text{Index(Room)} + \Delta\varphi_0 < 2\pi$.



TABLE XI
JUMP DEAD ZONE

| Description | Initial Space | Reduced, VBR | Reduced, CBR | Rem. |
|---|---|---|---|---|
| Direct Depiction / Inverse Depiction | 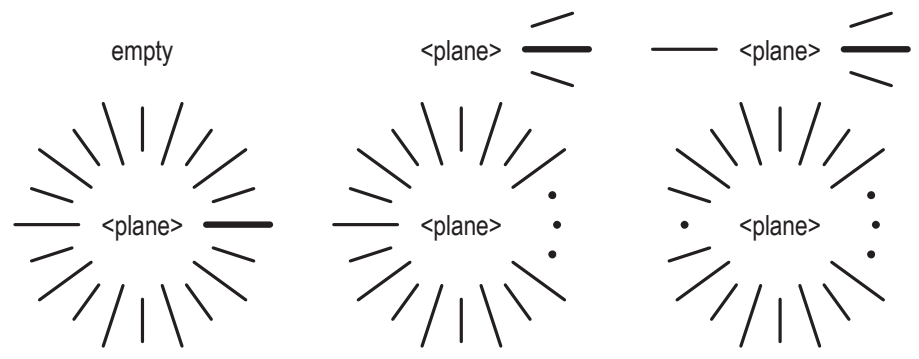 | | | Δφ; Dots are for visual clarification only. |
| Views... excluded | — | 1 + 2 = 3 | 2 + 2 = 4 | |
| accessible | $N + N = 2N$ | $2N - 3$ | $2N - 4$ | |



TABLE XII
DYNAMIC CODING OPTIONS

| Opt.→ ↓Desc. | $G_J$ | $G_J$+ | $G_{2J}$ | $G_{2J}$+ (covers Ɖ) | ··· |
|---|---|---|---|---|---|
| Planes' Jump Dead Zones | 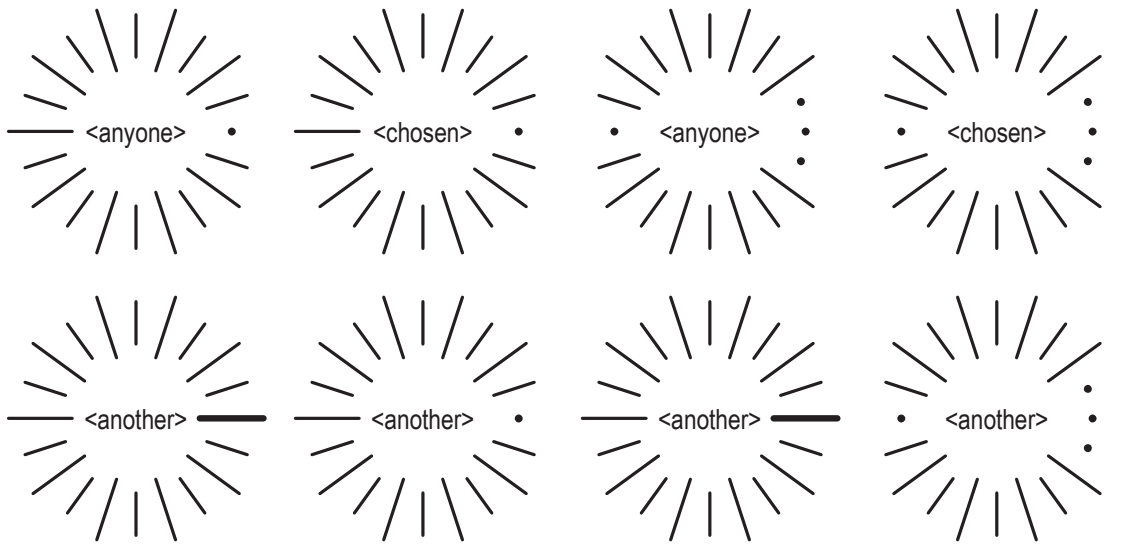 | | | | · · · |
| Space | $\{J_1 \cdots J_{N-1}\} \times \{K, J_1 \cdots J_{N-1}\}$ | $\{J_1 \cdots J_{N-1}\} \times \{J_1 \cdots J_{N-1}\}$ | $\{J_1 \cdots J_{N-2}\} \times \{K, J_1 \cdots J_{N-1}\}$ | $\{J_1 \cdots J_{N-2}\} \times \{J_1 \cdots J_{N-2}\}$ | · · · |



TABLE XIII
CODING GAIN AS A CHANNEL LIMIT

| Pts Q/P | RP: $\Sigma^{*/8}G$, in dB, v. Effective Size, in Points | | | | | Transport Dictionary: $\Sigma^{**}G$, in dB | | | | |
|---|---|---|---|---|---|---|---|---|---|---|
| 4 | $6.2^{/8\times8}$ | ——— | no $G_{\dots}$ option available | | | 1.14 | ——— | | | |
| | | ——— | no $G_{\dots}$+ option available | | | | ——— | | | |
| 4½ | $5.7^{/9\times9}$ | $6.7^{/8\times9}$ | ——— | | | 1.04 | 1.54 | ——— | | |
| | | $7.1^{/8\times8}$ | ——— | | | | 1.76 | ——— | | |
| 5 | $5.2^{/10\times10}$ | $6.1^{/9\times10}$ | $6.1^{/8\times9}$ | ——— | | 0.97 | 1.41 | 1.88 | ——— | |
| | | $6.5^{/9\times9}$ | $6.5^{/8\times8}$ | ——— | | | 1.60 | **2.29** | ——— | |
| 5½ | $4.9^{/11\times11}$ | $5.7^{/10\times11}$ | $5.7^{/9\times10}$ | $6.6^{/8\times9}$ | ——— | 0.90 | 1.30 | 1.73 | 2.17 | — |
| | | $6.0^{/10\times10}$ | $6.0^{/9\times9}$ | $7.4^{/8\times8}$ | ——— | | 1.48 | 2.09 | 2.76 | — |
| 6 | $4.5^{/12\times12}$ | $5.3^{/12\times11}$ | $5.3^{/11\times10}$ | $6.1^{/9\times10}$ | $6.1^{/8\times9}$ | 0.84 | 1.21 | 1.60 | 2.00 | 2.42 |
| | | $5.6^{/11\times11}$ | $5.6^{/10\times10}$ | $6.9^{/9\times9}$ | $6.9^{/8\times8}$ | | 1.37 | 1.93 | 2.53 | 3.18 |
| Opt. → | Static | $G_J$(+) | $G_{2J}$(+) | $G_{3J}$(+) | $G_{4J}$(+) | Static | $G_J\cdots$ | $G_{2J}\cdots$ | $G_{3J}\cdots$ | $G_{4J}\cdots$ |

## TRIGONOMETRY-BACKED SUPPOSITIONS

Partitioning a (plane) stellarial constellation [1] results into two (basic) pages [1], orthogonal to each other as well as of the same number of points within, and into four such pages with a couple of two instances of such a constellation, again orthogonal and of the same cardinality, see Table IX.

Repartitioning [1] the two by two (basic) pages of a couple of two instances of a given (plane) stellarial constellation sets up eight new, repartitioned pages [1], also orthogonal and of an equal number of points within each, see Table XIV.

Thus, in the sense of the coding space, a page, repartitioned and not, is a finite set of a regular size known a priory.

Every point of a page (with an ideal constellation) lies at a certain power orbit [1] and occupies a certain power angle [1], together—as a room [2]—unique within the page and within the two pages of a certain plane, see Table IX again.

Every two rooms of the same angle and of the same plane together form a view [2], unique with that plane, and a change of the current view—during the operation time of the design—looks like a jump, i.e., a transit [2], one among the number of possible, $N$, a part of those, $N_X : N_X < N$, may be excluded, see Table X, that, purposefully, defines the jump dead zone, see Table XI, necessary to commit the dynamic measure [2] contributing into the (page or dictionary) coding gain.[5]

[5]Following the definition of the gain of a measure [1], the minimal distance for a given dynamic [2] coding option is $\Delta_{\mathrm{J,min}} = \frac{\pi}{N} \times \min\{N_X; N - N_X\}$, therefore such a single-plane transit restriction has a limit of

$$\lim_{*\to\infty} G_J(G_{*J}) \quad = 20\log_{10}\frac{2\pi}{2\pi - \sqrt{1}\,\frac{\pi}{2}} \approx 2.498775 < 2.5 \text{ dB} \ ,$$

as well as such a double-plane transit restriction has a limit of

$$\lim_{*\to\infty} G_J(G_{*J}+) = 20\log_{10}\frac{2\pi}{2\pi - \sqrt{2}\,\frac{\pi}{2}} \approx 3.789347 < 3.8 \text{ dB} \ ,$$

respectively, achievable so at the sensible cost of half the views per a plane.

TABLE XIV
STATIC REPARTITIONED PAGES

RP's Room Disposition

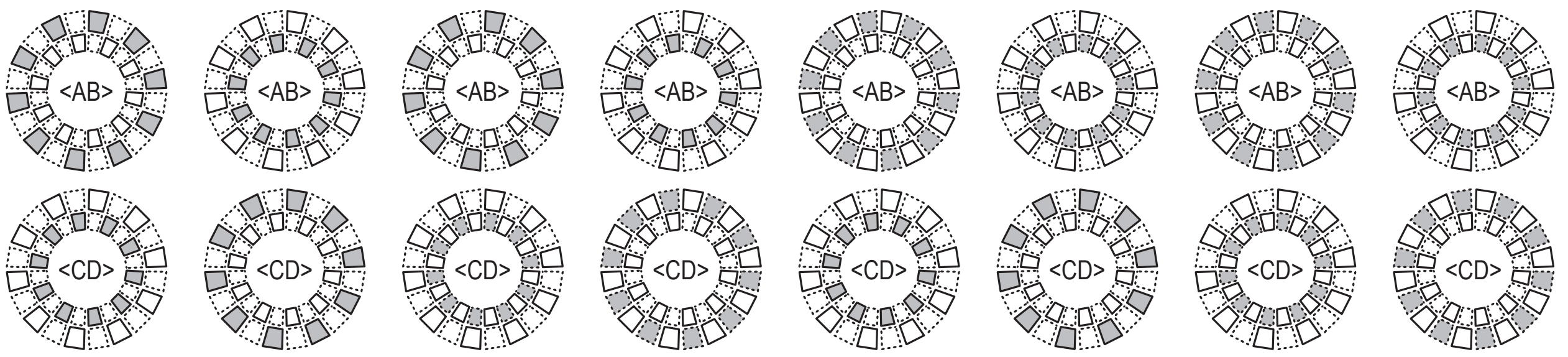

2 × 2 × 2 = 8 RPs in Dictionary

| | $P_0 = P0_R : P0_r$ | $P_1 = P0_r : P0_R$ | $P_2 = P0_R : P1_r$ | $P_3 = P0_r : P1_R$ | $P_4 = P1_R : P0_r$ | $P_5 = P1_r : P0_R$ | $P_6 = P1_R : P1_r$ | $P_7 = P1_r : P1_R$ | |
|---|---|---|---|---|---|---|---|---|---|
| RP's Room Codes | 1 AB 1<br>3 AB 1<br>5 AB 1<br>2 AB 2<br>4 AB 2<br>1 AB 3<br>3 AB 3<br>5 AB 3<br>2 AB 4<br>4 AB 4<br>×<br>2 cd 1<br>4 cd 1<br>1 cd 2<br>3 cd 2<br>5 cd 2<br>2 cd 3<br>4 cd 3<br>1 cd 4<br>3 cd 4<br>5 cd 4 | 2 ab 1<br>4 ab 1<br>1 ab 2<br>3 ab 2<br>5 ab 2<br>2 ab 3<br>4 ab 3<br>1 ab 4<br>3 ab 4<br>5 ab 4<br>×<br>1 CD 1<br>3 CD 1<br>5 CD 1<br>2 CD 2<br>4 CD 2<br>1 CD 3<br>3 CD 3<br>5 CD 3<br>2 CD 4<br>4 CD 4 | 1 AB 1<br>3 AB 1<br>5 AB 1<br>2 AB 2<br>4 AB 2<br>1 AB 3<br>3 AB 3<br>5 AB 3<br>2 AB 4<br>4 AB 4<br>×<br>1 cd 1<br>3 cd 1<br>5 cd 1<br>2 cd 2<br>4 cd 2<br>1 cd 3<br>3 cd 3<br>5 cd 3<br>2 cd 4<br>4 cd 4 | 2 ab 1<br>4 ab 1<br>1 ab 2<br>3 ab 2<br>5 ab 2<br>2 ab 3<br>4 ab 3<br>1 ab 4<br>3 ab 4<br>5 ab 4<br>×<br>2 CD 1<br>4 CD 1<br>1 CD 2<br>3 CD 2<br>5 CD 2<br>2 CD 3<br>4 CD 3<br>1 CD 4<br>3 CD 4<br>5 CD 4 | 2 AB 1<br>4 AB 1<br>1 AB 2<br>3 AB 2<br>5 AB 2<br>2 AB 3<br>4 AB 3<br>1 AB 4<br>3 AB 4<br>5 AB 4<br>×<br>2 cd 1<br>4 cd 1<br>1 cd 2<br>3 cd 2<br>5 cd 2<br>2 cd 3<br>4 cd 3<br>1 cd 4<br>3 cd 4<br>5 cd 4 | 1 ab 1<br>3 ab 1<br>5 ab 1<br>2 ab 2<br>4 ab 2<br>1 ab 3<br>3 ab 3<br>5 ab 3<br>2 ab 4<br>4 ab 4<br>×<br>1 CD 1<br>3 CD 1<br>5 CD 1<br>2 CD 2<br>4 CD 2<br>1 CD 3<br>3 CD 3<br>5 CD 3<br>2 CD 4<br>4 CD 4 | 2 AB 1<br>4 AB 1<br>1 AB 2<br>3 AB 2<br>5 AB 2<br>2 AB 3<br>4 AB 3<br>1 AB 4<br>3 AB 4<br>5 AB 4<br>×<br>1 cd 1<br>3 cd 1<br>5 cd 1<br>2 cd 2<br>4 cd 2<br>1 cd 3<br>3 cd 3<br>5 cd 3<br>2 cd 4<br>4 cd 4 | 1 ab 1<br>3 ab 1<br>5 ab 1<br>2 ab 2<br>4 ab 2<br>1 ab 3<br>3 ab 3<br>5 ab 3<br>2 ab 4<br>4 ab 4<br>×<br>2 CD 1<br>4 CD 1<br>1 CD 2<br>3 CD 2<br>5 CD 2<br>2 CD 3<br>4 CD 3<br>1 CD 4<br>3 CD 4<br>5 CD 4 | 10 × 10 = 100 pts/RP |

Alias

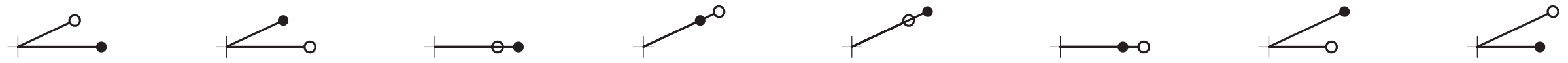

"clock"

TABLE XV
OPTIMAL DESIGN DETAILS

| $\varphi_0$ | $\Delta\varphi$ | $\Delta\varphi_0$ | Scheme | $2\Delta\varphi$-inv.-keeping | Per-RP ←Coding Space→ | $2\Delta\varphi$-inv.-inverting | $r : R$ | $^{*/8}G_\delta$ | $^{*/8}G_\varphi$ | $^{*/8}G_\rho$ | $^{*/8}G_J$ | $\Sigma^{*/8}G$ | $^{**}G_\delta$ | $^{**}G_J$ | $\Sigma^{**}G$ | $\lvert L_w \rvert$ |
|---|---|---|---|---|---|---|---|---|---|---|---|---|---|---|---|---|
| $\pi/20$ flat | $2\varphi_0$ 5 Pts/Q/P | $\varphi_0$ | 100/800+Ø static, 8 RPs/D. | $\{J^{\pm 2\Delta\varphi}, J^{\pm 4\Delta\varphi}, J^{\pm 6\Delta\varphi}, J^{\pm 8\Delta\varphi}\}^2$ dynamic, regular, 8 × 8 = 64 pts/RP | | $\{J^{\pm 3\Delta\varphi}, J^{\pm 5\Delta\varphi}, J^{\pm 7\Delta\varphi}, J^{\pm 9\Delta\varphi}\}^2$ dynamic, regular, 8 × 8 = 64 pts/RP | .637 | 2.0 + | 0.9 + | 2.3 + | 1.3 = | 6.5 dB | 1.0 + | 1.3 = | 2.3 dB | 0.1 dB |

Every word sent, there is one point associated with it on the hypersphere, whose coordinates are two accordant rooms, by one per a plane (as a projection of that hypersphere's space), see Table XIV again, the design picks up from the whole pool during its operation time, except those of the respective jump dead zone, if the design implements such an option.

Considering a jump dead zone consisting of a zero-transit-symmetric, current-view-centric angular range, that covers on the accordant $2N_\text{X} - 1$ views per a plane, plus up to one free view, that is diagonally opposite to the current view, typically and if this is necessary, we so construct the respective pattern for a given dynamic coding option [2], see Table XII.

Every option contributes into the whole coding gain that is the right reflection to reveal such a limit, see Table XIII.

Based on this, we suppose that the coding scheme variant of $N = 10$ views (points) with up to $N_\text{X} = 2$ nearest (to the current) of them dynamically excluded, both per a plane per a repartitioned page, is optimal because is close enough to the respective channel limits of the original coding [1].

## ANOTHER-BACKED SUPPOSITIONS

We suppose that there also are other (different) limits, e.g., those prevent us from obtaining an enough coding scheme [1] or an unambiguous outline [2] for a given design.

Comparing with the limits described earlier, we assume the rest limits, obvious and possible, are minor and so negligibly affect the optimal design, see Table XV.

## CONCLUSION

So, we highlighted the design limits related to the balanced MDI output power-ensuring, hyperspherical coding means, the major of those are geometry- and trigonometry-backed.

Being restricting the properties of the static design behavior only, the major geometry-backed limits thus relate to the both stellarial [1] and dynamic [2] design approaches.

Being restricting the properties of the rest, dynamic design behavior, the major trigonometry-backed limits relate only to the dynamic [2] design approach, that is expected.

In regard to a particular design, the major limits may draw its transmit and receive limits, which are geometry-backed, as well as its channel one, which is trigonometry-backed.

As both the goal and the prize, we balance the output power of that design via restricting those static (definition-time) and dynamic (operation-time) features of its line code.[6]

[6] We could call it an $n$-ball line code, $n$BLC in short, e.g., 4BLC, or simply a ball line code, BLC, when the dimension may be omitted losslessly for the sense of the discussion context. If this may be useful, we also could mention its other features sensitive in the context, e.g., 4BLC-PAM17.

# Extra Material, Including Advanced, Related to the Data Coding Means and Event Coding Means Multiplexed Inside the 1000BASE-T PMA Sublayer

Alexander Ivanov

***Abstract*—Via this paper, we're supplementing to the material of the development of the Gigabit Ethernet physical layer, type 1000BASE-T, done inside its PMA sublayer, to reveal extra design techniques, including general and advanced.**

## INTRODUCTION, DESCRIPTION, CONCLUSION

BASED on all the prior work comprising the stellarial [1], dynamic [2], and optimal [3] design approaches, we now discuss an extra approach that is developing on the latter one as its alternative in restricting an angular jump.

We begin with a static [1], MDI output power-balanced [1], hyperspherical [3] constellation set up using the trigonometry-based restriction [1] in a 4-D Euclidean geometry-based space [1], and then enable a dynamic [2] rule restricting [2] an inter-angular jump [2], now single because coupled over the paired manipulation planes [1], see Tables I then II,[1] instead of two self-independent as it was done in [2] and [3].

Other terms, it defines the respective dynamic constellation present completely within its static parent, being set up in the scope of the current room [2] of the current repartitioned page [1] for the current word time period, see Table III.

With a redundant coding space, see Table III again, we can use a linguistic multiplexing routine [1], see Table IV.

We implement the static hyperspherical constellation based on a couple of 2-D stellarial outlines [2], also including here those of two denser cases,[2] see Tables V and VI.[3]

Thanks to this all, the physical signaling of a design implementing such the constellation-in-constellation (CiC) encoding demonstrates attractive properties, see Table VII.[4]

TABLE I
PLANE INVARIANCE INVERSION ORDER

| Invariance Relations | | $\varphi({}^{\langle AB\rangle}\mathrm{Room}_{m+1})$ is... | $\varphi({}^{\langle CD\rangle}\mathrm{Room}_{m+1})$ is... | Order | Alias |
|---|---|---|---|---|---|
| inverted | inverted | non-2Δφ-invariant | non-2Δφ-invariant | 2 | I-I |
| inverted | directed | non-2Δφ-invariant | 2Δφ-invariant | 1 | I-D |
| directed | inverted | 2Δφ-invariant | non-2Δφ-invariant | 1 | D-I |
| directed | directed | 2Δφ-invariant | 2Δφ-invariant | 0 | D-D |
| at <AB> | at <CD> | ... for $\varphi({}^{\langle AB\rangle}\mathrm{Room}_{m})$ | ... for $\varphi({}^{\langle CD\rangle}\mathrm{Room}_{m})$ | of Inversion | |

TABLE II
JUMPS POSSIBLE INSIDE 4-POINT/P/Q COUPLED SPACE

| $G_{jump}$, dB versus invariance | | $\varphi({}^{\langle AB\rangle}\mathrm{Room}_{m+1}) - \varphi({}^{\langle AB\rangle}\mathrm{Room}_{m})$: π or 0 | ±7 Δφ or ±1 Δφ | ±6 Δφ or ±2 Δφ | ±5 Δφ or ±3 Δφ | ±4 Δφ |
|---|---|---|---|---|---|---|
| $\varphi({}^{\langle CD\rangle}\mathrm{Room}_{m+1}) - \varphi({}^{\langle CD\rangle}\mathrm{Room}_{m})$ | π or 0 | **—<br>D-D** | *0.6<br>I-D* | **1.2<br>D-D** | *1.8<br>I-D* | **2.5<br>D-D** |
| | ±7 Δφ or ±1 Δφ | *0.6<br>D-I* | 0.8<br>I-I | *1.3<br>D-I* | 1.9<br>I-I | *2.6<br>D-I* |
| | ±6 Δφ or ±2 Δφ | **1.2<br>D-D** | 1.3<br>I-D | **1.7<br>D-D** | 2.2<br>I-D | **2.8<br>D-D** |
| | ±5 Δφ or ±3 Δφ | *1.8<br>D-I* | 1.9<br>I-I | *2.2<br>D-I* | 2.7<br>I-I | *3.3<br>D-I* |
| | ±4 Δφ | **2.5<br>D-D** | *2.6<br>I-D* | **2.8<br>D-D** | *3.3<br>I-D* | **3.8<br>D-D** |

| Jumps' Count vs. Gain: $G_j$ | **D-D** | *D-I* / *I-D* | | I-I |
|---|---|---|---|---|
| 3.8 | 4 | — | — | — |
| 3.3 | — | 8 | 8 | — |
| 2.8 | 16 | — | — | — |
| 2.7 | — | — | — | 16 |
| 2.6 | — | 8 | 8 | — |
| 2.5 | 8 | — | — | — |
| 2.2 | — | 16 | 16 | — |
| 1.9 | — | — | — | 32 |
| 1.8 | — | 8 | 8 | — |
| 1.7 | 16 | — | — | — |
| 1.3 | — | 16 | 16 | — |
| 1.2 | 16 | — | — | — |
| 0.8 | — | — | — | 16 |
| 0.6 | — | 8 | 8 | — |
| — | 4 | — | — | — |

Despite of similarities in their behavior, see Table VIII, the coding options enabled from the optimal design approach [3], see Table IX,[5] look just a little but still yet more attractive in the sense of the complexity of the underlying coding scheme and the gain it promises us with, see again Tables VII and IX, but now in comparison with each other.

Recalling the fate of submission of many prior works to the peer reviewed journal, such a try with this one also promises no chance, probably.

Please sorry for the author has no time to find this work a new home, peer reviewed or not, except of arXiv, and just hopes there it meets its reader, one or maybe various, whom the author beforehand thanks for their regard.

A. Ivanov is with JSC Continuum, Yaroslavl, the Russian Federation.

[1]See Table I in conjunction and Table II in comparison with [3]'s Table X and with [3]'s Tables XI and XII as well as [2]'s Table XI, respectively.

[2]Arranging the vertices labeled "(!)" so as it is done in Table VI causes a decrease of about .05 dB in the coding gain, that we accept as negligible.

[3]See Tables V and VI in conjunction with [2]'s Tables III, IX, and X, also, see the mentioned tables in conjunction with [1]'s Tables XIII and XIV.

[4]See Table VII in conjunction with [1]'s Table XX.

[5]See Table IX in conjunction with [2]'s Table XII and [3]'s Table XIII.

TABLE III
ANGULAR CONSTELLATION OPTIONS

| $N_E$ | $N_R$ | 6 Pts/P/Q = 24 Views/Plane | | | 5.5 Pts/P/Q = 22 Views/Plane | | | 5 Pts/P/Q = 20 Views/Plane | | | 4.5 Pts/P/Q = 18 Views/Plane | | | 4 Pts/P/Q = 16 Views/Plane | | | Linguistic Multiplexing Variants |
|---|---|---|---|---|---|---|---|---|---|---|---|---|---|---|---|---|---|
| | | I-I | I-D/D-I | D-D | I-I | I-D/D-I | D-D | I-I | I-D/D-I | D-D | I-I | I-D/D-I | D-D | I-I | I-D/D-I | D-D | |
| ⋮ | ⋮ | · | · | · | ⋮ | ⋮ | ⋮ | · | · | · | ⋮ | ⋮ | ⋮ | · | · | · | |
| | — | $48^{\geq 2.418}$ | | | | | | | | $48^{\geq 2.198}$ | | | | $48^{\geq 1.913}$ | | | |
| | — | | | | | | | $52^{\geq 2.071}$ | $52^{\geq 2.005}$ | | | | | | | | |
| | — | | $56^{\geq 2.207}$ | | | | | | | | | | | | $56^{\geq 1.308}$ | | |
| | — | | | $60^{\geq 2.335}$ | $60^{\geq 1.974}$ | $60^{\geq 1.974}$ | $60^{\geq 1.974}$ | | | | $60^{\geq 1.485}$ | $60^{\geq 1.485}$ | $60^{\geq 1.485}$ | | | $60^{\geq 1.160}$ | |
| ≥64 | — | | | | $64^{\geq 1.861}$ | $64^{\geq 1.861}$ | $64^{\geq 1.861}$ | | | $64^{\geq 1.938}$ | | | | $64^{\geq 0.804}$ | $64^{\geq 0.561}$ | $64^{w/zeros}$ | |
| | 4 | | | | | | | | $68^{\geq 1.727}$ | | $68^{\geq 1.152}$ | $68^{\geq 1.152}$ | $68^{\geq 1.152}$ | | | | $(16\to17)\cdot(4\to4) = 68$ |
| | 8 | | | | $72^{\geq 1.803}$ | $72^{\geq 1.803}$ | $72^{\geq 1.803}$ | | | | $72^{\geq 1.023}$ | $72^{\geq 1.023}$ | $72^{\geq 1.023}$ | | | | $(8\to9)\cdot(8\to8) = 72$ |
| | 12 | | | | $76^{\geq 1.743}$ | $76^{\geq 1.743}$ | $76^{\geq 1.743}$ | | | | $76^{\geq 0.711}$ | $76^{\geq 0.711}$ | $76^{\geq 0.711}$ | | | | $(16\to19)\cdot(4\to4) = 76$ |
| | 16 | $80^{\geq 2.075}$ | $80^{\geq 2.029}$ | | | | | | | $80^{\geq 1.324}$ | $80^{\geq 0.496}$ | $80^{\geq 0.496}$ | $80^{\geq 0.496}$ | | | | $(4\to5)\cdot(16\to16) = 80$ |
| ≥81 | 17 | | | | | | | | | | $81^{w/zeros}$ | $81^{w/zeros}$ | $81^{w/zeros}$ | | | | $3^4 = (8\to9)\cdot(8\to9) = 81$ |
| | 20 | | | | $84^{\geq 1.555}$ | $84^{\geq 1.555}$ | $84^{\geq 1.555}$ | $84^{\geq 1.495}$ | | | | | | | | | $2^2 21^1 = (16\to21)^{root}\cdot(4\to4)^{affix} = 84$ |
| | 28 | | | $92^{\geq 1.791}$ | $92^{\geq 1.384}$ | $92^{\geq 1.384}$ | $92^{\geq 1.384}$ | | $92^{\geq 1.030}$ | | | | | | | | $2^1 3^2 5^1 = (4\to5)^{root}\cdot(8\to9)^{root}\cdot(2\to2)^{affix} = 90 < 2^2 23^1 = (16\to23)^{root}\cdot(4\to4)^{affix} = 92$ |
| | 32 | | $96^{\geq 1.637}$ | | $96^{\geq 1.273}$ | $96^{\geq 1.273}$ | $96^{\geq 1.273}$ | | | $96^{\geq 0.915}$ | | | | | | | $2^5 3^1 = (2\to3)^{root}\cdot(32\to32)^{affix} = 96$ |
| ≥100 | 36 | | | | $100^{\geq 1.195}$ | $100^{\geq 1.195}$ | $100^{\geq 1.195}$ | $100^{\geq 0.637}$ | $100^{\geq 0.446}$ | $100^{w/zeros}$ | | | | | | | $2^2 5^2 = (4\to5)^{root}\cdot(4\to5)^{root}\cdot(4\to4)^{affix} = 100$ |
| | 44 | | | $108^{\geq 1.584}$ | $108^{\geq 0.931}$ | $108^{\geq 0.931}$ | $108^{\geq 0.931}$ | | | | | | | | | | $2^3 13^1 = (8\to13)^{root}\cdot(8\to8)^{affix} = 104 < 2^2 3^3 = (2\to3)^{root}\cdot(8\to9)^{root}\cdot(4\to4)^{affix} = 108$ |
| | 48 | | $112^{\geq 1.414}$ | | $112^{\geq 0.828}$ | $112^{\geq 0.828}$ | $112^{\geq 0.828}$ | | | | | | | | | | $2^1 5^1 11^1 = (4\to5)^{root}\cdot(8\to11)^{root}\cdot(2\to2)^{affix} = 110 < 2^4 7^1 = (4\to7)^{root}\cdot(16\to16)^{affix} = 112$ |
| | 52 | | | | $116^{\geq 0.577}$ | $116^{\geq 0.577}$ | $116^{\geq 0.577}$ | | | | | | | | | | $2^1 3^1 19^1 = (2\to3)^{root}\cdot(16\to19)^{root}\cdot(2\to2)^{affix} = 114 < 2^2 29^1 = (16\to29)^{root}\cdot(4\to4)^{affix} = 116$ |
| | 56 | | $120^{\geq 1.160}$ | | $120^{\geq 0.404}$ | $120^{\geq 0.404}$ | $120^{\geq 0.404}$ | | | | | | | | | | $(16\to29)^{root}\cdot(4\to4)^{affix} = 116 < 3^2 13^1 = (8\to9)^{root}\cdot(8\to13)^{root} = 117 < 2^3 3^1 5^1 = (2\to3)^{root}\cdot(4\to5)^{root}\cdot(8\to8)^{affix} = 120$ |
| ≥121 | 57 | | | | $121^{w/zeros}$ | $121^{w/zeros}$ | $121^{w/zeros}$ | | | | | | | | | | $11^2 = (8\to11)^{root}\cdot(8\to11)^{root} = 121$ |
| | 60 | | | $124^{\geq 1.089}$ | | | | | | | | | | | | | $2^2 31^1 = (16\to31)^{root}\cdot(4\to4)^{affix} = 124$ |
| | 64 | $128^{\geq 1.227}$ | | | | | | | | | | | | | | | $5^3 = (4\to5)^{root}\cdot(4\to5)^{root}\cdot(4\to5)^{root} = 125 < 2^7 = (1\to2)^{root}\cdot(64\to64)^{affix} = 128$ |
| | 72 | | $136^{\geq 0.849}$ | | | | | | | | | | | | | | $2^2 3^1 11^1 = \cdots = 132 < 3^3 5^1 = (2\to3)^{root}\cdot(4\to5)^{root}\cdot(8\to9)^{root} = 135 < 2^3 17^1 = (1\to2)^{root}\cdot(16\to17)^{root}\cdot(4\to4)^{affix} = 136$ |
| | 76 | | | $140^{\geq 0.756}$ | | | | | | | | | | | | | $2^1 3^1 23^1 = (2\to3)^{root}\cdot(16\to23)^{root}\cdot(2\to2)^{affix} = 138 < 2^2 5^1 7^1 = (4\to5)^{root}\cdot(4\to7)^{root}\cdot(4\to4)^{affix} = 140$ |
| ≥144 | 80 | $144^{\geq 0.528}$ | $144^{\geq 0.370}$ | $144^{w/zeros}$ | | | | | | | | | | | | | $2^4 3^2 = (2\to3)^{root}\cdot(2\to3)^{root}\cdot(16\to16)^{affix} = 144$ |

(view-to-view transits per repartitioned page)

NOTE – The label "w/zeros" instead of a $G_j$ value means including point(s) with no $G_j$.

NOTE – The expression $N_j^{\geq G_j}$ means that there are exact $N_j$ different view-to-view transits ensuring a gain addition of $G_j$ or more (i.e., at least $G_j$) dB.

Linguistic Multiplexing Variants with $\Pi({}^iN_C \to {}^iN_E)/\mathrm{GCD}({}^iN_C, {}^iN_E)\cdot\mathrm{GCD}(\Pi {}^iN_C, \Pi {}^iN_E) \leq N_E$

TABLE IV
SUITABLE LINGUISTIC MULTIPLEXING PROFILES

| ${}^iN_E$ | ${}^iN_C$ | ${}^iN_R$ | ${}^in_e$ | ${}^ik$ | ${}^iN_E$ | ${}^iN_C$ | ${}^iN_R$ | ${}^in_e$ | ${}^ik$ | ${}^iN_E$ | ${}^iN_C$ | ${}^iN_R$ | ${}^in_e$ | ${}^ik$ | ${}^iN_E$ | ${}^iN_C$ | ${}^iN_R$ | ${}^in_e$ | ${}^ik$ |
|---|---|---|---|---|---|---|---|---|---|---|---|---|---|---|---|---|---|---|---|
| 3 | 2 | 1 | 2 | 1 | 5 | 4 | 1 | 4 | 2 | 9 | 8 | 1 | 6 | 3 | 21 | 16 | 5 | 3 | 4 |
| | | type 2→3 root | | | | | type 4→5 root | | | | | type 8→9 root | | | | | type 16→21 root | | |

TABLE V
5.5-POINT/P/Q STELLARIAL CONSTELLATION

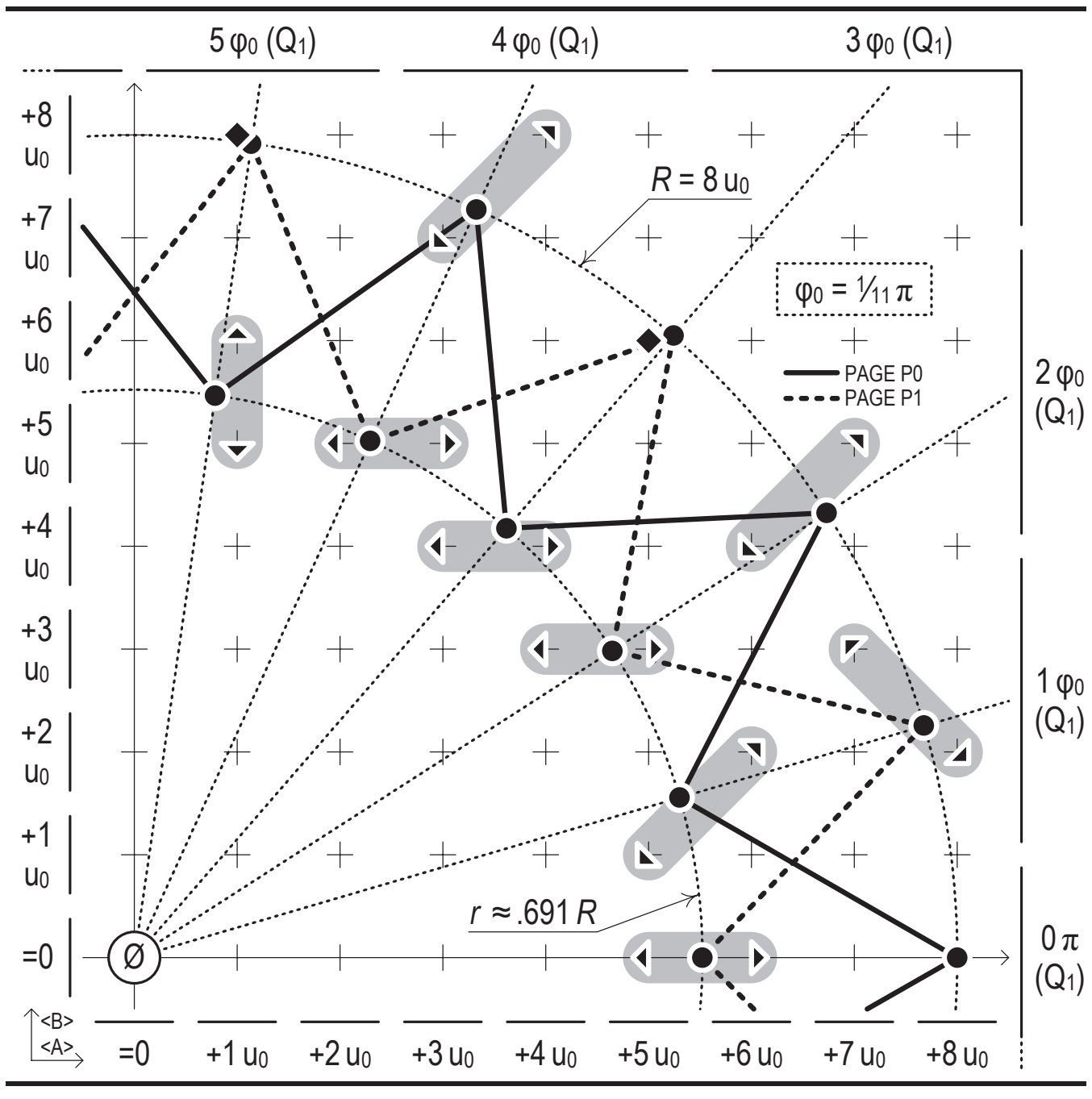

TABLE VI
6-POINT/P/Q STELLARIAL CONSTELLATION

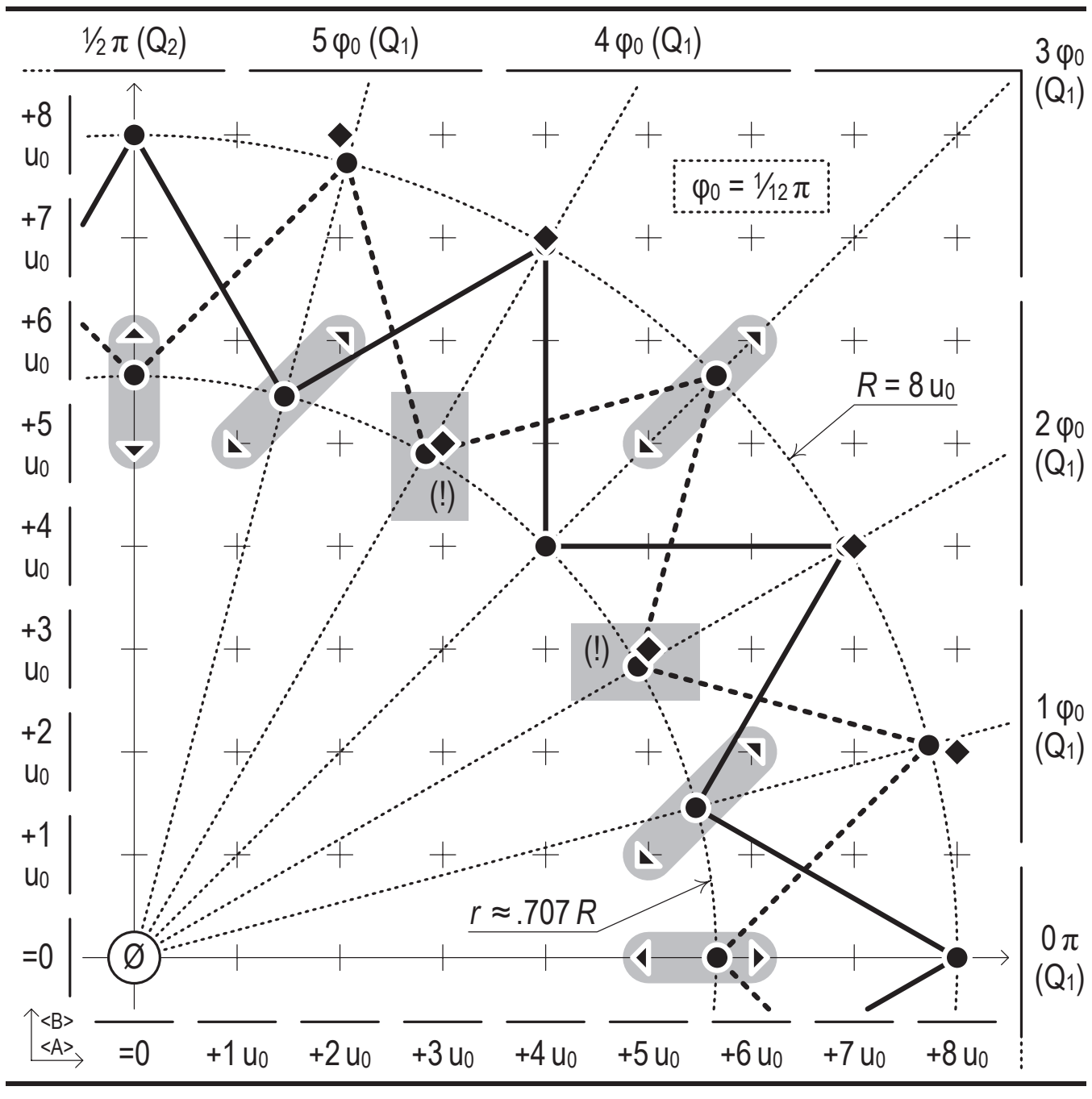

TABLE VII
MDI PROPERTIES WITH CIC DESIGN OPTIONS

| Constellation | $G_j$, dB | MDI Output Power, $W/W_{max}$ | | | | | Output Wobble Power, $W/W_{max}$ | | | | Output Change Power, $\frac{d}{dt}W/\frac{d}{dt}W_{max}$ | | | | | Coding Gain, dB | | |
|---|---|---|---|---|---|---|---|---|---|---|---|---|---|---|---|---|---|---|
| NOTE – Wmax is when all four lines drive ±Uref, never present by design. | | Mean | Dev. | Min | Max | Span | Mean | Dev. | Min | Max | Mean | Dev. | Min | Max | Span | Per-RP | Dict. | Losses due to Wobble |
| 4 Pts/P/Q, Ideal | w/zeros | 0.349 | 0 | | | | completely absent | | | | 0.174 | 0.086 | 0 | 0.349 | 0.349 | 6.2 | 1.1 | |
| 4 Pts/P/Q, Grid | | 0.358 | 0.024 | 0.324 | 0.406 | 0.082 | 0.001 | 0.001 | 0.00001 | 0.002 | 0.164 | 0.091 | 0 | 0.406 | 0.406 | | | −0.1 |
| 4.5 Pts/P/Q, Ideal | w/zeros | 0.357 | 0 | | | | completely absent | | | | 0.178 | 0.088 | 0 | 0.357 | 0.357 | 5.7 | 1.0 | |
| 4.5 Pts/P/Q, Grid | | 0.366 | 0.027 | 0.285 | 0.418 | 0.133 | 0.002 | 0.002 | 0.00011 | 0.006 | 0.183 | 0.092 | 0 | 0.418 | 0.418 | | | −0.2 |
| 4.5 Pts/P/Q, Ideal | ≥0.496 | 0.357 | 0 | | | | completely absent | | | | 0.178 | 0.087 | 0.003 | 0.353 | 0.350 | 6.2 | 1.5 | |
| 4.5 Pts/P/Q, Grid | | 0.366 | 0.027 | 0.285 | 0.418 | 0.133 | 0.002 | 0.002 | 0.00011 | 0.006 | 0.182 | 0.091 | 0.004 | 0.412 | 0.408 | | | −0.2 |
| 4.5 Pts/P/Q, Ideal | ≥0.711 | 0.357 | 0 | | | | completely absent | | | | 0.178 | 0.086 | 0.011 | 0.346 | 0.335 | 6.4 | 1.8 | |
| 4.5 Pts/P/Q, Grid | | 0.366 | 0.027 | 0.285 | 0.418 | 0.133 | 0.002 | 0.002 | 0.00011 | 0.006 | 0.181 | 0.089 | 0.008 | 0.397 | 0.390 | | | −0.2 |
| 4.5 Pts/P/Q, Ideal | ≥1.023 | 0.357 | 0 | | | | completely absent | | | | 0.178 | 0.084 | 0.012 | 0.344 | 0.332 | 6.7 | 2.1 | |
| 4.5 Pts/P/Q, Grid | | 0.366 | 0.027 | 0.285 | 0.418 | 0.133 | 0.002 | 0.002 | 0.00011 | 0.006 | 0.183 | 0.087 | 0.010 | 0.396 | 0.386 | | | −0.2 |
| 4.5 Pts/P/Q, Ideal | ≥1.152 | 0.357 | 0 | | | | completely absent | | | | 0.178 | 0.082 | 0.020 | 0.337 | 0.317 | 6.8 | 2.2 | |
| 4.5 Pts/P/Q, Grid | | 0.366 | 0.027 | 0.285 | 0.418 | 0.133 | 0.002 | 0.002 | 0.00011 | 0.006 | 0.183 | 0.085 | 0.014 | 0.383 | 0.369 | | | −0.2 |
| 5 Pts/P/Q, Ideal | w/zeros | 0.372 | 0 | | | | completely absent | | | | 0.186 | 0.092 | 0 | 0.372 | 0.372 | 5.2 | 1.0 | |
| 5 Pts/P/Q, Grid | | 0.367 | 0.033 | 0.266 | 0.426 | 0.160 | 0.005 | 0.008 | 0.00045 | 0.034 | 0.183 | 0.092 | 0 | 0.426 | 0.426 | | | −0.3 |
| 5 Pts/P/Q, Ideal | ≥0.446 | 0.372 | 0 | | | | completely absent | | | | 0.186 | 0.092 | 0.003 | 0.369 | 0.367 | 6.1 | 1.4 | |
| 5 Pts/P/Q, Grid | | 0.367 | 0.033 | 0.266 | 0.426 | 0.160 | 0.005 | 0.008 | 0.00045 | 0.034 | 0.183 | 0.092 | 0.004 | 0.422 | 0.418 | | | −0.3 |
| 5 Pts/P/Q, Ideal | ≥0.637 | 0.372 | 0 | | | | completely absent | | | | 0.186 | 0.090 | 0.009 | 0.363 | 0.354 | 6.1 | 1.6 | |
| 5 Pts/P/Q, Grid | | 0.367 | 0.033 | 0.266 | 0.426 | 0.160 | 0.005 | 0.008 | 0.00045 | 0.034 | 0.183 | 0.090 | 0.004 | 0.406 | 0.402 | | | −0.3 |
| 5 Pts/P/Q, Ideal | ≥0.915 | 0.372 | 0 | | | | completely absent | | | | 0.186 | 0.088 | 0.011 | 0.361 | 0.350 | 6.1 | 1.9 | |
| 5 Pts/P/Q, Grid | | 0.367 | 0.033 | 0.266 | 0.426 | 0.160 | 0.005 | 0.008 | 0.00045 | 0.034 | 0.183 | 0.089 | 0.004 | 0.406 | 0.402 | | | −0.3 |
| 5 Pts/P/Q, Ideal | ≥1.030 | 0.372 | 0 | | | | completely absent | | | | 0.186 | 0.087 | 0.017 | 0.355 | 0.338 | 6.5 | 2.0 | |
| 5 Pts/P/Q, Grid | | 0.367 | 0.033 | 0.266 | 0.426 | 0.160 | 0.005 | 0.008 | 0.00045 | 0.034 | 0.184 | 0.087 | 0.004 | 0.396 | 0.392 | | | −0.3 |
| 5 Pts/P/Q, Ideal | ≥1.324 | 0.372 | 0 | | | | completely absent | | | | 0.186 | 0.083 | 0.024 | 0.348 | 0.324 | 6.5 | 2.3 | |
| 5 Pts/P/Q, Grid | | 0.367 | 0.033 | 0.266 | 0.426 | 0.160 | 0.005 | 0.008 | 0.00045 | 0.034 | 0.183 | 0.084 | 0.004 | 0.396 | 0.392 | | | −0.3 |
| 5 Pts/P/Q, Ideal | ≥1.495 | 0.372 | 0 | | | | completely absent | | | | 0.186 | 0.082 | 0.024 | 0.348 | 0.324 | 7.2 | 2.4 | |
| 5 Pts/P/Q, Grid | | 0.367 | 0.033 | 0.266 | 0.426 | 0.160 | 0.005 | 0.008 | 0.00045 | 0.034 | 0.183 | 0.082 | 0.004 | 0.396 | 0.392 | | | −0.3 |
| 5.5 Pts/P/Q, Ideal | w/zeros | 0.369 | 0 | | | | completely absent | | | | 0.185 | 0.091 | 0 | 0.369 | 0.369 | 4.9 | 0.9 | |
| 5.5 Pts/P/Q, Grid | | 0.371 | 0.035 | 0.301 | 0.469 | 0.168 | 0.003 | 0.002 | 0.00035 | 0.007 | 0.186 | 0.094 | 0 | 0.469 | 0.469 | | | −0.2 |
| 5.5 Pts/P/Q, Ideal | ≥0.404 | ⋯ | | | | | | | | | 0.185 | 0.091 | 0.002 | 0.367 | 0.364 | 5.3 | 1.3 | |
| 5.5 Pts/P/Q, Grid | | ⋯ | | | | | | | | | 0.185 | 0.093 | 0.004 | 0.457 | 0.453 | | | |
| 5.5 Pts/P/Q, Ideal | ≥0.577 | ⋯ | | | | | | | | | 0.185 | 0.090 | 0.007 | 0.362 | 0.354 | 5.4 | 1.5 | |
| 5.5 Pts/P/Q, Grid | | ⋯ | | | | | | | | | 0.186 | 0.092 | 0.004 | 0.453 | 0.449 | | | |
| 5.5 Pts/P/Q, Ideal | ≥0.828 | ⋯ | | | | | | | | | 0.185 | 0.088 | 0.009 | 0.360 | 0.350 | 5.7 | 1.7 | |
| 5.5 Pts/P/Q, Grid | | ⋯ | | | | | | | | | 0.186 | 0.090 | 0.004 | 0.453 | 0.449 | | | |
| 5.5 Pts/P/Q, Ideal | ≥0.931 | ⋯ | | | | | | | | | 0.185 | 0.087 | 0.015 | 0.355 | 0.340 | 5.8 | 1.8 | |
| 5.5 Pts/P/Q, Grid | | ⋯ | | | | | | | | | 0.185 | 0.089 | 0.008 | 0.439 | 0.432 | | | |
| 5.5 Pts/P/Q, Ideal | ≥1.195 | ⋯ | | | | | | | | | 0.185 | 0.084 | 0.021 | 0.349 | 0.328 | 6.0 | 2.1 | |
| 5.5 Pts/P/Q, Grid | | ⋯ | | | | | | | | | 0.185 | 0.086 | 0.013 | 0.439 | 0.427 | | | |
| 5.5 Pts/P/Q, Ideal | ≥1.273 | ⋯ | | | | | | | | | 0.185 | 0.083 | 0.021 | 0.349 | 0.328 | 6.1 | 2.2 | |
| 5.5 Pts/P/Q, Grid | | ⋯ | | | | | | | | | 0.185 | 0.085 | 0.013 | 0.439 | 0.427 | | | |
| 5.5 Pts/P/Q, Ideal | ≥1.384 | ⋯ | | | | | | | | | 0.185 | 0.082 | 0.026 | 0.344 | 0.318 | 6.2 | 2.2 | |
| 5.5 Pts/P/Q, Grid | | ⋯ | | | | | | | | | 0.184 | 0.083 | 0.017 | 0.423 | 0.406 | | | |
| 5.5 Pts/P/Q, Ideal | ≥1.555 | ⋯ | | | | | | | | | 0.185 | 0.079 | 0.035 | 0.334 | 0.300 | 6.4 | 2.5 | |
| 5.5 Pts/P/Q, Grid | | ⋯ | | | | | | | | | 0.185 | 0.081 | 0.025 | 0.423 | 0.397 | | | |
| 5.5 Pts/P/Q, Ideal | ≥1.743 | ⋯ | | | | | | | | | 0.185 | 0.077 | 0.035 | 0.334 | 0.300 | 6.6 | 2.6 | |
| 5.5 Pts/P/Q, Grid | | ⋯ | | | | | | | | | 0.185 | 0.078 | 0.025 | 0.423 | 0.397 | | | |
| 5.5 Pts/P/Q, Ideal | ≥1.803 | ⋯ | | | | | | | | | 0.185 | 0.075 | 0.040 | 0.329 | 0.289 | 6.7 | 2.7 | |
| 5.5 Pts/P/Q, Grid | | ⋯ | | | | | | | | | 0.185 | 0.077 | 0.029 | 0.400 | 0.371 | | | |
| 5.5 Pts/P/Q, Ideal | ≥1.861 | ⋯ | | | | | | | | | 0.185 | 0.072 | 0.051 | 0.318 | 0.267 | 6.7 | 2.8 | |
| 5.5 Pts/P/Q, Grid | | ⋯ | | | | | | | | | 0.185 | 0.074 | 0.035 | 0.400 | 0.365 | | | |
| 6 Pts/P/Q, Ideal | w/zeros | ⋯ | | | | | | | | | 0.187 | 0.093 | 0 | 0.375 | 0.375 | 4.5 | 0.8 | |
| 6 Pts/P/Q, Grid | | ⋯ | | | | | | | | | 0.191 | 0.096 | 0 | 0.438 | 0.438 | | | |
| 6 Pts/P/Q, Ideal | ≥0.370 | ⋯ | | | | | | | | | 0.187 | 0.093 | 0.002 | 0.373 | 0.371 | 5.3 | 1.2 | |
| 6 Pts/P/Q, Grid | | ⋯ | | | | | | | | | 0.189 | 0.096 | 0.002 | 0.426 | 0.424 | | | |
| 6 Pts/P/Q, Ideal | ≥0.528 | ⋯ | | | | | | | | | 0.187 | 0.092 | 0.006 | 0.369 | 0.362 | 5.3 | 1.4 | |
| 6 Pts/P/Q, Grid | | ⋯ | | | | | | | | | 0.190 | 0.094 | 0.004 | 0.422 | 0.418 | | | |
| 6 Pts/P/Q, Ideal | ≥0.756 | ⋯ | | | | | | | | | 0.187 | 0.090 | 0.008 | 0.367 | 0.358 | 5.3 | 1.6 | |
| 6 Pts/P/Q, Grid | | ⋯ | | | | | | | | | 0.190 | 0.093 | 0.004 | 0.422 | 0.418 | | | |
| 6 Pts/P/Q, Ideal | ≥0.849 | ⋯ | | | | | | | | | 0.187 | 0.089 | 0.013 | 0.362 | 0.350 | 5.6 | 1.7 | |
| 6 Pts/P/Q, Grid | | ⋯ | | | | | | | | | 0.191 | 0.092 | 0.008 | 0.408 | 0.400 | | | |
| 6 Pts/P/Q, Ideal | ≥1.089 | ⋯ | | | | | | | | | 0.187 | 0.087 | 0.018 | 0.357 | 0.338 | 5.6 | 1.9 | |
| 6 Pts/P/Q, Grid | | ⋯ | | | | | | | | | 0.190 | 0.089 | 0.016 | 0.408 | 0.393 | | | |
| 6 Pts/P/Q, Ideal | ≥1.160 | ⋯ | | | | | | | | | 0.187 | 0.086 | 0.018 | 0.357 | 0.338 | 6.1 | 2.0 | |
| 6 Pts/P/Q, Grid | | ⋯ | | | | | | | | | 0.190 | 0.088 | 0.016 | 0.408 | 0.393 | | | |
| 6 Pts/P/Q, Ideal | ≥1.227 | ⋯ | | | | | | | | | 0.187 | 0.085 | 0.023 | 0.352 | 0.330 | 6.1 | 2.1 | |
| 6 Pts/P/Q, Grid | | ⋯ | | | | | | | | | 0.191 | 0.087 | 0.020 | 0.406 | 0.387 | | | |
| 6 Pts/P/Q, Ideal | ≥1.414 | ⋯ | | | | | | | | | 0.187 | 0.082 | 0.031 | 0.344 | 0.313 | 6.1 | 2.3 | |
| 6 Pts/P/Q, Grid | | ⋯ | | | | | | | | | 0.190 | 0.085 | 0.025 | 0.406 | 0.381 | | | |
| 6 Pts/P/Q, Ideal | ≥1.584 | ⋯ | | | | | | | | | 0.187 | 0.080 | 0.031 | 0.344 | 0.313 | 6.1 | 2.4 | |
| 6 Pts/P/Q, Grid | | ⋯ | | | | | | | | | 0.191 | 0.083 | 0.025 | 0.406 | 0.381 | | | |
| 6 Pts/P/Q, Ideal | ≥1.637 | ⋯ | | | | | | | | | 0.187 | 0.079 | 0.036 | 0.339 | 0.304 | 6.3 | 2.5 | |
| 6 Pts/P/Q, Grid | | ⋯ | | | | | | | | | 0.190 | 0.081 | 0.029 | 0.388 | 0.358 | | | |
| 6 Pts/P/Q, Ideal | ≥1.791 | 0.375 | 0 | | | | completely absent | | | | 0.187 | 0.076 | 0.046 | 0.329 | 0.282 | 6.3 | 2.6 | |
| 6 Pts/P/Q, Grid | | 0.382 | 0.027 | 0.293 | 0.438 | 0.145 | 0.001 | 0.001 | 0 | 0.006 | 0.189 | 0.078 | 0.039 | 0.379 | 0.340 | | | −0.1 |
| 6 Pts/P/Q, Ideal | ≥2.029 | 0.375 | 0 | | | | completely absent | | | | 0.187 | 0.074 | 0.046 | 0.329 | 0.282 | 6.9 | 2.9 | |
| 6 Pts/P/Q, Grid | | 0.382 | 0.027 | 0.293 | 0.438 | 0.145 | 0.001 | 0.001 | 0 | 0.006 | 0.190 | 0.076 | 0.039 | 0.379 | 0.340 | | | −0.1 |
| Reference (1000BASE-T) | | 0.313 | 0.151 | 0 | 1 | 1 | n/a | | | | 0.109 | 0.064 | 0 | 0.563 | 0.563 | 6.0 | 2.5 | n/a |

NOTE – For the 5.5 and 6 pts/P/Q cases depicted with omissions, the rest numbers are visually the same, at least for the selected precision, as of the respective options whose cases are depicted without such omissions; the real numbers behind the shown may vary.
NOTE – In the case of 6 pts/P/Q and $G_j$ ≥ 2.029 dB, we can use a variable bit rate (VBR) channel only, never a constant bit rate (CBR) one, but with an average of { 80 (I-I) + 80 (I-D) + 80 (D-I) + 64 (D-D) = }304 / 4 = 76 > 64 jumps per a mean repartitioned page (RP).

TABLE VIII
PER-PAGE CODING SPACE WITH DIFFERENT DESIGN OPTIONS

| Pts/Q/P | Static | $G_J$ | $G_J$+ | $G_{2J}$ | $G_{2J}$+ | $G_{3J}$ | $G_{3J}$+ | $G_{4J}$ | $G_{4J}$+ | $G_j \geq 0.3$ | $G_j \geq 0.4$ | $G_j \geq 0.5$ | $G_j \geq 0.6$ | $G_j \geq 0.7$ | $G_j \geq 0.8$ | $G_j \geq 0.9$ | $G_j \geq 1.0$ | $G_j \geq 1.1$ | $G_j \geq 1.2$ | · · · |
|---|---|---|---|---|---|---|---|---|---|---|---|---|---|---|---|---|---|---|---|---|
| 4 | 64 CBR | | | | | | | | | | | | | | | | | | | |
| 4½ | 81 CBR | 72 CBR | 64 CBR | | | | | | | | 80 CBR | | | 76 CBR | | | 72 CBR | 68 CBR | | |
| 5 | 100 CBR | 90 CBR | 81 CBR | 72 CBR | 64 CBR | | | | | | 96÷100 CBR or VBR | | 92÷100 CBR or VBR | | | 84÷96 CBR or VBR | 80÷92 CBR or VBR | | | ••• |
| 5½ | 121 CBR | 110 CBR | 100 CBR | 90 CBR | 81 CBR | 72 CBR | 64 CBR | | | | 120 CBR | 116 CBR | | | 112 CBR | 108 CBR | | 100 CBR | 96 CBR | ••• |
| 6 | 144 CBR | 132 CBR | 121 CBR | 110 CBR | 100 CBR | 90 CBR | 81 CBR | 72 CBR | 64 CBR | 140÷144 CBR or VBR | | 136÷144 CBR or VBR | | 128÷140 CBR or VBR | 124÷136 CBR or VBR | | 120÷128 CBR or VBR | 108÷128 CBR or VBR | 108÷128 CBR or VBR | ••• |
| Ref. | 64 | | | | | | | | | | | | | | | | | | | |
| | Nom. | Dynamic/Effective (...) | | | | | | | | ... (up to the respective number may be used, but not less than 64 pts/RP for each of eight RPs/Dictionary) | | | | | | | | | | |

TABLE IX
MDI PROPERTIES WITH $G_{*J}/G_{*J}+$ DESIGN OPTIONS

| Constellation | MDI Output Power, W/$W_{max}$ | | | | | Output Wobble Power, W/$W_{max}$ | | | | Output Change Power, $\frac{d}{dt}W/\frac{d}{dt}W_{max}$ | | | | | Coding Gain, dB | | |
|---|---|---|---|---|---|---|---|---|---|---|---|---|---|---|---|---|---|
| NOTE – Min, Max, and Span are for lower, upper, and inter-peak values. | Mean | Dev. | Min | Max | Span | Mean | Dev. | Min | Max | Mean | Dev. | Min | Max | Span | Per-RP | Dict. | Losses due to Wobble |
| 5 Pts/P/Q, Ideal | 0.372 | 0 | | | | completely absent | | | | 0.186 | 0.092 | 0 | 0.372 | 0.372 | 5.2 | 1.0 | |
| 5 Pts/P/Q, Grid | 0.367 | 0.033 | 0.266 | 0.426 | 0.160 | 0.005 | 0.008 | $0.000_{45}$ | 0.034 | 0.183 | 0.092 | 0 | 0.426 | 0.426 | | | −0.3 |
| 5 Pts/P/Q, Ideal, $G_J$ | 0.372 | 0 | | | | completely absent | | | | 0.191 | 0.091 | 0.003 | 0.372 | 0.369 | 6.1 | 1.4 | |
| 5 Pts/P/Q, Grid, $G_J$ | 0.367 | 0.033 | 0.266 | 0.426 | 0.160 | 0.005 | 0.008 | $0.000_{45}$ | 0.034 | 0.188 | 0.091 | 0.004 | 0.426 | 0.426 | | | −0.3 |
| 5 Pts/P/Q, Ideal, $G_J$+ | 0.372 | 0 | | | | completely absent | | | | 0.196 | 0.090 | 0.009 | 0.372 | 0.363 | 6.5 | 1.6 | |
| 5 Pts/P/Q, Grid, $G_J$+ | 0.367 | 0.033 | 0.266 | 0.426 | 0.160 | 0.005 | 0.008 | $0.000_{45}$ | 0.034 | 0.192 | 0.090 | 0.008 | 0.426 | 0.418 | | | −0.3 |
| 5 Pts/P/Q, Ideal, $G_{2J}$ | 0.372 | 0 | | | | completely absent | | | | 0.197 | 0.087 | 0.011 | 0.369 | 0.358 | 6.1 | 1.9 | |
| 5 Pts/P/Q, Grid, $G_{2J}$ | 0.367 | 0.033 | 0.266 | 0.426 | 0.160 | 0.005 | 0.008 | $0.000_{45}$ | 0.034 | 0.194 | 0.087 | 0.004 | 0.422 | 0.418 | | | −0.3 |
| 5 Pts/P/Q, Ideal, $G_{2J}$+ | 0.372 | 0 | | | | completely absent | | | | 0.207 | 0.080 | 0.036 | 0.363 | 0.328 | 6.5 | 2.3 | |
| 5 Pts/P/Q, Grid, $G_{2J}$+ | 0.367 | 0.033 | 0.266 | 0.426 | 0.160 | 0.005 | 0.008 | $0.000_{45}$ | 0.034 | 0.204 | 0.081 | 0.010 | 0.403 | 0.394 | | | −0.3 |
| 5.5 Pts/P/Q, Ideal | 0.369 | 0 | | | | completely absent | | | | 0.185 | 0.091 | 0 | 0.369 | 0.369 | 4.9 | 0.9 | |
| 5.5 Pts/P/Q, Grid | 0.371 | 0.035 | 0.301 | 0.469 | 0.168 | 0.003 | 0.002 | $0.000_{35}$ | 0.007 | 0.186 | 0.094 | 0 | 0.469 | 0.469 | | | −0.2 |
| 5.5 Pts/P/Q, Ideal, $G_J$ | 0.369 | 0 | | | | completely absent | | | | 0.189 | 0.090 | 0.002 | 0.369 | 0.367 | 5.7 | 1.3 | |
| 5.5 Pts/P/Q, Grid, $G_J$ | 0.371 | 0.035 | 0.301 | 0.469 | 0.168 | 0.003 | 0.002 | $0.000_{35}$ | 0.007 | 0.188 | 0.092 | 0.004 | 0.469 | 0.465 | | | −0.2 |
| 5.5 Pts/P/Q, Ideal, $G_J$+ | 0.369 | 0 | | | | completely absent | | | | 0.193 | 0.089 | 0.007 | 0.369 | 0.362 | 6.0 | 1.5 | |
| 5.5 Pts/P/Q, Grid, $G_J$+ | 0.371 | 0.035 | 0.301 | 0.469 | 0.168 | 0.003 | 0.002 | $0.000_{35}$ | 0.007 | 0.194 | 0.091 | 0.008 | 0.469 | 0.461 | | | −0.2 |
| 5.5 Pts/P/Q, Ideal, $G_{2J}$ | 0.369 | 0 | | | | completely absent | | | | 0.194 | 0.087 | 0.009 | 0.367 | 0.357 | 5.7 | 1.7 | |
| 5.5 Pts/P/Q, Grid, $G_{2J}$ | 0.371 | 0.035 | 0.301 | 0.469 | 0.168 | 0.003 | 0.002 | $0.000_{35}$ | 0.007 | 0.195 | 0.088 | 0.004 | 0.457 | 0.453 | | | −0.2 |
| 5.5 Pts/P/Q, Ideal, $G_{2J}$+ | 0.369 | 0 | | | | completely absent | | | | 0.204 | 0.081 | 0.029 | 0.362 | 0.333 | 6.0 | 2.1 | |
| 5.5 Pts/P/Q, Grid, $G_{2J}$+ | 0.336 | 0.036 | 0.301 | 0.469 | 0.168 | 0.003 | 0.002 | $0.000_{35}$ | 0.007 | 0.204 | 0.083 | 0.017 | 0.432 | 0.415 | | | −0.2 |
| 5.5 Pts/P/Q, Ideal, $G_{3J}$ | 0.369 | 0 | | | | completely absent | | | | 0.209 | 0.084 | 0.021 | 0.369 | 0.394 | 6.6 | 2.2 | |
| 5.5 Pts/P/Q, Grid, $G_{3J}$ | 0.371 | 0.035 | 0.301 | 0.469 | 0.168 | 0.003 | 0.002 | $0.000_{35}$ | 0.007 | 0.210 | 0.086 | 0.013 | 0.469 | 0.456 | | | −0.2 |
| 5.5 Pts/P/Q, Ideal, $G_{3J}$+ | 0.369 | 0 | | | | completely absent | | | | 0.233 | 0.074 | 0.064 | 0.369 | 0.306 | 7.4 | 2.8 | |
| 5.5 Pts/P/Q, Grid, $G_{3J}$+ | 0.336 | 0.036 | 0.301 | 0.469 | 0.168 | 0.003 | 0.002 | $0.000_{35}$ | 0.007 | 0.234 | 0.076 | 0.039 | 0.469 | 0.430 | | | −0.2 |
| 6 Pts/P/Q, Ideal | 0.375 | 0 | | | | completely absent | | | | 0.187 | 0.093 | 0 | 0.375 | 0.375 | 4.5 | 0.8 | |
| 6 Pts/P/Q, Grid | 0.382 | 0.027 | 0.293 | 0.438 | 0.145 | 0.001 | 0.001 | 0 | 0.006 | 0.191 | 0.096 | 0 | 0.438 | 0.438 | | | −0.1 |
| 6 Pts/P/Q, Ideal, $G_J$ | 0.375 | 0 | | | | completely absent | | | | 0.191 | 0.092 | 0.002 | 0.375 | 0.373 | 5.3 | 1.2 | |
| 6 Pts/P/Q, Grid, $G_J$ | 0.382 | 0.027 | 0.293 | 0.438 | 0.145 | 0.001 | 0.001 | 0 | 0.006 | 0.094 | 0.095 | 0.002 | 0.438 | 0.436 | | | −0.1 |
| 6 Pts/P/Q, Ideal, $G_J$+ | 0.375 | 0 | | | | completely absent | | | | 0.195 | 0.091 | 0.006 | 0.375 | 0.369 | 5.6 | 1.4 | |
| 6 Pts/P/Q, Grid, $G_J$+ | 0.382 | 0.027 | 0.293 | 0.438 | 0.145 | 0.001 | 0.001 | 0 | 0.006 | 0.198 | 0.094 | 0.006 | 0.438 | 0.432 | | | −0.1 |
| 6 Pts/P/Q, Ideal, $G_{2J}$ | 0.375 | 0 | | | | completely absent | | | | 0.196 | 0.089 | 0.008 | 0.373 | 0.364 | 5.3 | 1.6 | |
| 6 Pts/P/Q, Grid, $G_{2J}$ | 0.382 | 0.027 | 0.293 | 0.438 | 0.145 | 0.001 | 0.001 | 0 | 0.006 | 0.200 | 0.091 | 0.004 | 0.426 | 0.422 | | | −0.1 |
| 6 Pts/P/Q, Ideal, $G_{2J}$+ | 0.375 | 0 | | | | completely absent | | | | 0.205 | 0.084 | 0.025 | 0.369 | 0.343 | 5.6 | 1.9 | |
| 6 Pts/P/Q, Grid, $G_{2J}$+ | 0.382 | 0.027 | 0.293 | 0.438 | 0.145 | 0.001 | 0.001 | 0 | 0.006 | 0.208 | 0.086 | 0.020 | 0.407 | 0.388 | | | −0.1 |
| 6 Pts/P/Q, Ideal, $G_{3J}$ | 0.375 | 0 | | | | completely absent | | | | 0.210 | 0.086 | 0.018 | 0.375 | 0.357 | 6.1 | 2.0 | |
| 6 Pts/P/Q, Grid, $G_{3J}$ | 0.382 | 0.027 | 0.293 | 0.438 | 0.145 | 0.001 | 0.001 | 0 | 0.006 | 0.213 | 0.089 | 0.016 | 0.438 | 0.422 | | | −0.1 |
| 6 Pts/P/Q, Ideal, $G_{3J}$+ | 0.375 | 0 | | | | completely absent | | | | 0.232 | 0.077 | 0.055 | 0.375 | 0.320 | 6.9 | 2.5 | |
| 6 Pts/P/Q, Grid, $G_{3J}$+ | 0.382 | 0.027 | 0.293 | 0.438 | 0.145 | 0.001 | 0.001 | 0 | 0.006 | 0.236 | 0.080 | 0.049 | 0.438 | 0.389 | | | −0.1 |
| 6 Pts/P/Q, Ideal, $G_{4J}$ | 0.375 | 0 | | | | completely absent | | | | 0.216 | 0.081 | 0.031 | 0.373 | 0.342 | 6.1 | 2.4 | |
| 6 Pts/P/Q, Grid, $G_{4J}$ | 0.382 | 0.027 | 0.293 | 0.438 | 0.145 | 0.001 | 0.001 | 0 | 0.006 | 0.220 | 0.084 | 0.025 | 0.426 | 0.400 | | | −0.1 |
| 6 Pts/P/Q, Ideal, $G_{4J}$+ | 0.375 | 0 | | | | completely absent | | | | 0.245 | 0.066 | 0.094 | 0.369 | 0.275 | 6.9 | 3.2 | |
| 6 Pts/P/Q, Grid, $G_{4J}$+ | 0.382 | 0.027 | 0.293 | 0.438 | 0.145 | 0.001 | 0.001 | 0 | 0.006 | 0.250 | 0.069 | 0.070 | 0.407 | 0.337 | | | −0.1 |